\documentclass[oneside,12pt]{article}
\usepackage[english]{babel}
\usepackage{lmodern}
\usepackage[T1]{fontenc} 
\usepackage{leftidx} 
\usepackage{enumitem}   
\usepackage{float}
\usepackage{graphics,latexsym,amsfonts}    
\usepackage{amssymb}   
\usepackage{amssymb,amsthm,hyperref,mathtools} 
\usepackage{cite}
\usepackage{mathrsfs} 
\usepackage{diagbox}    
 
\usepackage[dvipsnames]{xcolor}
\usepackage{ulem}   
\hypersetup{
    colorlinks, 
    linkcolor={red!50!black},  
    citecolor={blue!50!black}, 
    urlcolor={blue!80!black} 
}
\usepackage[longnamesfirst]{natbib}  

\usepackage[displaymath, mathlines]{lineno}  % elsevier
\usepackage{microtype} 
\usepackage{mathabx}  
\renewcommand{\emph}{\textsl}

\def\reference#1{\href{#1}{Cliquer ici pour voir une r\'ef\'erence.}} 
\usepackage{siunitx}
\usepackage[toc,page]{appendix}
\usepackage{epigraph}
\usepackage{picture}
\usepackage{sectsty}  

\usepackage{pstricks} 
\usepackage{pst-plot}
\usepackage{pst-plot}
\usepackage{pst-func}
\usepackage{pst-node}
\usepackage{pst-math}
\usepackage{setspace}
\usepackage{listings}
\usepackage{epstopdf}
\usepackage{fancyhdr}
\usepackage{multirow}
\usepackage[section]{placeins}
\usepackage{rotating} 
\usepackage{graphicx} 
\usepackage{epstopdf}
\usepackage{graphics,latexsym,amsfonts}
\usepackage[affil-it]{authblk}
\usepackage{lipsum}
\sectionfont{\normalfont\scshape\centering}
\subsectionfont{\normalfont\itshape\centering}
\providecommand{\U}[1]{\protect \rule{.1in}{.1in}}
{\normalfont\itshape}
\usepackage[top=1.25in,bottom=1.25in,left=1.25in]{geometry}
\def\vs{\vspace{-0,1cm}}

\def\L{\mathscr{L}}
\def\X{\mathscr{X}}
\def\es{\varnothing}

\usepackage[longnamesfirst]{natbib}

\makeatletter
 
\newtheoremstyle{mytheoremstyle} % name
    {\topsep}                    % Space above
    {\topsep}                    % Space below
    {\itshape}                   % Body font
    {}                           % Indent amount
    {\scshape}                   % Theorem head font
    {.}                          % Punctuation after theorem head
    {.5em}                       % Space after theorem head
    {}  % Theorem head spec (can be left empty, meaning ‘normal’)

\theoremstyle{mytheoremstyle}
\newtheorem{theorem}{Theorem} % reset theorem numbering for each chapter
\newtheorem{proposition}{Proposition}
\newtheorem{lemma}{Lemma}
\newtheorem{corollary}{Corollary}

\newtheoremstyle{mydefinitionstyle} % name
    {\topsep}                    % Space above
    {\topsep}                    % Space below
    {}                   % Body font
    {}                           % Indent amount
    {\scshape}                   % Theorem head font
    {.}                          % Punctuation after theorem head
    {.5em}                       % Space after theorem head
    {}  % Theorem head spec (can be left empty, meaning ‘normal’)
\theoremstyle{mydefinitionstyle}
\newtheorem{definition}{Definition} % definition numbers are dependent on theorem numbe
\newtheorem{example}{Example} % same for example numbers

\newtheorem{remark}{Remark}
\newtheorem*{theorem*}{Theorem*}

\makeatletter

\setlist[description]{font=\normalfont\scshape}
\title{Semantics meets attractiveness:\\
Choice by salience\thanks{
The authors wish to thank Davide Carpentiere for several suggestions. 
Alfio Giarlotta gratefully acknowledges the support of \``Ministero dell'Istruzione, dell'Universit\`a e della Ricerca (MIUR) -- PRIN 2017'', project \textit{Multiple Criteria Decision Analysis and Multiple Criteria Decision Theory}, grant 2017CY2NCA.
Additional acknowledgements will be mentioned in the final draft.}}  
\author{\textsc{Alfio Giarlotta}\thanks{Department of Economics and Business, University of Catania, Italy. alfio.giarlotta@unict.it},
\textsc{Angelo Petralia}\thanks{Collegio Carlo Alberto \& University of Turin, Italy. angelo.petralia@carloalberto.org %\textit{(corresponding author)}
}, \textsc{Stephen Watson}\thanks{Department of Mathematics and Statistics, York University, Canada. swatson@yorku.ca}} 

\usepackage{fancyhdr} 

\begin{document} 

\date{}
\sloppy
%\linenumbers
\maketitle

\renewcommand\abstractname{\normalfont\scshape}
\vspace{-1cm}
\begin{abstract}
\noindent
We describe a context-sensitive model of choice, in which the selection process is shaped not only by the attractiveness of items but also by their semantics (`salience').  
Items are ranked according to a binary relation of salience, and a linear order is associated to each item.
The selection of a unique element from a menu is justified by one of the linear orders associated to the most salient items in the menu.   
%The restricted model is independent from most bounded rationality approaches, although it is a special case of `choice with limited attention' of \cite{MasatliogluNakajimaOzbay2012}. 
The general model provides a structured explanation for any behavior, and allows us to model a notion of `moodiness' of the decision maker, typical of choices requiring as many distinct rationales as items. 
%We prove that 
Asymptotically, all choices are moody. 
%Our approach is a semantic refinement of the `rationalization by multiple rationales' of \cite{KalaiRubinsteinSpiegler2002}, because it reveals the underlying structure of the family of rationales.   
%Our model shares important features of both the rationalization by multiple rationales of \cite{KalaiRubinsteinSpiegler2002} and the sequential rationalization of \cite{ManziniMariotti2007}.  
We single out a model of linear salience, in which the order encoding semantics is transitive and complete. %called WARP(S). 
Choices rationalizable by linear salience can only exhibit non-conflicting violations of WARP.
%Linear salience is a rational specification of the well-known model of choice with limited attention of \citet[AER]{MasatliogluNakajimaOzbay2012}.  
Numerical estimates show the sharp selectivity of this testable model. 
%We provide a multiple characterization of linear salience  based on the asymmetry of revealed salience relation, nonconflicting violations of WARP, and a behavioral axiom, called WARP(S).  
% Choice rationalizability is characterized by the asymmetry of a relation of revealed salience second order rationality: violations of Axiom$\:\alpha$ (equivalently, WARP), if any, are pairwise nonconflicting. 

\medskip


\medskip
\end{abstract}

%%%%%%%%%%%%%%%%%%%%%%%%%%%%%%%%%
%%%%%%%%%%%%%%%%%%%%%%%%%%%%%%%%%
%%%%%%%%%%%%%%%%%%%%%%%%%%%%%%%%%
%%%%%%%%%%%%%%%%%%%%%%%%%%%%%%%%%

\section*{Introduction} \label{SECTION:intro}

In this paper we describe an approach to individual choice, in which the salience of some alternatives forges the decision maker's (DM's) judgement.
Our main assumption is that each alternative can be looked at from two different points of view:\vs\vs 
\begin{itemize}
	\item[(1)] `semantics', related to the information provided by the item;\vs\vs
	\item[(2)] `attractiveness', related to the possibility of being selected.\vs\vs 
\end{itemize} 
These two aspects are typically unrelated: for instance, the item frog's legs in a restaurant menu may be unattractive to me (and so I will never select it), and yet it catches my attention, delivering important information about the chef's skills (and so convincing me to order an item that I would otherwise avoid). 
 The informative content of special items in a menu is emphasized by \cite{Sen1993}:\vs\vs 
\begin{quote}\small\textsl{What\;is\;offered\;for\;choice\;can\;give\;us\;information\;about\;the\;underlying\;situation, and can thus influence our preferences over the alternatives, as we see them.}\vs\vs
\end{quote}

We describe the semantics of alternatives by means of a binary relation of \textsl{salience}, which provides an ordinal evaluation of how intriguing an item is when compared to a different one.  
%Note that there may be various reasons for which some alternatives have a higher epistemic value than others: for instance, the presence of a certain item in a restaurant menu may convey the idea that the chef is very skilled, or the quality of food raw materials is excellent, or the management cares about presenting modern and unconventional dishes, etc.  
Note that some items may display a similar salience (indifference), whereas some others may carry semantically dissimilar salience (incomparability). 

The idea that special items in a menu may affect individual judgements is not new; what is the new is how this feature is modeled. 
\cite{Kreps1979} characterizes \textsl{preferences for flexibility}, in which any menu is weakly preferred to its subsets, and the union of two menus may be strictly preferred to each of them. 
From an opposite perspective, \cite{GulPsendorfer} describe \textsl{preferences for commitment}, in which a DM may strictly prefer a proper submenu to a menu in order to avoid temptation. 
\cite{MasatliogluOk2005} design a rational choice model with \textsl{status quo bias}: in each menu, the choice is affected by the item selected as the default option. 
All these models suggest the influence of special items on the choice process, but they do not explicitly refer to the informativeness of alternatives.

In psychology, the effects of salient information on judgement is first documented  by \cite{TaylorFiske}, who, rephrasing~\cite{TverskyKahneman1974}, write:\vs\vs

\begin{quote}\small \textsl{Instead of reviewing all the evidence that bears upon a particular problem, people frequently use the information which is most salient or available to them, that is, that which is most easily brought to mind.}\vs\vs  		
\end{quote}
Along this path, \cite{BordaloGennaioliShleifer2012,BordaloGennaioliShleifer2013} describe a DM whose attention is captured by the salience of the attributes that evaluate alternatives. 
The authors argue that attention may only be focused on some specific aspects of the environment (e.g., quality and price), and the DM inflates the relative weights attached to the more salient attributes in the process of choosing among alternatives. 
%Empirical evidence --in the form of a decoy effect, compromise effect, and context-dependent willingness to pay-- supports their approach. 

In our model of \textsl{choice by salience}, we use multiple rationales (linear orders) to explain choice behavior, where each rationale is labeled by an item of the ground set.  
Salience is encoded by a binary relation that describes how the DM's attention is focused on items. 
Thus, differently from \cite{BordaloGennaioliShleifer2013}, we define a notion of salience for items, rather than for attributes; moreover, we only give an ordinal priority of consideration rather than a cardinal evaluation of salience. 

More formally, a salient justification for a choice on $X$ consists of a pair $\langle \succsim, \L \rangle$, where $\succsim$ is the salience order on $X$, and $\L = \{\rhd_x : x \in X\}$ is a family of linear orders on $X$.  
Salience guides choice by pointing at the linear orders that may be used to justify the selection from a menu $A$: first the DM identifies the most salient elements of $A$, and then she rationalizes $A$ by choosing one among the linear orders associated to the maximally salient elements of $A$.     

To illustrate how the model of choice by salience works, we use a famous example due to \cite{LuceRaiffa}.    
 
\begin{example}[\it Luce and Raiffa's dinner] \label{EX:Luce&Raiffa}
	Thea selects a main course from a restaurant menu. 
She prefers steak ($s$) over chicken ($c$), provided that steak is appropriately cooked; moreover, she is not interested in exotic dishes such as frog's legs ($f$). 
We observe that Thea chooses chicken over steak when they are the only available items, but selects steak if also frog's legs are in the menu.  
This happens because having frog's legs in the menu is perceived by Thea as a sign that the chef knows how to grill a steak. 
Formally, if $X=\lbrace c,f,s\rbrace$ is the set of items, Thea's preferences are described by the linear order $s \rhd c \rhd f$, and her observed choice is $cf\underline{s},\, \underline{c}s,\,f \underline{s},\,\underline{c}f\,$, where the item selected from each menu is underlined. 
This choice is not rationalizable by a single binary relation, % (in particular, the linear order $\rhd$ does not suffice),
 because it violates Axiom$\:\alpha$ \citep{Chernoff1954}.  

%Linear 
Salience explains Thea's choice behavior by means of two binary rationales.     
To that end, let $\succsim$ be the (transitive and complete) salience order on $X$ defined by $f\succ c$, $f \succ s$, and $c \sim s$, where $\succ$ means `is strictly more salient', and $\sim$ stands for `has the same salience as'. 
% $f$ is more salient than $s$ and $c$, and $s$ and $c$ are being equally salient.   
%This generates a partition of $X$ in two classes of salience, the `upper' $U=\{f\}$, and the `lower' $L=\{s,c\}$. 
Furthermore, let $\L = \{\rhd_c,\rhd_f,\rhd_s\}$ be the family of linear orders on $X$ such that $c \rhd_{c} s \rhd_{c} f$, $s \rhd_{f} c \rhd_{f} f$, and $\rhd_s \!= \rhd_c$. 
%(The equality between $\rhd_c$ and $\rhd_s$ follows from normality, since $c \sim s$.)   
%Thus, as a \textit{set}, $\L$ has two elements, the upper rationale $\rhd_f$ and the lower rationale $\rhd_c = \rhd_s$.  
Selection from any menu $A$ is then explained by maximizing the linear order in $\L$ indexed by the most salient item of $A$.  
For instance, for $A = \{c,f,s\}$, the most salient item in $A$ is $f$, and the maximization of $\rhd_{f}$ justifies the selection of $s$.  
Similarly, $s$ and $c$ are the most salient items in $A = \{s,c\}$, hence maximizing $\rhd_s \!= \rhd_{c}$ explains the selection of $c$. 
\end{example}

Luce and Raiffa's dinner is also used by \cite{KalaiRubinsteinSpiegler2002} to illustrate their choice model of \textsl{rationalization by multiple rationales} \textsl{(RMR)}. 
According to their approach, the DM is allowed to use several rationales (linear orders) to justify her choice: she selects from each menu the unique element that is maximal according to one (\textit{any}) of these preferences.  
The family of linear orders carries no structure, and the selection of a rationalizing order among the available ones is independent of the menu itself. 
In fact, they write (p.\,2287):\vs\vs   
\begin{quote} \small 
	\textsl{We fully acknowledge the crudeness of our approach. 
	The appeal of the RMR proposed for ``Luce and Raiffa's dinner'' does not emanate only from its small number of orderings, but also from the simplicity of describing in which cases each of them is applied. [...] 
	More research is needed to define and investigate ``structured'' forms of rationalization.}\vs\vs 
\end{quote}
Our approach based on salience reveals the hidden structure of the set of rationales.

Choice by salience is also related to those bounded rationality models that use `sequentiality' to explain behavior, e.g., (i) the \textsl{sequential rationalization} of \cite{ManziniMariotti2007}, (ii) the model of \textsl{choice with limited attention} of \cite{MasatliogluNakajimaOzbay2012}, (iii) the \textsl{theory of rationalization} of \cite{CherepanovFeddersenSandroni2013}, and (iv) the model of \textsl{list-rational choice} due to \cite{Yildiz2016}.  
The underlying general principle of all these models is the same: the DM's selection from each menu is performed by successive rounds of contraction of the menu, eventually selecting a single item.  
Specifically, a menu is shrunk by either (i) maximizing two or more acyclic binary relations always considered in the same order, or (ii) applying a suitable choice correspondence (\textsl{attention filter}) first and a linear order successively, or (iii) applying a choice correspondence satisfying Axiom$\,\alpha$ (\textsl{psychological constraint}) first and a linear order successively, or (iv) sequentially comparing (and eliminating) pairs of items through an asymmetric binary relation.

The model of choice by salience draws a bridge between the two different categories of bounded rationality approaches described in the two preceding  paragraphs, namely the non-testable RMR model and the mentioned sequential models: we achieve this goal by separately encoding semantics (via the salience order) and attractiveness (via the  rationales assigned to alternatives).  
Moreover, our approach explains well-known \textsl{behavioral anomalies}, such as the decoy effect, % the introduction of an asymmetrically dominated alternative in a menu may change the DM's mind, and induce her to select the dominating alternatives, possibly contradicting features of rationality.
the compromise effect, and the handicapped avoidance.
% that is, a tendency to select an intermediate alternative in a menu rather than an extreme one.
%

On a more technical side, the only assumption that we make about the salience order is the satisfaction of a minimal feature of rationality, namely the \textsl{acyclicity} of its asymmetric part.  
The level of refinement of this binary relation is not fixed \textit{a priori}; in fact, it depends on the DM's preference structure and the context of the choice problem. 
%As a consequence, the salience order may range from being extremely selective (a linear order) to being almost irrelevant in the process.  
In this paper, we describe the general approach based on salience, and then a specification of it, called `linear'. 

In the general model of choice by salience, no additional assumption is made.  
This flexibility -- which is purely \textsl{endogenous}, insofar as determined by the DM's attention structure -- entails rationalizability of any observed choice behavior.
It can be shown that there exist choices requiring as many distinct rationales as the number of items in the ground set: we label all these choices as expressive of a DM's `moody behavior'. 
We show that moodiness is rare on a small number of alternatives.
However, and possibly not surprisingly,\footnote{On the other hand, the proof of this fact is surprisingly technical: see Appendix~B.} this feature becomes the norm for large sets.
In fact, as the number of items diverges to infinity, the fraction of moody choices tends to one. 

In the linear model, we require that (1) the salience order is transitive and complete, and (2) 
%This yields a linear ordering on the family of equivalence classes of salience, which are ranked from the most salient to the least salient.   
%Upon imposing a condition of `normality' --that is, 
all linear orders indexed by indifferent items are equal. %--, we  may associate a linear rationale to each indifference class, and so every menu is rationalized by maximizing a single rationale. 
This linear variant is independent from most existing models of bounded rationality, being however a special case of the \textsl{choice with limited attention} of \cite{MasatliogluNakajimaOzbay2012}, being characterized by a property of the correspondent attention filter. 
%We show that that choices rationalizable by linear salience are characterized (among those with limited attention) by a more stringent structure of attention filters.  

\smallskip

The paper is organized as follows. 
Section~\ref{SECTION:preliminaries} collects preliminary notions.  
In Section~\ref{SECTION:general_model} we describe the general approach of choice by salience, showing that moodiness exists (Theorem~\ref{THM:exists_chaos}) and asymptotically prevails (Theorem~\ref{THM:chaos_rules}).
In Section~\ref{SECTION:restricted_model} we discuss the linear model, and provide a multiple characterization of it (Theorem~\ref{THM:main}).  
Section~\ref{SECTION:relation with literature} compares our approach to the existing literature.
Section~\ref{SECTION:conclusion} collects final remarks and possible directions of research.
All proofs are in Appendix~A, with the exception of the long proof of Theorem~\ref{THM:chaos_rules}, which is in Appendix~B.   
Appendix~C shows that linear salience is independent of some models of bounded rationality.  

%%%%%%%%%%%%%%%%%%%%%%%%%%%%%%%%%%%%%%%%%%%%%%%%
%%%%%%%%%%%%%%%%%%%%%%%%%%%%%%%%%%%%%%%%%%%%%%%%
%%%%%%%%%%%%%%%%%%%%%%%%%%%%%%%%%%%%%%%%%%%%%%%%
%%%%%%%%%%%%%%%%%%%%%%%%%%%%%%%%%%%%%%%%%%%%%%%%

\section{Preliminaries}\label{SECTION:preliminaries}

For readers' convenience, here we collect all basic notions about choice and preference. 
A finite nonempty set $X$ of alternatives (\textsl{ground set}) is fixed throughout. 
We denote by $\X$ the family of all nonempty subsets of $X$, and call any $A$ in $\X$  a \textsl{menu}.
Elements of a menu are often referred to as \textsl{items}. 
A \textsl{choice correspondence} on $X$ is a map $\Gamma \colon \X \rightarrow \X$ that selects some items (at least one) from each menu, that is, $\es \neq \Gamma(A) \subseteq A$ for any $A\in\X$.
A \textsl{choice function} is a choice correspondence in which a unique item is selected from each menu; thus, we may identify it with a map $c \colon \X\rightarrow X$ such that $c(A)\in A$ for any $A\in\X$.
Here we mostly deal with choice functions, and only occasionally refer to correspondences; thus, unless confusion may arise, we use `choice' in place of `choice function'.\footnote{To further distinguish choice functions from choice correspondences, we use lower case Roman letters for the former, and upper case Greek letters for the latter.} 
To simplify notation, we often omit set delimiters and commas: for instance, $A \cup x$ stands for $A \cup \{x\}$, $A -x$ for $A \setminus \{x\}$, $c(xy)$ for $c(\{x,y\})$, etc. 

Next, we introduce preferences. 
Recall that a binary relation $R$ on $X$ is: 

- \textsl{reflexive} if $x R x$, for all $x \in X$;

- \textsl{asymmetric} if $x R y$ implies $\neg(y R x)$, for all $x,y \in X$;

- \textsl{symmetric} if $x R y$ implies $y R x$, for all $x,y \in X$;

- \textsl{antisymmetric} if $x R y$ and $y R x$ implies $x = y$, for all $x,y \in X$;

- \textsl{transitive} if $x R y$ and $y R z$ implies $x R z$, for all $x,y,z \in X$;

%- \textsl{negatively transitive} if $\neg(x R y)$ and $\neg(y R z)$ implies $\neg(x R z)$, for all $x,y,z \in X$; 

- \textsl{acyclic} if $x_1 R x_2 R \ldots R x_n R x_1$ holds for no $x_1,x_2, \ldots, x_n \in X$, with $n \geqslant 3$;\footnote{\label{FOOTNOTE:asymmetry_vs_acyclicity}Sometimes a binary relation is called \textsl{acyclic} if there is no cycle of length $\geqslant 2$ (see. e.g., \citealp{MasatliogluNakajimaOzbay2012}): according to this terminology, asymmetry is a special case of acyclicity. We prefer to keep the properties of asymmetry and acyclicity explicitly distinct, using the former term for the absence of cycles of length two, and the latter term for the absence of cycles of length at least three.}

- \textsl{complete} if either $x R y$ or $y R x$ (or both) holds, for all distinct $x,y \in X$.

%Observe that asymmetry and negative transitivity together imply transitivity. 
\noindent The symbol $\succsim$ denotes a reflexive binary relation on $X$, and is here interpreted as a \textsl{weak preference} on the set of alternatives.  
The following derived relations are associated to a weak preference $\succsim$ ($x,y$ range over $X$): 

- \textsl{strict preference} $\succ$, defined by $x \succ y$ if $x \succsim y$ and $\neg(y \succsim x)$;

- \textsl{indifference} $\sim$, defined by $x \sim y$ if $x \succsim y$ and $y \succsim x$;

- \textsl{incomparability} $\perp$, defined by $x \perp y$ if $\neg(x \succsim y)$ and $\neg(y \succsim x)$.

\noindent Note that $\succ$ is asymmetric, $\sim$ is symmetric, and $\succsim$ is the disjoint union of $\succ$ and $\sim$. 
%
%Moreover, for any disjoint $S,T \in \X$, the notation $S \succ T$ stands for $s \succ t$ for all $s \in S$ and $t \in T$; in particular, when $S =\{s\}$ or $T=\{t\}$, we write $s \succ T$ or $S \succ t$.  
%Note that, given a relation of strict preference $\succ$, we can derive from it a binary relation of \textit{indifference} $\sim$, defining $x \sim y$ if neither $x \succ y$ nor $y \succ x$ holds, for all $x,y \in X$. 
%In this case, we denote by $\succsim$ the (reflexive and complete) binary relation obtained as the disjoint union of $\succ$ and $\sim$.  
%
A weak preference $\succsim$ on $X$ is a \textsl{suborder} if $\succ$ is acyclic, a \textsl{preorder} if it is transitive, a \textit{partial order} if it is transitive and antisymmetric, a \textsl{total preorder} if it is a preorder with empty incomparability, and a \textit{linear order} if it is a complete partial order.     
%(Note that the indifference of a preorder is an equivalence relation.) % which coincides with equality in the special case of a linear order.  
We denote by $\rhd$ (the strict part of) a linear order (asymmetric, transitive, and complete).

The theory of revealed preferences pioneered by \cite{Samuelson1938} studies when a binary relation suffices to explain choice behavior by maximization. % by appealing to the maximization paradigm.  
Given a suborder $\succsim$ on $X$ and a menu $A \in \X$, the set of $\succsim$-\textsl{maximal} elements of $A$ is \vs\vs 
$$
\max(A,\succsim)=\{x \in X : y \succ x \text{ for no } y \in A\} \neq \es\,.\footnote{Note that $\max(A,\succsim) \neq \es$ because $X$ is finite and $\succ$ is acyclic.}\vs\vs %Observe also that if $\succsim$ is a linear order $\rhd$, then the set $\max(A,\rhd)$ has size one, and it can equivalently be written as $\max(A,\rhd) = \{x \in X : x \rhd y \text{ for all } y \in A \setminus \{x\}\}$.} \vs \vs 
$$ 
A choice $c \colon \X \to X$ is \textsl{rationalizable} if there exists a suborder (in fact, a linear order) $\rhd$ on $X$ such that $c(A) \in \max(A,\rhd)$ for any $A \in \X$.  
As customary, we abuse notation, and write $c(A) = \max(A,\rhd)$ in place of $c(A) \in \max(A,\rhd)$.

The rationalizability of a choice function\footnote{For a choice \textit{correspondence}, rationalizability is characterized by Axioms$\:\alpha$ and$\;\gamma$ \citep{Sen1971}.} is characterized by the property of \textsl{Contraction Consistency} due to \cite{Chernoff1954}, also called \textsl{Independence of Irrelevant Alternatives} by \cite{Arrow1963}, or \textsl{Axiom}$\:\alpha$ by \cite{Sen1971}.   
This property states that if an item is chosen in a menu, then it is also chosen in any submenu containing it:\vs
\begin{description}
	\item[Chernoff Property (Axiom$\:\alpha$):\!]
  for all $A,B\in \X$ and $x \in X$, if $x \in A \subseteq B$ and $c(B)=x$, then $c(A)=x$.\vs 
\end{description} 
For a (finite) choice function, Axiom$\:\alpha$ is equivalent to the \textsl{Weak Axiom of Revealed Preference} \citep{Samuelson1938}, which says that if an alternative $x$ is chosen when $y$ is available, then $y$ cannot be chosen when $x$ is available:\vs
\begin{description}
	\item[WARP:\!] for all $A,B \in \X$ and $x,y \in X$, if $x,y \in A \cap B$ and $c(A) = x$, then $c(B) \neq y$.
\end{description}

\section{Choice by salience} \label{SECTION:general_model} 

%Our approach to bounded rationality uses salience to encode semantics, and multiple rationales to encode attractiveness. 
Here we describe the general model of choice by salience, which explains any observed behavior and allows to detect moody choice behavior.  
In Section~\ref{SECTION:restricted_model}, we shall derive a testable model of choice by imposing rational constraints on salience.  

%%%%%%%%%%%%%%%%%%%%%%%%%%%%%%%%%%%%%%%%
%%%%%%%%%%%%%%%%%%%%%%%%%%%%%%%%%%%%%%%%
\subsection{\bf The general approach}
 
%and provide a methodology that encompasses some of bounded rationality models present in the literature.  
%Its ingredients are: (1) a suborder on $X$, and (2) a family of rationalizing linear orders indexed by the elements of $X$. 
%Henceforth, $c \colon \X \to X$ is a choice.   
%Note that the possible incompleteness/intransitivity of salience is an \textit{endogenous} feature of the DM's perception: for instance, she may consider some items pairwise incomparable. 
%The only assumption of coherence we make is that the strict salience order be asymmetric and acyclic.\footnote{Cf.\ Footnote~\ref{FOOTNOTE:asymmetry_vs_acyclicity}.}

%We still need another piece of notation before introducing the main notion of this paper.
%Given an equivalence relation $\equiv$ on $X$, we denote by $\S$ the associated partition of $X$, and by $\S_A$ the subset of $\S$ containing all $\equiv$-equivalence classes intersecting $A$, that is, 
%$$
%\S_A := \{S \in \S : S \cap A \neq \es\} .
%$$
%The set $\S_A$ of $\equiv$-equivalence classes is the \textsl{minimum cover} of $A$ with respect to $\equiv$.\footnote{The notion of `minimum cover' of a menu by equivalence classes naturally comes up in the study of \textsl{congruence relations} on a choice space: see \cite{CantoneGiarlottaWatson2019}.}   

\begin{definition} \label{DEF:GRS}
%	Let $c \colon \X \to X$ be a choice. 
	A \textsl{rationalization by salience} of $c \colon \X \to X$ is a pair $\langle \succsim,\L\rangle$, where  
	\begin{itemize}
		\item[(S1)] $\succsim$ is a suborder on $X$ (the \textsl{salience order}), and \vs\vs 
		\item[(S2)] $\L = \big\{\rhd_x : x \in X \big\}$ is a family of linear orders on $X$ (the \textsl{rationales}),\vs\vs 
	\end{itemize} 
	such that for any $A \in \X$, we have $c(A) = \max\left(A,\rhd_x\right)$ for some $x \in \max \left(A,\succsim \right)$. 
	In this case, we call $\langle \succsim,\L\rangle$ an \textsl{RS} for $c$, and $c$  an \textsl{RS choice}.% (Accordingly, we shall also speak of the \textsl{RS model}.) 
\end{definition}
	
%The two components in the definition of a RS can be explained as follows.
%First, the salience of available alternatives draws the DM's attention. 
%Some of these alternatives can be pairwise incomparable, whereas some others may be ranked according to a strict relation of salience, which is asymmetric and acyclic.
%Linear orders are associated to all alternatives in the menu.
Given a menu $A$, the DM's attention is captured by the most salient items, and an element is chosen in $A$ by maximizing one of the rationales indexed by these items. 
This approach is flexible, because it allows for an incompleteness/intransitivity of the salience order, according to an endogenous feature of the DM's perception.  
For instance, some items may display an incomparable salience, and thus suggest different preferences to apply in the decision.
Moreover, transitivity may fail, even for the relation of strict salience. 
This flexibility yields non-testability:
%This flexibility yields a justification for any choice behavior:

\begin{lemma} \label{LEMMA:extreme_GRS}
		 Any choice is rationalizable by salience.   
\end{lemma}

The general approach of choice by salience is connected to the RMR model of~\cite{KalaiRubinsteinSpiegler2002}. 
Recall that a set $\{\rhd _{1}, \ldots, \rhd_p\}$ of linear orders on $X$ is a \textsl{rationalization by multiple rationales (RMR)} of $c$ if, for all $A\in \X$, the equality $c(A)=\max(A,\rhd_i)$ holds for some $i$ in $\{1,\ldots,p\}$. 
In other words, an RMR is a \textit{set} of rationales such that any menu can be justified by maximizing one of them. 
Similarly to the RS model, the RMR model is non-testable, because it rationalizes any choice.
Thus, \cite{KalaiRubinsteinSpiegler2002} classify choices according to the minimum size of an RMR.
Specifically, they prove that any choice on $n$ elements needs at most $n-1$ rationales (Proposition~1), and as $n$ goes to infinity, all choices need the maximum number of rationales (Proposition~2).  %thus, the `degree of rationality' of a choice ranges from $1$ to $n-1$. % (we call these choices \textsl{context-free moody}).\footnote{The reason why we say `context-free moody' instead of `moody' will become apparent in Section \ref{SECTION:general_model}.}
Note that in the RMR model, choice behavior is collectively explained by rationales, with no need of an explicit connection between each menu and the linear order rationalizing it.  
On the compelling necessity of having a `structured' multiple rationalization, \citet[p.\,2287]{KalaiRubinsteinSpiegler2002} write:\vs\vs
\begin{quote} \small 
	\textsl{As emphasized in the introduction, our approach is ``context-free''. We agree with \cite{Sen1993} that if ``motives, values or conventions'' are missing from our description of the alternatives, then we'd better correct our model, whether or not IIA is violated.}\vs\vs
\end{quote}

The RS model refines the RMR model by revealing the internal structure of the set of rationales. %and relating them to the decision context. 
Moreover, the derived partition into `equivalence classes of rationality' is very selective: in fact, choices requiring the maximum number of rationales according to salience are more rare than choices requiring the maximum number of rationales according to the RMR model, especially for a small set of alternatives (see Section~\ref{SECTION:general_model}\ref{SUBSECTION:moodiness}).   
Finally, differently from the RMR approach, we can derive models of salience with empirical content by requiring the salience relation to satisfy suitable properties (see Section~\ref{SECTION:restricted_model}).

\subsection{\bf Moodiness} \label{SUBSECTION:moodiness}

Some choices do require the maximum number of rationales to encode attractiveness. 
  
\begin{definition}\label{DEF:moody}
A choice $c$ is \textsl{moody} if for any RS $\langle \succsim,\L \rangle$ of $c$,  $\rhd_x\neq \rhd_y$ whenever $x \neq y$. 
(Thus, a moody choice on $X$ always demands $\vert X \vert$-many distinct rationales.)
\end{definition}

The situation described by Definition~\ref{DEF:moody} is somehow pathological: it is peculiar of a DM who justifies whatever choice behavior she may exhibit by `local' explanations, that is, an \textit{ad hoc} rationale for each case.\footnote{A different notion of moody choice is used by \cite{ManziniMariotti2010}.}
In relation to Definition~\ref{DEF:moody}, one may wonder whether moody choices exist. 
This query is by no means trivial.  
Let us explain why. 

For the RMR model, \citet[Proposition~1]{KalaiRubinsteinSpiegler2002} show that any choice on a ground set of size $n$ can be always rationalized by $n-1$ linear orders. 
The crucial point here is that the RMR model imposes no constraints on the linear order that can be used to rationalize a \textit{specific} menu.

On the contrary, any RS $\langle \succsim,\L\rangle$ requires the rationales in $\L$ to be \textit{directly} connected to the menus they rationalize: each linear order $\rhd_x$ in $\L$ carries a label, and a menu $A$ can only be rationalized by an order whose label is a maximally salient items of $A$.
%Even in the limit case of a total absence of salience judgements (i.e., $\succsim \:= \{(x,x,): x \in X\}$), there is still a condition to be verified: $\rhd_x \in \L$ can rationalize $A \in \X$ only if $x \in A$.          
This necessary condition implies that the proof of Proposition~1 in \cite{KalaiRubinsteinSpiegler2002} does not carry over the RS approach. 
However, similarly to the RMR model, we still have: 

\begin{theorem}\label{THM:exists_chaos}
There are moody choices.	
\end{theorem}
 
The proof of Theorem~\ref{THM:exists_chaos} is non-trivial: it uses the notion of a \textsl{flipped choice}, which is defined on a linearly ordered set $X$, and is such that  the selection of elements in a menu systematically `oscillates' from the best item to the worst item. 
In Appendix~A, we describe this construction in detail, and show that any RS for a flipped choice on $39$ elements always needs 39 distinct rationales. 
We are not aware of smaller ground sets that give rise to such a pathology.
Thus, it appears that moody choice behavior arises only when a large number of items is involved, which in turn justifies a classification that labels `strongly irrational' all moody choices.

Theorem~\ref{THM:exists_chaos} raises a new query, concerning the ubiquity of moody choices when the size of the ground set grows larger and larger.
Similarly to what Proposition~2 in \cite{KalaiRubinsteinSpiegler2002} states for the RMR model, we have:  

\begin{theorem}\label{THM:chaos_rules}
%Asymptotically, all choices are moody. That is, 
The fraction of moody choices tends to one as the number of items in the ground set goes to infinity.   
\end{theorem}
  
The proof of Theorem~\ref{THM:chaos_rules} is rather involved, in fact it requires elements of Ramsey Theory.  
We present it in Appendix B, where we prove a more general result (Theorem~\ref{THM:tail_fail_weakly_hereditary_asymptotically_fails}), from which Theorem~\ref{THM:chaos_rules} follows as a corollary.  
The proofs of Theorems~\ref{THM:exists_chaos}, \ref{THM:chaos_rules},  and~\ref{THM:tail_fail_weakly_hereditary_asymptotically_fails} suggest that moodiness only arises for rather large sets of alternatives. 
This is compatible with empirical evidence: any DM who is presented with too many items tends to loose focus, and  ends up randomly selecting one of them; moreover, the larger the ground set, the more likely this randomness/irrationality surfaces. 
The fact that moody behavior only appears for large datasets also suggests that models employing too many rationales are empirically not desirable.

By virtue of Theorem~\ref{THM:chaos_rules}, if we partition the family of all finite choices into `classes of rationality' (that is, according to the minimum number of rationales needed for an RS), the class of moody choices does eventually collect almost all choices.
However, moodiness remains quite a rare phenomenon for a small number of alternatives. 
This consideration gives empirical content to the partition based on salience: the larger the difference between the number of items and that of rationales, the more rational the choice behavior.

%We point out that 
An analogous conclusion can hardly be drawn for the partition generated by the RMR model.
%In fact, choices that need the maximum number of rationales cannot be labeled as moody.  
For instance, \textit{all} choices on $n=3$ items are boundedly rationalizable by many known models, such as \textsl{choice with limited limited attention} \citep{MasatliogluNakajimaOzbay2012}, \textsl{categorize-then-choose}  \citep{ManziniMariotti2012}, \textsl{basic rationalization theory} \citep{CherepanovFeddersenSandroni2013}, and \textsl{overwhelming choice} \citep{LlerasMasatliogluNakajimaOzbay2017}.   
However, some of these choices need $n-1=2$ rationales.   
The situation is similar on a ground set of size $n=4$.  
Here the fraction of choices satisfying any of the models mentioned above is between $\frac{1}{3}$ and $\frac{3}{8}$,\footnote{For the computation of these fractions, see \citet[Lemma~8]{GiarlottaPetraliaWatson2022a}.} and yet many of these boundedly rationalizable choices require the maximum number $n-1=3$ of rationales. 
We conclude that the last class of the partition generated by the RMR model is hardly expressive of a form of `strong irrationality', whereas this feature can only be detected by employing a context-dependent approach.

\section{A testable model of salience} \label{SECTION:restricted_model} 

%The general model of choice by salience described in Section \ref{SECTION:general_model} allows the salience suborder to be incomplete and/or intransitive. 
Upon imposing rational constraints on salience, testable models of choice arise.
 
%%%%%%%%%%%%%%%%%%%%%%%%%%%%%%%%%%%%%%%%%%%%%%%
%%%%%%%%%%%%%%%%%%%%%%%%%%%%%%%%%%%%%%%%%%%%%%%

\subsection{\bf Linear salience} 

We describe a specification of the general model, in which the salience suborder satisfies the two basic tenets of economic rationality: transitivity and completeness.

\begin{definition} \label{DEF:rationalization_by_linear_salience} 
A \textsl{rationalization by linear salience (RLS)} of a choice $c \colon \X \to X$ is a pair $\langle \succsim,\L\rangle$, where\vs\vs  
%
%\footnote{As we shall see in Section~\ref{SECTION:general_model}, there are three ingredients in the definition of a rationalization by salience: (i) an equivalence relation of salience, (ii) a salience suborder on the set of equivalence classes, and (iii) a family of linear orders, one per salience class. However, in the special case that salience classes are linearly ordered, steps (i) and (ii) collapse in a single one, namely the definition of a weak order on the ground set of alternatives (which naturally induces a partition into equivalence classes of salience and a linear order on the quotient set). See Definition~\ref{DEF:linear_partition_associated_to_weak_order} and Section~\ref{SECTION:general_model}.}  
\begin{itemize} 
	\item[(LS1)] $\succsim$ is a total preorder on $X$ (the \textsl{salience order}),\vs \vs 	
	\item[(LS2)] $\L = \{ \rhd_x : x \in X\}$ is a family of linear orders on $X$ (the \textsl{rationales}), and\vs\vs 
	\item[(LS3)]$\rhd_x$ equals $\rhd_y$ whenever $x \sim y$ (the \textsl{normality condition}),\vs\vs 
\end{itemize} 
such that, for any $A \in \X$, $c(A)=\max\left(A,\rhd_x\right)$ for some $x \in \max(A,\succsim)$. 
%Hereafter, a choice having a rationalization by linear salience is called an \textsl{RLS choice}.
\end{definition}

The term `linear' is justified by the joint action of axioms LS1 and LS3: see Remark~\ref{REM:representation_of_RS_by_classes} and Lemma~\ref{LEMMA:equivalence_of_rationalization_by_salience}(ii) below. 
For any menu, the DM's attention is captured by the most salient items in it.
This leads her to make her selection by maximizing the (uniquely determined) rationale suggested by those items.\footnote{Note that Definition~\ref{DEF:rationalization_by_linear_salience} is sound because of the normality condition LS3.}
According to condition LS1, salience classes form a partition of the ground set, and they are linearly ordered by importance.\footnote{This ordering assumption has been already considered in more structured models of salience \citep{BordaloGennaioliShleifer2012,BordaloGennaioliShleifer2013}  as a key feature of DM's sensory perception.}
Condition LS3 says that equally salient items suggest identical criteria to apply in the selection process.\footnote{\label{FOOTNOTE:limited_number_switches}We could also make the less restrictive assumption that  preferences attached to equally informative alternatives be `very close' to each other, in the sense that a limited numbers of binary switches are allowed. In technical terms, this accounts to ask that linear orders associated to indifferent items must have a bounded \textsl{Kendal tau distance} \citep{Kendall1938}, or, more generally, a bounded distance according to a semantically meaningful notion of `metric for preferences' \citep{NishimuraOk2022}. This is a topic for future research.}

\begin{remark}\label{REM:representation_of_RS_by_classes}
	The rational structure of the salience order and the normality condition yield an alternative formulation of an RLS choice.
	In fact, the elements of $\L$ can be indexed by the equivalence classes of salience, rather than by the elements of the ground set. 
	Specifically, the total preorder $\succsim$ on $X$ generates a partition $\mathscr{S}_\succsim$ of  $X$ into equivalence classes of salience, which are linearly ordered by $\succ$ as follows: %where each $S \in \mathscr{S}_\succsim$ contains all and only those items of $X$ having the same salience. 
for all $S,T \in \mathscr{S}_\succsim$, let $S \succ T$ if $s \succ t$ for some (equivalently, for all) $s \in S$ and $t \in T$. 
	Now the normality condition LS3 allows us to rewrite the family of rationalizing preferences in LS2 by $\L = \{\rhd_S : S \in \mathscr{S}_\succsim \}$. 
	This representation has the obvious advantage of being more compact.
	However, we still prefer to use the original formulation given in Definition~\ref{DEF:rationalization_by_linear_salience}, because it is more intuitive. %and easier to be generalized.    
\end{remark}

Any rationalizable choice is RLS: take $X \times X$ as salience order (that is, all items are indifferent from a semantic point of view), %\footnote{As customary, a weak preference is \textsl{trivial} when its strict part is empty. It follows that a total preorder on $X$ is trivial if and only if it is equal to the whole Cartesian product $X \times X$.} 
and let $\L$ be the family composed of the unique linear order that explains choice by maximization.  
The next result provides alternatives formulations of rationalizability by linear salience; its proof is straightforward, and is left to the reader.

\begin{lemma} \label{LEMMA:equivalence_of_rationalization_by_salience} 
The following statements are equivalent for any choice $c \colon \X \to X$:\vs\vs 
\begin{itemize}
	\item[\rm (i)] $c$ is RLS; \vs\vs  
	\item[\rm (ii)] there are a linear order $\rhd$ on $X$ and a set $\{\rhd_{x} : x\in X\}$ of linear orders on $X$ such that $c(A)=\max\left(A,\rhd_{\max(A,\rhd)}\right)$ for any $A \in \X$; \vs\vs  
	%\item[\rm (iii)] there are a choice correspondence $\Phi \colon \X \to \X$ satisfying WARP (the \textsl{focusing filter}) and a set $\{\rhd_{\Phi(A)} : A\in \X\}$ of linear orders on $X$ such that $c(A)=\max\left(A,\rhd_{\Phi(A)}\right)$ for any $A \in \X$;\footnote{A choice correspondence $\Phi \colon \X \to \X$ satisfies WARP when for all $A,B \in \X$ and $x,y \in X$, if $x,y \in A \cap B$, $x \in \Phi(A)$, and $y \in \Phi(B)$, then $x \in \Phi(B)$. By the Fundamental Theorem of Revealed Preference Theory --see~\cite{Arrow1959} and~\cite{Sen1971}-- a choice correspondence satisfies WARP if and only if it is rationalizable by a total preorder.} \vs\vs 
	\item[\rm (iii)] there are a choice correspondence $\Phi \colon \X \to \X$ satisfying WARP (called a \textsl{focusing filter}) and a set $\{\rhd_{x} : x\in \X\}$ of linear orders on $X$ such that $c(A)=\max\left(A,\rhd_{x}\right)$ for some (equivalently, for all) $x \in \Phi(A)$;\footnote{A choice correspondence $\Phi \colon \X \to \X$ satisfies WARP when for all $A,B \in \X$ and $x,y \in X$, if $x,y \in A \cap B$, $x \in \Phi(A)$, and $y \in \Phi(B)$, then $x \in \Phi(B)$. By the \textit{Fundamental Theorem of Revealed Preference Theory} -- see~\cite{Arrow1959} and~\cite{Sen1971} -- a choice correspondence satisfies WARP if and only if it is rationalizable by a total preorder.}\vs\vs  
	\item[\rm (iv)] there are a choice function $d \colon \X \to X$ satisfying WARP and a set $\{\rhd_{x} : x\in X\}$ of linear orders on $X$ such that $c(A)=\max\left(A,\rhd_{d(A)}\right)$ for any $A \in \X$.
\end{itemize}
\end{lemma}

Lemma~\ref{LEMMA:equivalence_of_rationalization_by_salience}(ii) (and (iv)) provides an apparently simpler notion  of choice by linear salience. 
However, we still prefer Definition~\ref{DEF:rationalization_by_linear_salience}, because it emphasizes that items with the same salience should be associated to the same rationales (or, at least, to very similar rationales: see Remark~\ref{REM:representation_of_RS_by_classes} and Footnote~\ref{FOOTNOTE:limited_number_switches}). 
This is relevant also in view of the possibility to relaxing the completeness and the transitivity of the relation of salience, thus obtaining a more permissive (testable) model of choice.  
%Note that, upon taking the largest (in a set-theoretic sense) salience order in an RLS, the number of its equivalence classes coincides with the minimum number of \textit{distinct} linear orders that are needed for a rationalization by salience, and this gives a `direct' cardinal estimation of the bounded rationality of the observed choice. %% I found this comment a bit criptic, we can discuss about it.  %(a feature which would be obscured, were we to employ the notion given in Lemma~\ref{LEMMA:equivalence_of_rationalization_by_salience}(ii) instead). 

Lemma~\ref{LEMMA:equivalence_of_rationalization_by_salience}(iii) points out an alternative description of the behavioral process entailed by a linear salience approach.
The DM's salience is described by a focusing filter, which assigns to any menu those items that draw her attention.
These items will induce the DM to use a specific criterion to make her choice.
The focusing filter must satisfy WARP: if an item is among the most salient in a given menu, the same must happen in any submenu.
This condition is a consequence of DM's ability to rank items according to their salience.
  
In Sections~\ref{SECTION:restricted_model}\ref{SUBSECTION:CLLA} and~\ref{SECTION:relation with literature}, we shall extensively discuss the relationship of our linear model with several approaches of bounded rationality already present in the literature. 
Lemma~\ref{LEMMA:equivalence_of_rationalization_by_salience} already allows us to point out a few differences of this kind.
For instance, \cite{BordaloGennaioliShleifer2012} adopt the notion of \textsl{salience function}, which can be seen as a cardinal version of a focusing filter.  
A choice correspondence, called an \textsl{attention filter}, is also involved in the model of \cite{MasatliogluNakajimaOzbay2012}; however, its properties and behavioral interpretation are different from those of a focusing filter.
%In their seminal paper, \cite{ManziniMariotti2007} develop a theoretical framework in which the DM's choice is justified by the sequential maximization of two or more asymmetric preference relations, which are always applied in the same order to rationalize all menus.  
%Nevertheless, in our model this feature of sequentiality is shaped by a different philosophy, because the rationale justifying the selection from a menu depends on the menu itself. 
%In fact, Lemma~\ref{LEMMA:equivalence_of_rationalization_by_salience}(iii) clarifies that a DM who selects by salience first chooses the most salient items in the menu (using the salience --or \textit{focusing}-- choice correspondence $\sigma$), and then makes her final selection from the menu by maximizing the linear order prompted by the most salient items in it. 
%
There is also an apparent analogy with the approach of \cite{CherepanovFeddersenSandroni2013}, who consider a DM shrinking the set of feasible items using a choice correspondence $\Psi$ satisfying Axiom$\:\alpha$ (see Section 4.1 of the mentioned paper), before applying a suitable rationale (which is an asymmetric binary relation, or, in some cases, a linear order).\footnote{A choice \textit{correspondence} $\Psi \colon \X\rightarrow \X$ satisfies Axiom$\:\alpha$ when, for any $x \in X$ and $A,B\in\X$, if $x \in A \subseteq B$ and $x\in \Psi(B)$, then $x\in \Psi(A)$. Note that for choice correspondences, WARP is stronger than Axiom$\:\alpha$, being equivalent to the join satisfaction of Axiom$\:\alpha$ and Axiom$\:\beta$ \citep{Sen1971}.}
However, the focusing filter $\Phi$ in Lemma~\ref{LEMMA:equivalence_of_rationalization_by_salience}(iii) plays a role that is different from that of $\Psi$.  

%%%%%%%%%%%%%%%%%%%%%%%%%%%%%%%%%%%%%%%%%%%%%%%
%%%%%%%%%%%%%%%%%%%%%%%%%%%%%%%%%%%%%%%%%%%%%%%

\subsection{\bf Minimal switches and conflicting menus} 

Here we describe some possible features of `irrationality'.  

\begin{definition} \label{DEF:minimal_violations_of_alpha}
	For any choice $c \colon \X \to X$, a \textsl{switch} is an ordered pair $(A,B)$ of menus such that $A \subseteq B$ and $c(A) \neq c(B) \in A$.\footnote{\cite{CherepanovFeddersenSandroni2013} refer to such a pair of menus as \textsl{anomalous}.} 
	A switch $(A,B)$ is \textsl{minimal} if $\vert B \setminus A \vert = 1$. 
%	Whenever clear from context, we shall simply refer to $\langle A,A \cup \{x\}\rangle$ as a \textsl{violation of} $\alpha$. 
Equivalently, a minimal switch is a pair $(A,A \cup x)$ of menus such that $c(A) \neq c(A \cup x) \neq x$.
%\footnote{Recall that $A \cup x$ is an abbreviation for $A \cup \{x\}$.}
%We shall write $A \leftrightharpoons Ax$ to mean that $(A,A \cup \{x\})$ is a minimal switch. 
\end{definition} 

%For the sake of clarity, sometimes we slightly abuse notation, and underline the selected element in each menu, also omitting brackets.  
%For instance, the switch given by $\langle x\underline{y},\underline{x}yz \rangle$ stands for $\langle A,B \rangle$, where $A=\{x,y\}$, $B =\{x,y,z\}$, $c(A)=y$, and $c(B)=x$.  
Switches are violations of Axiom\,$\:\alpha$ (equivalently, WARP). 
A minimal switch $(A,A \cup x)$ arises whenever if the DM chooses $y$ from a menu $A$, and a new item $x$ is added to $A$, then the item selected from the larger menu $A \cup x$ is neither the old nor the new.  
When the ground set $X$ is finite, switches can always be reduced to a minimal ones: 
 
\begin{lemma} \label{LEMMA:minimal_violations_of_alpha}
%Any choice violating Axiom$\:\alpha$ minimally violates it. 
Let $c \colon \X \to X$ be a choice. 
For any switch $(A,B)$, there are a menu $C \in \X$ and an item $x \in X$ such that $A\subseteq C \subseteq C \cup x \subseteq B$ and $(C,C \cup x)$ is a switch. 
%$\langle C, C\cup \lbrace x\rbrace \rangle$ is a switch.
\end{lemma}
%
%  other formulations dismissed
%
%there is a violation of $\alpha$ by a choice, then there is also a minimal violation of it. 
%Any choice violating Axiom$\:\alpha$ minimally violates it. there is a minimal violation of $\alpha$. are $A \in \X$ and $x\in X$ such that $\langle A,A \cup \{x\} \rangle$ violates $\alpha$. %$c\left(A\,\cup\,\lbrace x\rbrace\right)\in A$ and $c\left(A\,\cup\,\lbrace x\rbrace\right)\neq c(A)$.
%
%
%\begin{proof} 
%Suppose $c \colon \X \to X$ is non-rationalizable, and so Axiom$\:\alpha$ fails to hold. 
%Thus, there exist $A,B \in \X$ such that $A\subseteq B$ and $c(A) \neq c(B)\in A$. 
%Take $A$ to be maximal and $B$ to be minimal, in the following sense: we cannot make $A$ larger, fixing $B$, and we cannot make $B$ smaller, fixing $A$.
%If $\vert B\setminus A\vert=1$, then the claim holds. 
%Otherwise, there is $D \in \X$ such that $A \subsetneq D \subsetneq B$. 
%If $c(D)\neq c(B)$, then the maximality of $A$ is contradicted.
%On the other hand, if $c(D)=c(B)$, then the minimality of $B$ is contradicted.
%\end{proof} 

By Lemma~\ref{LEMMA:minimal_violations_of_alpha}, the existence of minimal switches characterizes non-rationalizable choices. 
Suitable pairs of minimal switches identify a strong type of pathology: 
 
\begin{definition} \label{DEF:rational_violations_of_alpha}
	Two distinct menus $A,B \in \X$ are \textsl{conflicting} %denoted by $A \leftrightharpoons B$, 
	if there are $a \in A$ and $b \in B$ such that both $(A,A \cup b)$ and $(B,B \cup a)$ are switches. 
	%
%	We say that $c$ is \textsl{strongly discordant} if there are $A,B \in \X$, $a \in A$, and $b \in B$ such that $A \leftrightharpoons Ab$ and $B \leftrightharpoons Ba$; otherwise, $c$ is \textsl{weakly discordant}.  
%	
%	Let $x,y \in X$ be distinct items. 
%	We say that a menu $A \in \X$ \textsl{prompts a $(x,y)$-switch} for a choice $c$ if $y\in A$ and $\langle A, A \cup \{x\}  \rangle$ is a switch. 
%	 Then $c$ \textsl{rationally violates $\alpha$} if there are no distinct menus $A,B \in \X$ that prompt, respectively, a $(x,y)$-switch and a $(y,x)$-switch; otherwise, $c$ \textsl{irrationally violates $\alpha$}.  
\end{definition}

%\textbf{\magenta WARNING:} {\blue Angelo, I changed names, notation (no more subtractions from menu, but only additions), etc. etc. Please check everything VERY CAREFULLY, and change proofs accordingly if needed.}

Theorem~\ref{THM:main} in Section~\ref{SECTION:restricted_model}\ref{SUBSECTION:characterization} states that RLS choices display no conflicting menus. % `second order rational' if it is weakly discordant. %violations of Chernoff property are exclusively of the weak type described in Definition~\ref{DEF:rational_violations_of_alpha}. 
%Let us clarify why.
%Axiom $\alpha$ fails for any non-rationalizable choice.
%This failure may be strong or weak.  
%In other words, our approach only considers pathological the category of choices that display at least a pair of menus.  because they exhibit pairs of `conflicting menus'. 
%Theorem~\ref{THM:main} in this paper shows that rationalization by linear salience is characterized by the absence of this pathology.  
%either (i) the total absence of violations of $\alpha$, or (ii) the presence of rational violations of $\alpha$.  
%The possible presence of irrational violations of $\alpha$ suggest a form of pathological choice behavior, which is detected by the impossibility of rationalizing a choice by salience. 
%This fact gives our approach an empirical content. 
%
%Finally, we state a simple necessary condition for being an RLS choice:  
%
We conclude this section with a necessary condition for RLS choices. 

\begin{lemma} \label{LEMMA:NC_for_being_RS}
	Let $c \colon \X \to X$ be an RLS choice, and $\succsim$ the associated salience order.  
	For any $A \in \X$ and $x \in X$, if $(A,A \cup x)$ is a switch, then $x \succ a$ for all $a \in A$.  
\end{lemma}

In words, if a new element $x$ is added to a menu $A$, and the item chosen in the enlarged menu $A \cup x$ is neither the old nor the new,  then $x$ is more salient than any element in $A$.  
This is exactly what happens in Luce and Raiffa's dinner (Example~\ref{EX:Luce&Raiffa}), when the item $f$ (frog's legs) is added to the menu $\{c,s\}=\{\text{chicken, steak}\}$. 

%\begin{proof}
%	Let $A \in \X$ and $x \in X \setminus A$ be such that $c(A) =y$ and $c(A \cup \{x\}) = z \neq x,y$.
%	Since $c$ is rationalizable by salience, there are linear orders $\rhd_1$ and $\rhd_2$ on $X$ such that 
%	$$
%	\max(A,\rhd_1) = c(A) = y \neq z = c(A \cup \{x\}) = \max(A \cup \{x\}\,\rhd_2)\,.      
%	$$
%	Note that since $z \neq x$, the two linear orders $\rhd_1$ and $\rhd_2$ must be distinct. %we have $z \in A$ and $z \rhd_2 a$ for all $a \in A \setminus \{z\} \cup \{x\}$.  
%	Let $\{S_1,\ldots,S_n\}$ be the ordered partition of $A \cup \{x\}$ in salience classes, that is, $S_i \neq \es$ for all $i =1,\ldots,n$, $S_i \cap S_j = \es$ for $i \neq j$, $A \cup \{x\} = S_1 \cup \ldots \cup S_n$, and $S_1 \succ \ldots \succ S_n$.  	
%	Toward a contradiction, suppose there is $a' \in A$ such that $a'\succ x$. 
%	Thus, there must be $a \in A$ such that $a \in S_1$, which in turn implies that $a$ also belongs to the class of most salient items in $A$. 
%	However, the last fact yields that $\rhd_1$ coincides with $\rhd_2$, a contradiction. 
%\end{proof}

%%%%%%%%%%%%%%%%%%%%%%%%%%%%%%%%%%%%%%%%%%%%%%%
%%%%%%%%%%%%%%%%%%%%%%%%%%%%%%%%%%%%%%%%%%%%%%%

\subsection{\bf Revealed salience}\label{SUBSECTION:revealed_salience}

Any choice can be associated with an irreflexive relation revealed by minimal switches. 
 
\begin{definition}\label{DEF:revealed salience}
	Given $c \colon \X \to X$, define a relation $\vDash $ of \textsl{revealed salience} on $X$ by\vs\vs
	$$
	x \vDash y \quad \iff \quad \text{there is a menu $A$ containing $y$ such that $(A,A \cup x)$ is a switch}\vs\vs
	$$
	for any distinct $x,y \in X$.
	Hereafter, we write $x \vDash  A$ if $(A,A \cup x)$ is a switch, because the latter fact implies $x \vDash  a$ for all $a \in A$.\footnote{Indeed, $\vDash $ arises as a \textsl{hyper-relation} on $X$, that is, a subset of $X \times \X$.
	The hyper-relation $\vDash$ compares items to menus by declaring $x \vDash  A$ if $(A,A \cup x)$ is a switch. Hyper-relations have proven useful in rational choice, often providing a rather general perspective: see the pioneering papers by~\cite{AizermanMalishevski1981} and \cite{Nehring1997}, as well as the recent work by \cite{ChambersYenmez2017} and \cite{Stewart2020}. However, in our approach, hyper-relations would increase technicalities without getting any crucial leverage. Thus we define $\vDash $ as a binary relation.}    
%We also write $A \vDash x$ if $a \vDash x$ holds for any $a \in A$. 
\end{definition}
  
Essentially, $\vDash $ infers salience from observed data: if adding $x$ to a menu $A$ causes a switch, then $x$ is revealed to be more salient than any item in $A$.\footnote{Revealed salience is called \textsl{revealed conspicuity} in the reference-dependence theory of \cite{KibrisMasatliogluSuleymanov2021}.} %Our approach is semantically and technically different.}  
The next three remarks illustrate a connection between revealed salience and the binary relations revealed by three existing bounded rationality approaches. 
However, we point out that the rationale inspiring revealed salience is different from those described below. 

\begin{remark} \label{REM:revealed_Rev_in_rationalization_theory}
	The relation $\vDash$ evokes the relation $\textsf{Rev}$ defined in the \textsl{theory of rationalization} of \citet[p.\:780]{CherepanovFeddersenSandroni2013}: for any distinct $x,y \in X$, $x \,\textsf{Rev}\, y$ holds if there is a special violation of WARP, that is, an ordered pair $(A,B)$ of menus such that $x,y \in A \subseteq B$, $c(A) = x$, and $c(B) = y$.  
		%In words, an anomalous choice on a pair of menus reveals an item to be more attractive than a different one. 
	Since violations of WARP can be reduced to minimal switches (Lemma~\ref{LEMMA:minimal_violations_of_alpha}), $x \,\textsf{Rev} \,y$ holds if and only if there is $z \in X$ distinct from $x$ and $y$, and $A \in \X$ such that $c(A) = x$ and $c(A \cup z) = y$, which yields $z \vDash x$. 
%	This implies that $A$ prompts a $(z,x)$-switch, %of $\alpha$, 
	We conclude that $x \, \textsf{Rev} \, y$ implies $z \vDash x$ for some $z$ distinct from $x$ and $y$. 
	Thus, the two revealed relations $\textsf{Rev}$ and $\vDash$ describe different types of attitudes: $\textsf{Rev}$ looks at the attractiveness of items, whereas $\vDash$ is related to their semantics. %: thus, despite the apparent similarities, the two approaches are conceptually different. 
%	(See the Introduction for the distinction between `attractiveness' and `semantics'.)  
\end{remark}

\begin{remark} \label{REM:revealed_P_in_CLA}
	Revealed salience $\vDash$ is a a weak converse\footnote{The \textsl{converse} $R^c$ of a relation $R$ on $X$ is defined by $x R^c y$ if $y R x$, for all $x,y \in X$.} of the relation $P$ associated to a \textsl{choice with limited attention (CLA)} \cite[p.\:2191]{MasatliogluNakajimaOzbay2012}.
	Recall that for any distinct $x,y\in X$, $xPy$ holds if there is $A\in\X$ such that $x=c(A \cup y) \neq c(A)$. 	
	%In our terminology, there is a menu $A$ containing $x$ such that $(A,A \cup y)$ is a switch, and so $y\vDash x$ holds.
	It follows that $x P y$ implies $y \vDash x$, but the reverse implication does not hold (see Section~\ref{SECTION:restricted_model}\ref{SUBSECTION:CLLA} for details). %  and so $\vDash$ is strictly weaker than the converse of $P$. 
	Again, as $\textsf{Rev}$, the relation $P$ operates at a different level than $\vDash$, being related to the attractiveness of items (by Theorem~1 in the mentioned paper, the transitive closure of $P$ reveals preferences). 
\end{remark} 

\begin{remark}\label{REM:revealed_P_in_temptation}
	 The relation $\vDash$ is the converse of the relation $\widetilde{P}$ defined in \cite{RavidStevenson2021}, where $x\widetilde{P}y$ holds if there is a menu $A$ containing $x$ such that $(A,A\cup x)$ is a switch.
	The authors show that the asymmetry and the acyclicity of $\widetilde{P}$ are necessary conditions of their model.
	In Theorem \ref{THM:main} we refine their results, and show that the asymmetry of $\vDash$ (hence of $\widetilde{P}$) is necessary and sufficient for an RLS choice. 
%	(We will make use of the relation $\widetilde{P}$ in Section~\ref{SECTION:restricted_model}\ref{SUBSECTION:CLLA}.) 
\end{remark}

In the path to characterize RLS choices by the asymmetry and the acyclicity of revealed salience, it is worth mentioning the following crucial fact: 
%According to asymmetry, if the introduction of an item $x$ to a menu $A$ causes a violation of Axiom$\:\alpha$ by producing a shift of DM's preference, then the same does not happen when an item $y \in A$ is added to a different menu containing $x$.
%Acyclicity states that revealed salience cannot exhibit cycles: salience must be ordered, and the DM's attention and perception is coherent with this order.
%In fact, Theorem~\ref{THM:main} below characterizes RLS choices by revealed salience being asymmetric and acyclic. 
%Surprisingly, we have In the path to establish this characterization, we preliminarily mention a (possibly surprising) fact: for revealed salience, asymmetry is stronger than acyclicity.   

\begin{lemma} \label{LEMMA:asymmetry implies acyclicity}
	For any choice, if revealed salience is asymmetric, then it is also acyclic.
\end{lemma}

In words, the absence of revealed cycles of length two suffices to prove the absence of revealed cycles of any length.  
Lemma~\ref{LEMMA:asymmetry implies acyclicity} is important in applications, because checking asymmetry is computationally faster than checking acyclicity. 

The converse of Lemma~\ref{LEMMA:asymmetry implies acyclicity} fails to hold:   

\begin{example}[\it An acyclic but not asymmetric revealed salience]  
\label{EX:acyclic_non-asymmetric_revealed_salience} \vs 
	Let $X = \{x,y,z\}$, and define a choice $c \colon \X \to X$ by $xy\underline{z},\, \underline{x}y,\, \underline{x}z,\, \underline{y}z\,.$ 
	This choice is non-rationalizable, because doubletons are rationalized by the linear order $x \rhd y \rhd z$, but the $\rhd$-worst item $z$ is selected in $X$. 
	Revealed salience is acyclic but not asymmetric, because we have $x \vDash y$, $x \vDash z$, $y \vDash x$, and $y \vDash z$.  
	(For instance, $(yz,xyz)$ and $(xz,xzy)$ are minimal switches, which respectively yield %violations of $\alpha$ 
	$x \vDash  y$ and $y \vDash  x$.)
	%(For instance, $\langle \underline{y}z, xy\underline{z}\rangle$ and $\langle \underline{x}z, xy\underline{z}\rangle$ are switches %violations of $\alpha$ 
	%yielding $x \vDash  y$ and $y \vDash  x$, respectively.)
	Note also that the two menus $\{x,y\}$ and $\{y,z\}$ are conflicting.  
\end{example}

\subsection{\bf Characterization} \label{SUBSECTION:characterization}
The absence of conflicting menus -- or, alternatively, the asymmetry of revealed salience -- characterizes our model of linear salience.

\begin{theorem} \label{THM:main}
The following statements are equivalent for a choice $c$:\vs\vs
\begin{itemize}
	\item[\rm (i)] $c$ is RLS;\vs\vs
	\item[\rm (ii)] revealed salience is asymmetric;\vs\vs 
	\item[\rm (iii)] there are no conflicting menus.\vs\vs 
	%\item[\rm (iv)] WARP(S) holds for $c$.\vs\vs 
\end{itemize}\vs
\end{theorem}

Let us quickly sketch how to construct a rationalization by linear salience from an asymmetric revealed salience $\vDash$. 
By Lemma~\ref{LEMMA:asymmetry implies acyclicity}, $\vDash$ is acyclic, hence it is a suborder.    
Pick any total preorder $\succsim$ on $X$ that extends the transitive closure of $\vDash \,$: this will be our salience order.  
The linear rationales $\rhd_x$ on $X$ are obtained, for each $x \in X$, by a classical revealed preference argument: first get a partial order $>_x$ by declaring $y$ revealed better than $z$ if there is a menu $A$ such that $x$ is one of the most salient items in $A$ and $y$ is chosen in $A$; then, let $\rhd_x$ be any linear extension of $>_x$.\footnote{The elicitation of $>_x$ from the observed choice helps us to explain data. If an item $y$ is selected in a menu $A$ in which another item $x$ captures the DM's attention, then $y$ is better than any other alternative in $A$, according to the preference suggested by $x$.}
%The elicitation of $>_x$ (and $\rhd_x$) from the observed choice is the second component of the analysis 

%\begin{remark}
%	The proof of Theorem~\ref{THM:main} reveals a large degree of freedom in the selection of the rationalization by salience of an RLS choice.
%	However, this abundance is apparent, since under the alternative representation discussed in Remark \ref{REM:representation_of_RS_by_classes}, the salience order is uniquely determined.
%	Thus, this 
%\end{remark} 

%The proof of Theorem~\ref{THM:main} reveals a great deal of freedom, and prompts a natural extension of our approach to cases in which the salience order is less structured, as we do in the next section. 

%%%%%%%%%%%%%%%%%%%%%%%%%%%%%%%%%%%%%%%%%%
%%%%%%%%%%%%%%%%%%%%%%%%%%%%%%%%%%%%%%%%%%

\subsection{\bf Choices with salient limited attention}\label{SUBSECTION:CLLA} 
 
The objective of this section is twofold. 
Our first goal is to prove that linear salience is a special case of the well-known model of choice with limited attention due to  \citet{MasatliogluNakajimaOzbay2012}. 
Our second goal is to provide a descriptive characterization of RLS choices in terms of special types of attention filters.

%In our general model of choice by salience, there is no formal preselection and shrinking of the original menu.  
%In fact, the two successive stages of the process of choice by salience are based on `orthogonal' features, namely semantics (salience) and attractiveness (linear orders). 
%Maybe surprisingly, it turns out that the linear specification of our general approach is a special case of the model of choice with limited attention. 

\begin{definition} \citep{MasatliogluNakajimaOzbay2012} \rm \label{DEF:CLA}
A choice $c \colon \X \to X$ is \textsl{with limited attention} \textsl{(CLA)} if $c(A) = \max(\Gamma(A),\rhd)$ for all $A \in \X$, where\vs\vs
\begin{itemize}
	\item[(a)] $\rhd$ is a linear order (\textsl{rationale}) on $X$, and\vs\vs
	\item[(b)] $\Gamma \colon \X \to \X$ is a choice correspondence (\textit{attention filter}) such that for any $B \in \X$ and $x \in X$, $x \notin \Gamma(B)$ implies $\Gamma(B) = \Gamma(B -x)$.
\end{itemize}
\end{definition}

The DM selects an item from a menu maximizing a linear order on the subset of elements that attract her attention.  
Upon defining a binary relation $P$ on $X$ by\vs\vs 
\begin{equation} \label{EQ:P_in_CLA} 
	x P y \quad \Longleftrightarrow \quad \text{ there is $A \in \X$ such that $x=c(A) \neq c(A -y)$}\vs\vs
\end{equation}
for all distinct $x,y \in X$, \citet[Lemma~1 and Theorem~3]{MasatliogluNakajimaOzbay2012} prove that $c$ is CLA if and only if $P$ is both asymmetric and acyclic. 

To accomplish our first goal, let $\widetilde P$ be the converse of revealed salience $\vDash$. 
A simple computation shows that for all distinct $x,y \in X$, we have\vs\vs\vs 
\begin{equation} \label{EQ:Q_in_RLS}\vs
		x \widetilde P y \quad\Longleftrightarrow \quad \text{there is a $A \in \X$ such that } x \in A \text{ and } y \neq c(A) \neq c(A -y).\vs\vs 			
\end{equation}
Thus $\widetilde P$ extends $P$.
Since $c$ is RLS if and only if $\widetilde P$ is asymmetric (and acyclic), and considering the choice of Example~\ref{EX:acyclic_non-asymmetric_revealed_salience} (which is CLA but not RLS), we get: 
 
\begin{lemma} \label{LEMMA:RLS_is_CLA}
	Any RLS choice is a CLA. The converse is false. 
\end{lemma}

%The converse of Lemma~\ref{LEMMA:RLS_is_CLA} is false: the anomalous choice in Example~\ref{EX:acyclic_non-asymmetric_revealed_salience} is CLA (because all choices on $3$ items are with limited attention) but not RLS (by Theorem~\ref{THM:main}).  

To accomplish our second goal, we first identify a family of choices with limited attention characterized by special types of attention filters. %and then show that this family coincides with the family of RLS choices. 

\begin{definition} \label{DEF:CSLA} \rm 
	A choice $c \colon \X \to X$ is \textsl{with salient limited attention} \textsl{(CSLA)} if $c(A) = \max(\Gamma(A),\rhd)$ for all $A \in \X$, where\vs\vs
\begin{itemize}
	\item[(a)] $\rhd$ is a linear order on $X$ (\textsl{%mental
	 rationale}), and   \vs\vs
	\item[(b)$'$] $\Gamma \colon \X \to \X$ is a choice correspondence (\textsl{salient attention filter}) such that for all $B \in \X$ and $x \in X$, $x \neq \min(B,\rhd),\max(\Gamma(B),\rhd)$ implies $\Gamma(B) - x = \Gamma(B -x)$.  
\end{itemize}
\end{definition} 

Condition (b)$'$ in Definition~\ref{DEF:CSLA} is stronger than condition (b) in Definition~\ref{DEF:CLA}: if $x \notin \Gamma(B)$, then $\Gamma(B) = \Gamma(B) - x = \Gamma(B-x)$, and so any salient attention filter for $c$ is an attention filter.\footnote{Note also that Definition~\ref{DEF:CSLA} makes explicit the dependence of the salient attention filter from the DM's rationale. This dependence is implicit in the CLA model, but becomes explicit in the process of constructing an attention filter from the given rational: see the proof of Theorem~3 in \citet[p.\,2202]{MasatliogluNakajimaOzbay2012}.}  
%
%Similarly to a CLA, in a choice with salient limited attention, the process of selecting an item from a menu $A$ is guided by the maximization of a linear order $\rhd$.
%Moreover, this maximization is restricted to a subset $\Gamma(A)$ of $A$, which comprises all items of $A$ that catch the DM's attention. 
%The difference between a CLA and a CSLA lies in the features of the filter $\Gamma$. 
In a CLA, for any item $x$ in $A$ that does not catch the DM's attention (that is, $x \notin \Gamma(A)$), the filter $\Gamma$ does not discern between the original menu and the menu deprived of the irrelevant item (that is, the equality $\Gamma(A) = \Gamma(A -x)$ holds).  
In a CSLA, this indiscernibility feature is extended to all items of $A$ that are different from the best element in $\Gamma(A)$ and the worst element in $A$: among the items brought to her attention, the DM focuses only on the (salient) items holding an extreme position in her judgement, either maximum or minimum. 
(Note that if $\min(B,\rhd)\not\in\Gamma(A)$, then the DM does not consider the minimum.)
This feature is coherent with the salience theory of choice under risk as in \cite{BordaloGennaioliShleifer2012}: the DM's evaluation of lotteries is affected by extreme payoffs, which makes her risk-lover when upsides are high, and risk-averse if downsides are high.

 As announced, we have:\vs

\begin{proposition} \label{PROP:CSLA=RLS}
	RLS is equivalent to CSLA. 
\end{proposition}
 
Note also that Proposition \ref{PROP:CSLA=RLS} implies that CSLA holds if and only if the revealed preference $\widetilde{P}$ is asymmetric (and acyclic).
A CSLA representation of an RLS choice offers also an alternative interpretation of choice data.
In fact, we have:\footnote{The proof of this fact is left to the reader.}

\begin{lemma}
Let $c\colon \X\to X$ be CSLA, and $(\Gamma,\rhd)$ any associated explanation of it. 
If there are $A \in \X$ and $x,y\in A$ such that $y \neq c(A)\neq c(A-y)$ (i.e., $x\widetilde{P}y$), then $x\rhd y$, $y\in \Gamma(A)$, and $y$ is equal to $\min(\Gamma(A),\rhd)$.
\end{lemma}

 In other words, for a CSLA, if removing $y$ from a menu $A$ containing $x$ causes a switch, then we can deduce not only that the DM prefers $x$ to $y$ and pays attention to $y$ at $A$, but also that $y$ is the least preferred item among those brought to her attention in $A$.

%In fact, it is easy to show that if $c\colon\X\to X$ is CSLA and $x\widetilde{P}y$ for some $A\in X$ containing $x$ and $y$, we have that $x\rhd y$, $y\in\Gamma(A)$, and $y=\min(\Gamma,\rhd)$ or $y=\max(\Gamma,\rhd)$.\footnote{The easy proof is left to the reader.}
%Thus, if removing $y$ from a menu $A$ containing $x$ causes a switch, we can deduce not only that the DM prefers $x$ to $y$ and pays attention to $y$ at $A$, but also that $y$ is  either the most or the less preferred item  between those brought to her attention.

%%%%%%%%%%%%%%%%%%%%%%%%%%%%%%%%%%%%%%%%%%%%%%%%%
%%%%%%%%%%%%%%%%%%%%%%%%%%%%%%%%%%%%%%%%%%%%%%%%%

\subsection{\bf Numerical estimates}    
 
Here we show that linear salience yields a selective choice model, even when the number of items in the ground set is rather small.   
To that end, we evaluate the fraction of RLS choices for some sizes of the ground set.
All estimates are obtained by using the techniques introduced in \cite{GiarlottaPetraliaWatson2022a}, and specifically analyzed in~\citet{GiarlottaPetraliaWatson2022b}: we refer the reader to those papers for details. 

\begin{definition} \label{DEF:subchoice}
	A \textsl{subchoice} of a choice $c \colon \X \to X$ is any choice $c_{\upharpoonright A} \colon \mathscr A \to A$, with $A \in \X$ and $\mathscr{A} := \{B \in \X : B \subseteq A\}$, defined by $c_{\upharpoonright A}(B)=c(B)$ for all $B \in \mathscr{A}$.
\end{definition}

\begin{definition} \label{DEF:choice_isomorphism}
	Two choices $c \colon \X \to X$ and $c' \colon \X' \to X'$ are \textsl{isomorphic} if there is a bijection $\sigma \colon X \to X'$ such that $\sigma(c(A)) = c'(\sigma(A))$ for any $A\in\X$. %In this case, we say that $\sigma$ is an \textsl{isomorphism} between $c$ and $c'$.	
\end{definition}

\begin{definition}\label{DEF:HT_choice}
A \textsl{property} $\mathscr{P}$ \textsl{of choices} is a set of choices closed under isomorphism. %\footnote{Equivalently, a property of choices is a formula of second order monadic logic, which involves quantification over elements and sets, has a symbol for choice, and is invariant under choice isomorphisms.}  
We denote by $T(n)$, $T(n,\mathscr{P})$, and $F(n,\mathscr{P}) = \frac{T(n,\mathscr{P})}{T(n)}$, respectively, the total number of choices on $n$ elements, the total number of choices on $n$ elements satisfying property $\mathscr{P}$, and the fraction of choices on $n$ elements satisfying property $\mathscr{P}$.  
\end{definition}

The ratio $F(n,\mathscr{P}) = \frac{T(n,\mathscr{P})}{T(n)}$ can be computed only considering choices on $n$ elements that are pairwise non-isomorphic, because all isomorphism classes have exactly the same size ($= n!$): see \citet[Lemma~4]{GiarlottaPetraliaWatson2022a}. 
   
\begin{definition}\label{DEF:hereditary_property}
 A property $\mathscr{P}$ of choices is \textsl{hereditary} whenever if $\mathscr{P}$ holds for any choice, then it also holds for any of its subchoices.\footnote{Thus, $\mathscr P$ is hereditary if for all choices $c \colon \X \to X$, $c \in \mathscr{P}$ implies $c_{\upharpoonright A} \in \mathscr{P}$ for all $A \in \X$.}   
\end{definition} 

%\begin{lemma}[\cite{GiarlottaPetraliaWatson2022a}, Theorem~1] \label{LEMMA:help1_for_estimates}
%	 If $\mathscr{P}$ is a hereditary property of choices, then $\lim_{n \to \infty} F(n,\mathscr{P}) =0$. 
%\end{lemma}

\begin{lemma}[\cite{GiarlottaPetraliaWatson2022a}, Corollary~5] \label{LEMMA:help2_for_estimates}
If $\mathscr P$ is a hereditary property that contains at most $q$ pairwise non-isomorphic choices on four elements, then the following upper bounds to $F(n,\mathscr P)$ hold:\vs\vs 
\begin{table}[H] 
\begin{center} 
\small
\begin{tabular}{|c||c|c|c|c|c|}
\hline
$n$ & 4 & 16 & 20 & 28 & 32 \\ 
\hline\hline
$F(n,\mathscr P)$ &\footnotesize $\!=\!(q/864)\!$ & \footnotesize $\!\leqslant\! (q/864)^{20}\!$ & \footnotesize $\!\leqslant\! (q/864)^{29}\!$ & \footnotesize $\!\leqslant\! (q/864)^{57}\!$ & \footnotesize $\!\leqslant\! (q/864)^{72}\!$ \\ 
\hline
\end{tabular}     
\end{center}
\end{table}
\end{lemma} 
\vs\vs\vs\vs\vs\vs

It is not difficult to show that:\footnote{The proof is similar to that of Lemma~8 in \cite{GiarlottaPetraliaWatson2022a}.}

\begin{lemma} \label{LEMMA:estimates_for_RLS} \label{LEMMA:help3_for_estimates}
	The class of RLS choices is hereditary. 	
	Moreover, there are exactly $40$ pairwise non-isomorphic RLS choices on four elements.\vs
\end{lemma} 

In comparison, there are exactly 864 pairwise non-isomorphic choices on four items, of which 324 are CLA \citep[Lemma~8]{GiarlottaPetraliaWatson2022a}, and only $1$ is rationalizable. 
Lemmata~\ref{LEMMA:help2_for_estimates} and~\ref{LEMMA:help3_for_estimates} readily yield the numerical estimates we were after, which explicitly show the sharp selectivity of the RLS model: 

\begin{corollary}
    %Asymptotically, no choice is RSL.  
	The following upper bounds hold for the fractions $F(n,\mathscr P)$ of choices on $n= 4,16,20,28$ elements, which are, respectively, rationalizable, RLS, or CLA:\vs\vs 
\begin{table}[h!]
\begin{center}	
\begin{tabular}{|c||*{5}{c|}}
\hline
\backslashbox{$\mathscr P$}{$n$}
\small
&\makebox[2em]{\small 4}&\makebox[2em]{\small 16}&\makebox[2em]{\small 20}&\makebox[2em]{\small 28}&\makebox[2em]{\small 32} \\
\hline\hline  
\small WARP &\footnotesize$= 0.0011$& \footnotesize $\leqslant 10^{-58}$ & \footnotesize $\leqslant 10^{-85}$ & \footnotesize $\leqslant 10^{-167}$ & \footnotesize $\leqslant 10^{-211}$ \\ 
\hline
\small RLS & \footnotesize $= 0.046$ & \footnotesize $\leqslant 10^{-26}$ & \footnotesize $\leqslant 10^{-38}$ & \footnotesize $\leqslant 10^{-76}$ & \footnotesize $\leqslant 10^{-96}$ \\
\hline
\small CLA& \footnotesize $=0.37$ & \footnotesize $\leqslant 10^{-8}$ & \footnotesize $\leqslant 10^{-12}$ & \footnotesize $\leqslant 10^{-24}$ & \footnotesize $\leqslant 10^{-30}$ \\ 
\hline
\end{tabular}
\end{center}
\vs\vs\vs\vs
\end{table}\vs\vs\vs\vs\vs
\end{corollary}

%%%%%%%%%%%%%%%%%%%%%%%%%%%%%%%%%%%%%%%%%%%%%%%%%
%%%%%%%%%%%%%%%%%%%%%%%%%%%%%%%%%%%%%%%%%%%%%%%%%
%%%%%%%%%%%%%%%%%%%%%%%%%%%%%%%%%%%%%%%%%%%%%%%%%
%%%%%%%%%%%%%%%%%%%%%%%%%%%%%%%%%%%%%%%%%%%%%%%%%

\section{Additional relations with literature}\label{SECTION:relation with literature}

Here we compare choice by linear salience with several models of bounded rationality. 
We also show how a salience approach can accommodate some anomalies that have been extensively studied in the choice literature.\vs\vs

%%%%%%%%%%%%%%%%%%%%%%%%%%%%%%%%%%%%%%%%%%%%%%%%%
%%%%%%%%%%%%%%%%%%%%%%%%%%%%%%%%%%%%%%%%%%%%%%%%%

\subsection{Bounded rationality models}

Choice by linear salience is connected to the sequential rationalization of \cite{ManziniMariotti2007}.
%Recall that a choice function $c$ is \textsl{sequentially rationalizable} if there exists an ordered list $\L=(\succ_{1}, \ldots,\succ_{n})$ of asymmetric relations such that for each $ A\in\X$, upon defining recursively $M_{0}(A):=A$ and $M_{i}(A):=\max(M_{i-1}(A),\succ_{i})$ for $i=1,...,n$, the equality $c(A)=M_{n}(A)$ holds. 
According to their approach, in any menu the DM sequentially applies asymmetric rationales in a fixed order. 
In our model, sequentiality is shaped by a different philosophy, because the rationale justifying selection depends on the menu. 
%In fact, a DM who selects by salience first chooses the most salient items in the menu (using the salience order $\succsim$), and then makes her final selection from the menu by maximizing the linear order prompted by the most salient items in it.
%This bounded rationality method provides a sound justification for several non-rationalizable choices.
%However, not all choices are sequentially rationalizable, even allowing an unboundedly large number of rationales.
%Similarly, not all choices are rationalizable by linear salience, no matter how many linear orders we may use at the second stage.
As expected, these two procedures may yield very different results: see Appendix~C. % we show that there are choices that are sequentially rationalizable but not rationalizable by salience, and, vice versa, choices that are rationalizable by salience but not sequentially rationalizable.  

A sequential contraction of menus is also used in the theory of rationalization due to \cite{CherepanovFeddersenSandroni2013}. 
Here the DM discards from a menu all those items that are not allowed by a \textsl{psychological constraint} (a choice correspondence satisfying Axiom$\:\alpha$), and then maximizes a fixed linear order to select an item.  
Our Lemma~\ref{LEMMA:equivalence_of_rationalization_by_salience} shows that rationalizability by salience implies the existence of a choice correspondence $\Phi$ (a focusing filter) which satisfies WARP. 
Although the focusing filter may be seen as a special psychological constraint (since WARP implies Axiom$\:\alpha$ for choice correspondences), its interpretation is radically different in our model: in fact, $\Phi$ only picks the most salient alternatives, but causes no reduction of the selectable items.    
We show the independence of the two models in Appendix~C.
%
%Our model, as previously showed, does not justify the behavior showed in Example~ \ref{EX:acyclic_non-asymmetric_revealed_salience}.
%
%On the other hand, some choices rationalizable by salience are not consistent with any theory of behavior.
%Let $c \colon\X\to X$ be a choice function on $X=\lbrace w,x,y,z\rbrace$, defined as follows:
%
%\begin{equation}\label{No binary chain cycle violation}
%wx\underline{y}z \,,\; wx\underline{y} \,,\; wx\underline{z} \,,\; w\underline{y}z \,,\; \underline{x}yz\,,\; \underline{w}x \,,\; w\underline{y} \,,\; w\underline{z}\,,\; x\underline{y} \,,\; \underline{x}z \,,\; y\underline{z}\,. 
%\end{equation}
%Note that $x\:\textsf{Rev}\:y$, since $c(\{x,y,z\})=x$ and $c(\{x,y\})=y$.
%Moreover, $y\:\textsf{Rev}\:z$, since $c(\{w,x,y,z\})=~y$ and $c(\{y,z\})=z$.
%We have also that $z\:\textsf{Rev}\:x$, since $c(\{w,x,z\})=z$ and $c(\{x,z\})=x$, contradicting No Binary Chain Cycle. 
%Thus choice \eqref{No binary chain cycle violation} is not rationalizable according to theory of Rationalization.
%However, the weak order $w\,\succ\,z\succ\,\lbrace x,y\rbrace$, %the associated partition $\mathcal{S}\,_{\trianglerighteq}=
%\lbrace \lbrace w\rbrace,\lbrace z\rbrace,\lbrace x,y\rbrace\rbrace,$ 
%and the family of preferences  %$(\succ_{S})_{S\in\mathcal{S\,_{\unrhd}}}$:
%\begin{gather*}
 %y\rhd_{w}z\rhd_{w}w\rhd_{w}x\,,
 %\quad
% w\rhd_{z}x\rhd_{z}z\rhd_{z}y\,,
% \quad
% y\rhd_{xy}x\rhd_{xy}w\rhd_{xy}z\,,
%\end{gather*} 
%rationalizes by salience choice \eqref{No binary chain cycle violation}.

\cite{RavidStevenson2021} analyze the impact of (bad) temptations on individual choices.
In their model, the DM maximizes a function which is strictly increasing with respect to the utility of each item, and the difference between the item's temptation and the maximal temptation available in the menu.
It can be shown that choice behaviors explained by temptation can be justified by our linear salience approach (and vice versa), capturing however a rather different positive model of behavior.
Furthermore, the characterization of the model of \cite{RavidStevenson2021} relies on the \textsl{Axiom of Revealed Temptation (ART)},\footnote{ART requires each menu $A$ to contain at least one item $x$ such that WARP is obeyed on the collection of subsets of $A$ that contain $x$.} whereas our CLS model is characterized by the asymmetry of revealed salience or by the absence of conflicting menus.

In a concurrently written paper \cite{KibrisMasatliogluSuleymanov2021} propose a theory of reference point formation, applied to risk, time, and social preferences by \cite{Lim2021}.
Their model, which is characterized by the \textsl{Single Reversal Axiom} (SRA),\footnote{SRA: for all $S,T\in\X$ and distinct $x,y\in X$ such that $\{x,y\}\subseteq S \cap T$, if $x\neq c(S)\neq c(S-x)$, then either $c(T)=y$ or $c(T-y)=c(T)$.} is equivalent to the linear salience model; however, both the treatment of the topic and the underlying motivation are very different from ours. 
%In this work we mainly focus on salience and semantics of alternatives, and we mention reference dependence as an application of this method.
%First, we provide multiple representations of choice by linear salience, in which make use of focusing maps.
%Moreover, the characterization of ordered reference dependence is based on the \textsl{Single Reversal Axiom}(SRA),\footnote{SRA: for each $S,T\in\X$, and distinct $x,y\in \X$ with $\{x,y\}\subseteq S\cap T$, if $x\neq c(S)\neq c(S-x)$, then either $c(T)=y$ or $c(T-y)=c(T)$.} whereas our multiple characterization of linear salience (Theorem~\ref{THM:main}) hinges on the asymmetry of the salience ascertained from the choice, as well as the lack of conflicting menus. 
Furthermore, our approach allows us to prove that linear salience is a special case of limited attention, in which only salient items matter.
Last but not least, the RLS model is only one of the many specification of a general approach based on salience, whose flexibility may allow one to obtain a better fit to the DM's attention structure.\vs

%%%%%%%%%%%%%%%%%%%%%%%%%%%%%%%%%%%%%%
%%%%%%%%%%%%%%%%%%%%%%%%%%%%%%%%%%%%%%

\subsection{Anomalies}

Rationalization by linear salience can explain the following phenomena: (1) attraction effect, (2) compromise effect, and (3) avoidance of the handicapped. 
\smallskip

An \textsl{attraction effect} (or \textsl{decoy effect}) takes place when there is an increase of the probability to choose an item as soon as an asymmetrically dominated item is added to the menu.  
Originally studied by~\cite{HuberPaynePuto1982}, this phenomenon is modeled in a context of reference dependence and product differentiation by \cite{OkOrtolevaRiella2007,OkOrtolevaRiella2011,OkOrtolevaRiella2015}.
To illustrate it, consider a consumer who chooses between two goods $x,y$ with two distinct attributes.
Good $y$ is better that good $x$ on attribute~1, but $x$ overcomes $y$ on attribute~2. 
The consumer selects $y$ from $\lbrace x,y\rbrace$ (giving priority to attribute~1), but chooses $x$ from $\lbrace x,y,z\rbrace$, where $z$ is dominated by $x$ (but not by $y$) in both dimensions.
The new item $z$ acts as a decoy, enhancing the features of $x$ and inducing the consumer to favor attribute~2.

\begin{example}\label{EX:decoy}
Let $c$ be the choice on $X=\lbrace x,y,z\rbrace$ defined by $\underline{x}yz,\, x\underline{y},\,\underline{x}z,\,y\underline{z}\,.$
%(Note that this choice behavior displays a cyclic selection at a binary level.)
This choice is rationalizable by salience by $\langle\succsim, \L\rangle$, where $\succsim$ is defined by $z\succ x$, $z\succ y$, and $x\sim y$, and $\L$ is the set $\{\rhd_x,\rhd_y,\rhd_z\}$, with $\rhd_x=\rhd_y$, $y\rhd_{x} x\rhd_{x} z$, and $x\rhd_{z} z\rhd_{z} y$.
Here $\rhd_z$ reflects the ranking of items by the second attribute, whereas $\rhd_{x}=\rhd_{y}$ ranks items in accordance with the first attribute. 
Observe also that $z$ is more salient than both $x$ and $y$, and this provokes a shift of DM's preferences in the whole menu. 
%In fact, the presence of binary cycles does not constitute \textit{per se} an irrational violation of Axiom$\:\alpha$. 
\end{example}
 
%This is a typical case in which choice by salience is able to accommodate cyclic selections at a binary level. 

Choice by linear salience also explains the \textsl{compromise effect}, which accounts for an increase of the probability of selecting an item appearing as `intermediate' rather than `extreme' in a menu.
Compromise effect was first investigated by \cite{Simonson1989}, whose experiments show that a brand may gain market share when it becomes a compromise option in a choice set.\footnote{Later on, this phenomenon has been analyzed in various theoretical frameworks: see, e.g.,  \citet{KivetzNetzerSrinivasan2004}.} 
To illustrate it, consider a consumer who chooses among distinct versions of the same good, say $w,x,y,z$. 
According to their quality $q$, these items are ranked by $z>_{q} y>_{q} x>_{q} w$. 
A higher quality entails a lower affordability in price $p$, which yields the reverse ordering $w>_{p} x >_{p} y>_{p}z$.
A top-quality item is unlikely to be selected in a menu, whereas an intermediate alternative may be chosen. 
%For instance, in the menu $\{w,x,y\}$, item $y$ is the finest (and most expensive) good, and so the consumer is  discouraged to buy it, selecting instead a cheaper option, say $x$.
%Instead, in the menus $\{w,x,y,z\}$ and $\{x,y,z\}$, the variant $y$ appears to be a good compromise between the cheapest alternatives and the first-class version, and this induces the consumer to select it. 
%Linear salience can explain this anomaly as follows. 

\begin{example}\label{EX:compromise effect}
Let $c\colon \X\to X$ be the choice on $X= \lbrace w,x,y,z \rbrace$ defined by $wx\underline{y}z$, $w\underline{x}y$, $w\underline{x}z$, $w\underline{y}z$, $x\underline{y}z$, $\underline{w}x$, $w\underline{y}$, $w\underline{z}$, $\underline{x}y$, $\underline{x}z$, and $\underline{y}z$.
This choice is RLS: salience $\succsim$ is $z \succ y\succ w,x$ and $x\sim w$, whereas $\L$ is the family of linear orders $\{\rhd_w,\rhd_x,\rhd_y,\rhd_z\}$, where $z\rhd_{w} y\rhd_{w} x\rhd_{w} w$, $\rhd_x=\rhd_w$, $x\rhd_{y} y \rhd_{y} z \rhd_{y} w$, and $y\rhd_{z\!} x\rhd_{z\!} z\rhd_{z\!} w$. 
In any menu, the top quality good is the most salient item, and acts as a warning for the consumer, inducing her to accept an intermediate option.  
\end{example}

Finally, we show that the model of linear salience provides a sound explanation for the so-called \textsl{avoidance of the handicapped}.  
According to this behavioral pattern, tested by \cite{SnyderKleckTrentaMentzer} and mentioned in \cite{CherepanovFeddersenSandroni2013}, people masquerade motives behind their choice.
In the original experiment, three options are given: watching movie 1 alone ($x$), watching movie 2 alone ($y$), and watching movie 1 with a person in a wheelchair ($z$).
Several subjects, who must choose between $x$ and $z$, go for $z$.  
When their alternatives are $y$ and $z$, many subjects select $y$, apparently displaying a preference for movie 2 over movie 1. % (we assume that the same choice is performed when all three alternatives are available).
However, between $x$ and $y$ several subjects choose $x$, revealing a preference for movie 1 over movie 2.
The truth is that some subjects prefer movie 1 to movie 2, but they also want to avoid the handicapped, and are embarrassed by their motivation. 
Thus, in displaying a preference for watching movie 2 alone rather than watching movie 1 with the handicapped, they are hiding her real motive behind a \textit{false} preference for movie 2 over movie 1.\footnote{The handicapped avoidance is a between-subject experiment, so it does not actually allow us to observe people's choice functions.
However, choice by linear salience provides a sound interpretation of the behavior inferred from this experimental evidence.}
%Example~\ref{EX:decoy} (with $x$ and $y$ exchanged) describes a justification by salience of this phenomenon.

\begin{example}
	Define $c \colon \X \to X$ on $X=\lbrace x,y,z\rbrace$ by $x\underline{y}z,\, \underline{x}y,\,x\underline{z},\,\underline{y}z$. 
	(Note that $c$ is isomorphic to the choice in Example~\ref{EX:decoy}.) %with $x$ and $y$ exchanged.
	An RLS for $c$ is $\langle \succsim, \L \rangle$, where $z\succ x \sim y$, and $\L = \{\rhd_x,\rhd_y,\rhd_z\}$ is such that $\rhd_x=\rhd_y$, $x \rhd_{x} y \rhd_{x} z$, and $y \rhd_{z} z \rhd_{z} x$.
	The presence of the handicapped makes $z$ the most salient item, and induces the DM to hide her motives behind the preference of movie $2$ over movie $1$, as described by $\rhd_z$.
	When $z$ is not available, the subject shows her true preference, which ranks movie 1 over movie 2, and movie 1 with the handicapped is the least desirable option.\vs\vs 
\end{example}

%%%%%%%%%%%%%%%%%%%%%%%%%%%%%%%%%%%%%%%%%%%%%%%%%%%%%%
%%%%%%%%%%%%%%%%%%%%%%%%%%%%%%%%%%%%%%%%%%%%%%%%%%%%%%
%%%%%%%%%%%%%%%%%%%%%%%%%%%%%%%%%%%%%%%%%%%%%%%%%%%%%%
%%%%%%%%%%%%%%%%%%%%%%%%%%%%%%%%%%%%%%%%%%%%%%%%%%%%%%

\section{Concluding remarks} \label{SECTION:conclusion}

%\textbf{\magenta I do not like the first two paragraphs, they are a boring summary. Instead, we must emphasize the sequentiality of our approach, whose two stages are inspired by radically different philosophies. (Compare compensatory vs noncompensatory ahahah) This feature tells apart our models from others, whose  sequentiality is `homogeneous', resulting in a two-stage shrinking of the menu. We do not have two successive shrinking stages, but a focus and a shrinking.) (Also, write better the last paragraph.)}
The aim of this paper is to provide a general framework for context-sensitive behaviors, explaining how salience of items affects individual choice. 
%Our approach creates a link between (i) `universal' models (which explain any choice behavior and partition the family of all finite choices into classes of rationality) and (ii) `non-universal' models (which explain only a selected fraction of choices). 
%Let us explain why.
Choice by salience semantically extends the RMR model of \cite{KalaiRubinsteinSpiegler2002} by providing a structured explanation of choice behavior. 
 The classes of rationality prompted by RMR are refined by means of a partition of all choices in $n$ classes, where the last one encodes a notion of moodiness. 
For small ground sets, moodiness does not affect almost all choices: as a consequence, the partition in $n$ classes of rationality is empirically significative.  
We conjecture that all choices that can be explained by an existing (testable) model of bounded rationality are never moody. % contrary to what happens for context-free moody choices (see the last paragraph of Section~\ref{SECTION:general_model}).
 
The testable model of linear salience identifies a special class of choices with limited attention of \cite{MasatliogluNakajimaOzbay2012}, in which only non-conflicting violations of WARP are admitted.
On the other hand, choice by linear salience is independent from many other models of bounded rationality, such as the sequential rationalization of \cite{ManziniMariotti2007}, the theory of rationalization of \cite{CherepanovFeddersenSandroni2013}, and the model categorize-then-choose of \cite{ManziniMariotti2012}.    
In fact, the feature of sequentiality in a linear salience approach displays a crucial difference from existing models: salience does not reduce the set of available items, instead it endows the DM with a sound criterion to select a rationale to be maximized. 

The analysis of this paper hinges on a deterministic representation of salience, which implies that the perceived salience of items remains constant across menus. 
Possible extensions should consider a stochastic approach to salience, attained by considering a probability distribution over  different relations of salience. 
Moreover, although the assumption on the salience ordering is quite consolidated in the literature, the composition of menus may affect the role of the items in DM's perception, creating cycles of any length.
Thus, another possible direction of research is to design weaker properties of salience, which consider reversals of salience caused by different combinations of alternatives in distinct menus. 

%%%%%%%%%%%%%%%%%%%%%%%%%%%%%%%%%%%%%%%%%%%%%%%%%%%%%%
%%%%%%%%%%%%%%%%%%%%%%%%%%%%%%%%%%%%%%%%%%%%%%%%%%%%%%
%%%%%%%%%%%%%%%%%%%%%%%%%%%%%%%%%%%%%%%%%%%%%%%%%%%%%%
%%%%%%%%%%%%%%%%%%%%%%%%%%%%%%%%%%%%%%%%%%%%%%%%%%%%%%

\section*{Appendix A: Proofs}

\noindent \underline{\textbf{\large Proof of Lemma~\ref{LEMMA:extreme_GRS}}}. 
(i) Let $c \colon \X \to X$ be a choice function, where $\vert X \vert =n$. 
	Take $\succsim\: = \{(x,x) : x \in X\}$ (minimal), and define $n$ linear orders $\rhd_x$, with $x \in X$, with the property that $x$ is the top element of $\rhd_x$. 
	Clearly, for any $A \in \X$, $c(A) = \max (A, \rhd_x)$ for some $x \in A$.  
		%	(ii) Let $c \colon \X \to X$ be a rationalizable choice. 
%	Let $\succsim$ be any total preorder on $X$. 
%	Take the unique linear order that rationalizes $c$. 
%	Conversely, if $\langle \L,\succsim \rangle$ is a constant GRS on $X$, then, for any $x,y\in X$, $\rhd_x=\rhd_y=\rhd$.  
%	Take the choice function $c \colon \X \to X$ defined by $c(A)=\max(A,\rhd)$ for all $A \in \X$.
	\qed
\bigskip

\noindent \underline{\textbf{\large Proof of Theorem~\ref{THM:exists_chaos}}}.
We define a special type of choice function.
%(We keep the notation used in Lemma \ref{LEMMA:asymmetry implies acyclicity}.) 
 %\begin{definition}
%To prove this result, we assume that $X$ is linearly ordered
%Suppose $c$ is a choice function on  a set $X$. % = \{a,b,x_1,x_2,\ldots x_n\}$.
%If there are linear orders $<_x$ for each $x \in X$, not all different, 
%and, for any $A \subset X$, $c(A)$ is the maximum element of $A$ in 
%some $<_x$ for some $x \in A$,
%then we say $c$ is 2-EGRS.
%\end{definition}
\begin{definition}\label{DEF:flipped_choice} 
Let $\lessdot$ be a linear order on $X$, with $\vert X\vert\geq 6$.
A choice $c\colon \X\to X$ is  \textsl{flipped (w.r.t.\;$\lessdot$)} if for any $a,b,d,e,f,g\in X$ such that $a \lessdot b \lessdot d \lessdot e \lessdot f \lessdot g$,\vs\vs\vs
\[
c(ab) = a\,,\quad c(abd) = d\,, \quad c(abde) = b\,, \quad c(abde\!f) = e\,, \quad c(abde\!f\!g) = a\,.\vs\vs\vs  
\]
Thus, $c$ is flipped if the chosen items are the worst (on 2 items), the best (on 3), the second worst (on 4), the second best (on 5), and again the worst (on 6). 
 \end{definition} 
 
Then Theorem~\ref{THM:exists_chaos} is an immediate consequence of the following fact: 
 
 \begin{lemma} \label{LEMMA:flipping_generates_chaos}
 Any flipped choice on $39$ elements is moody.
 \end{lemma}
 
 \noindent \textsc{Proof.}
 Let $c \colon \X\to X$ be a flipped choice on the linearly ordered set $(X,\lessdot)$, where $\vert X\vert=39$. 
 Toward a contradiction, suppose $c$ is non-moody.
 Thus, there is an RS $\langle \succsim, \L \rangle$  for $c$ such that $\rhd_a = \rhd_b \in \L$ for some distinct $a,b \in X$; denote this linear order by $\rhd_{ab}$.   
 We can assume that $\succsim$ is $X \times X$, that is, $\succsim$ poses no constraints in the selection of the rationalizing linear order.   
 Without loss of generality, suppose $a \lessdot b$.

 Since $c$ is flipped, we have $c(ab)=a$, and so $a \rhd_{ab} b$.
 By the pigeon principle, there is $Y\subseteq X$, with $\vert Y \vert \geqslant \lfloor \frac{39-2}{3} \rfloor + 1= 13$, such that $a,b \notin Y$, and at least one of the following conditions holds:\vs\vs\vs
 \begin{itemize}
 	\item[(1)] $a \lessdot Y \lessdot b$, or\vs\vs\vs
 	\item[(2)] $Y \lessdot a \lessdot b$, or\vs\vs\vs
    \item[(3)] $a \lessdot b \lessdot Y$,\vs\vs\vs
 \end{itemize} 
 where $a \lessdot Y \lessdot b$ means $a \lessdot y \lessdot b$ for all $y \in Y$ (and a similar meaning have $a \lessdot b \lessdot Y$ and $Y \lessdot a \lessdot b$).   
Again by the pigeon principle, there is $Z \subseteq Y$, with $\vert Z \vert \geqslant 5$, such that at least one of the following cases happens:\vs\vs\vs
 \begin{itemize}
 	\item[(A)] $Z\rhd_{ab} a\rhd_{ab} b$, or\vs\vs\vs
 	\item[(B)] $a\rhd_{ab} Z\rhd_{ab} b$, or\vs\vs\vs
    \item[(C)] $a \rhd_{ab} b \rhd_{ab} Z$.\vs\vs\vs
 \end{itemize} 
A numbered case and a lettered case can overlap: we denote these cases by A1, A2, A3, B1, B2, B3, C1, C2, and C3, respectively. 
List the elements of $Z = \left\{z_1, z_2, \ldots , z_{\vert Z \vert} \right\}$ in increasing order according to $\lessdot$, that is, $z_1 \lessdot z_2 \lessdot \ldots \lessdot z_{\vert Z \vert}$.  
In what follows we examine all nine possible cases, and obtain a contradiction in each of them. \vs\vs
\begin{description}
\item[Case A1:] By definition of flipped choice, $c(a z b)=b$ holds for any $z \in Z$.
	It follows that $b \rhd_{z} z$ and $b \rhd_{z} a$ for any $z \in Z$.
The definition of flipped choice yields $c(a z_{h} z_{i} z_{j} z_{k} b)=a$ for any $h < i < j <k$. 
Thus, we must have either (i) $a \rhd_{ab} b$ and $a\rhd_{ab} z$ for any $z \in \{z_h,z_i,z_j,z_k\}$, or (ii) $a\rhd_{z} b$ and $a\rhd_{z} z'$ for some $z \in \{z_h,z_i,z_j,z_k\}$ and all $z' \in \{z_h,z_i,z_j,z_k\}$.
However, both (i) and (ii) are false.\vs\vs 
\item[Case A2:] Since $c(z a b) = b$ for any $z \in Z$, we have that $b \rhd_{z} z$ and $b \rhd_{z} a$ for any $z \in Z$.
Since $c(z_{h} z_{i} z_{j} a b) = a$ for any $h < i < j$, we must have either (i) $a \rhd_{ab} b$ and $a \rhd_{ab} z$ for all $z \in \{z_h,z_i,z_j\}$, or (ii) $a \rhd_{z} b$ and $a\rhd_{z} z'$ for some $z \in \{z_h,z_i,z_j\}$ and all $z' \in \{z_h,z_i,z_j\}$.
However, both (i) and (ii) are false.\vs\vs
\item[Case A3:] Since $c(a b z_{i} z_{j}) = b$ for any $i < j$, we have $b \rhd_{z} a$ for all but at most one $z \in Z$.
Since $c(a b z_{h} z_{i} z_{j} z_{k})= a$ for any $h < i < j < k$, we have either (i) $a \rhd_{ab} b$ and $a\rhd_{ab} z$ for all $z \in \{z_h,z_i,z_j,z_k\}$, or (ii) $a \rhd_{z} b$ and $a \rhd_{z} z'$ for some $z \in \{z_h,z_i,z_j,z_k\}$ and for all $z' \in \{z_h,z_i,z_j,z_k\}$.
Note that (i) is always false, hence $a \rhd_{z} b$ holds for some $z \in \{z_h,z_i,z_j,z_k\}$, say $z = z_h$. 
Since $\vert Z \vert \geqslant 5$, we can repeat the same argument using four items of $Z$ distinct from $z_h$, and conclude that $a \rhd_{z} b$ holds for at least two distinct $z \in Z$. 
However, this is impossible.\vs\vs 
\item[Case B1:] Since $c(a z_i b) = b$ for any $z_i \in Z$, we have\vs\vs\vs
\begin{equation}\label{COND:caseb1_first_step}
	b \rhd_{z_i} a \quad\text{ and }\quad  b \rhd_{z_i} z_i \vs\vs\vs 
\end{equation}	
for all $z_i \in Z$.
%Since $c(z_i,z_j)=z_i$ for any $i<j$, we must have that 
%\begin{equation}\label{COND:caseb1_second_step}
%z_i\rhd_{z_j}z_j\,\,\text{or}\,\, z_i\rhd_{z_i}z_j
%\end{equation} 
%for any $z_i,z_j\in Z$.
Since $c(a z_{i} z_{j} b) = z_{i}$ for any $i < j$, either (i)  $z_{i} \rhd_{z_i} a$, $z_{i} \rhd_{z_i} b$, and $z_{i} \rhd_{z_i} z_j$,  or (ii) $z_{i} \rhd_{z_j} a$, $z_{i} \rhd_{z_j} b$, and $z_{i} \rhd_{z_j} z_j$ holds.
Since (i) is impossible by condition~\eqref{COND:caseb1_first_step}, we get\vs\vs\vs  
\begin{equation}\label{COND:caseb1_third_step}
z_{i} \rhd_{z_j}a\,,\quad z_{i} \rhd_{z_j}b\,, \;\;\text{ and }\;\; z_{i} \rhd_{z_j} z_j \vs\vs\vs
\end{equation}
for all $i < j$.
Moreover, since $c(z_i z_j b)=b$ for any $i<j$, either (i) $b\rhd_{z_j} z_i$ and $b\rhd_{z_j} z_j$, or (ii) $b\rhd_{z_i}z_i$ and $b\rhd_{z_i} z_j$ holds.
Since (i) is impossible by condition \eqref{COND:caseb1_third_step}, we conclude\vs\vs 
\begin{equation}\label{COND:caseb1_fourth_step}
	b\rhd_{z_i} z_i \quad\text{ and }\quad b\rhd_{z_i} z_j \vs\vs
\end{equation}
for all $i < j$.
Finally, $c(a z_i z_j z_k b)=z_k$ for any $i<j<k$ yields 

- either (i) $z_k\rhd_{ab}a$, $z_k\rhd_{ab}b$, $z_k\rhd_{ab}z_i$, and $z_k\rhd_{ab}z_j$, 

- or (ii) $z_k\rhd_{z_i}a$, $z_k\rhd_{z_i}b$, $z_k\rhd_{z_i}z_i$, and $z_k\rhd_{z_i}z_j$, 

- or (iii) $z_k\rhd_{z_j}a$, $z_k\rhd_{z_j}b$, $z_k\rhd_{z_j}z_i$, and $z_k\rhd_{z_j}z_j$, 

- or (iv) $z_k\rhd_{z_k}a$, $z_k\rhd_{z_k}b$, $z_k\rhd_{z_k}z_i$, and $z_k\rhd_{z_k}z_j$.

Now we get a contradiction, because (i) is impossible by assumption, (ii) and (iii) are impossible by condition \eqref{COND:caseb1_fourth_step}, and (iv) is impossible by condition \eqref{COND:caseb1_first_step}.\vs\vs 
%
%Furthermore, $z_i \rhd_{z_i} b$ contradicts the hypothesis. 
%Therefore, we must have $z_i \rhd_{z_j} a$, $z_i \rhd_{z_j} b$, and $z_i \rhd_{z_j} z_j$ for all $i < j$.  \textbf{\magenta (need completion)}
%
%
\item[Case B2:] Since $c(z_i a b) = b$ for any $z_i \in Z$, we have\vs\vs\vs
\begin{equation}\label{COND:caseb2_first_step}
	b \rhd_{z_i} a \quad\text{ and }\quad  b \rhd_{z_i} z_i \vs\vs\vs 
\end{equation}	
for all $z_i \in Z$.
Since $c(z_{i} z_{j} a b) = z_{j}$ for any $i < j$, either (i)  $z_{j} \rhd_{z_j} a$, $z_{j} \rhd_{z_j} b$, and $z_{j} \rhd_{z_j} z_i$,  or (ii) $z_{j} \rhd_{z_i} a$, $z_{j} \rhd_{z_i} b$, and $z_{j} \rhd_{z_i} z_i$ holds.
Since (i) is impossible by condition \eqref{COND:caseb2_first_step}, we get\vs\vs\vs 
\begin{equation}\label{COND:caseb2_third_step}
z_{j} \rhd_{z_i}a\,,\quad z_{j} \rhd_{z_i}b\,, \;\;\text{ and }\;\; z_{j} \rhd_{z_i} z_i\vs\vs\vs
\end{equation}
for all $i < j$.
Furthermore, since $c(z_i z_j b)=b$ for any $i<j$, either (i) $b\rhd_{z_i} z_i$ and $b\rhd_{z_i} z_j$, or (ii) $b\rhd_{z_j}z_i$ and $b\rhd_{z_j} z_j$ holds.
Since (i) is impossible by condition~\eqref{COND:caseb2_third_step}, we conclude\vs\vs
\begin{equation}\label{COND:caseb2_fourth_step}
	b\rhd_{z_j} z_i \quad\text{ and }\quad b\rhd_{z_j} z_j \vs\vs
\end{equation}
for all $i < j$.
Finally, since $c(z_h z_i z_j z_k a b)=z_h$ for any $h < i < j < k$, we get 

- either (i) $z_h\rhd_{z_h} z_i$, $z_h\rhd_{z_h} z_j$, $z_h \rhd_{z_h}z_k$, $z_h \rhd_{z_h}a$, and $z_h \rhd_{z_h} b\,$,

- or (ii) $z_h \rhd_{z_i} z_i$, $z_h\rhd_{z_i} z_j$, $z_h \rhd_{z_i}z_k$, $z_h \rhd_{z_i}a$, and $z_h \rhd_{z_i} b\,$,

- or (iii) $z_h\rhd_{z_j} z_i$, $z_h\rhd_{z_j} z_j$, $z_h \rhd_{z_j}z_k$, $z_h \rhd_{z_j}a$, and $z_h \rhd_{z_j} b\,$,

- or (iv) $z_h\rhd_{z_k} z_i$, $z_h\rhd_{z_k} z_j$, $z_h \rhd_{z_k}z_k$, $z_h \rhd_{z_k}a$, and $z_h \rhd_{z_k} b\,$,

- or (v) $z_h\rhd_{ab} z_i$, $z_h\rhd_{ab} z_j$, $z_h \rhd_{ab}z_k$, $z_h \rhd_{ab}a$, and $z_h \rhd_{ab} b\,$.

However, (i)--(iv) contradict \eqref{COND:caseb2_third_step}, whereas (v) contradicts the hypothesis.\vs\vs 
\item[Case B3:] Since $c(a b z) = z$ for all $z \in Z$, we must have $z \rhd_{z} a$ and $z \rhd_{z} b$ for all $z \in Z$. 
Since $c(bz) = b$ for any $z\in Z$, we have (i) $b \rhd_{z} z$, or (ii) $b\rhd_{ab} z$ for all $z\in Z$.
However, both (i) and (ii) are false.\vs\vs
\item[Case C1:] Since $c(a z b) = b$ for all $z \in Z$, we must have $b \rhd_{z} a$ and $b \rhd_{z} z$ for all $z \in Z$.
Since $c(zb) = z$ for all $z\in Z$, we have either (i) $z\rhd_{ab}b$, or (ii) $z\rhd_z b$ for all $z\in Z$.
However, both (i) and (ii) are false.\vs\vs
\item[Case C2:] Since $c(z a b) = b$ for all $z \in Z$, we get $b \rhd_{z} a$ and $b \rhd_{z} z$ for all $z \in Z$.
        Since $c(zb)=z$ for all $z\in Z$, we have either (i) $z\rhd_{ab}b$, or (ii) $z\rhd_z b$ for all $z\in Z$.
        However, both (i) and (ii) are false.\vs\vs 
%Since $c(z_{i} z_{j} a b) = z_{j}$ for any $i < j$, we conclude that $z_{j} \rhd_{z} a$, $z_{j} \rhd_{z} b$, and $z_j \rhd_{z} z_i$ for some $z \in \{z_{i},z_{j}\}$. Since $z_j \rhd_{z_j} b$ cannot happen, we must have $z_{j} \rhd_{z_i} a$, $z_{j} \rhd_{z_i} b$, and $z_j \rhd_{z_i} z_i$ for any $i < j$. 
%Since $c(z_1 z_2 z_3 z_4 a b) = z_1$, 
%
%
\item[Case C3:] Since $c(a b z) = z$ for all $z \in Z$, $z \rhd_{z} a$ and $z \rhd_{z} b$ hold for all $z \in Z$.
Since $c(a b z_{i} z_{j}) = b$ for any $i < j$, we get $b \rhd_{z} a$, $b \rhd_{z} z_{i}$, and $b \rhd_{z} z_{j}$ for some  $z \in \{z_{i},z_{j}\}$, a contradiction.\vs  
\end{description}
\vs
This completes the proof of Lemma~\ref{LEMMA:flipping_generates_chaos}, and therefore of Theorem~\ref{THM:exists_chaos}. 
\qed
\bigskip

%%%%%%%%%%%%%%%%%%%%%%%%%%%%%%%%%%%%%%%%%
%%%%%%%%%%%%%%%%%%%%%%%%%%%%%%%%%%%%%%%%%

\noindent \underline{\textbf{\large Proof of Lemma~\ref{LEMMA:minimal_violations_of_alpha}}}.
Suppose there are $A,B \in \X$ such that $(A,B)$ is a switch, hence $c(A) \neq c(B) \in A$.  
If $\vert B\setminus A\vert=1$, the claim holds.
Thus, assume $\vert B\setminus A\vert > 1$, hence there are $x\in X$ and  $C \in \X$  such that $A \subsetneq C \subsetneq C \cup x = B$.
If $(C,B)$ is a (minimal) switch, then we are done again. 
Thus, suppose $(C,B)$ is not a switch.
\smallskip

\textsc{Claim:} \textit{$(A,C)$ is a switch.} 
By hypothesis, either (i) $c(B) = x$ or (ii) $c(B) = c(C)$ holds. 
Since $(A,B)$ is a switch, case (i) cannot happen, hence $c(B) = c(C) = b \neq x$. 
It follows that $b \in A \setminus c(A)$, because otherwise $(A,B)$ would fail to be a switch, contradicting the hypothesis. 
This proves that $(A,C)$ is a switch. 

\smallskip

Thus, the original violation of Axiom$\:\alpha$ witnessed by the switch $(A,B)$ takes place within the smaller pair $(A,C)$, where $C = B \setminus \{x\}$. 
If $(A,C)$ is minimal, then we are done.
Otherwise, we repeat the above argument, and show that there are $y \in X$ and $D \in \X$ such that $A\subsetneq D \subsetneq D \cup y = C$, and either $(A,D)$ or $(D,C)$ is a switch. 
In the latter case, we are done.
In the former case, $(A,D)$ is a switch, and we can continue as above.
Since $X$ is finite, we eventually obtain what we are after. 
(Note that the assumption of the finiteness of $X$ is essential in proving Lemma~\ref{LEMMA:minimal_violations_of_alpha}.
\qed 
\bigskip

\noindent \underline{\textbf{\large Proof of Lemma~\ref{LEMMA:NC_for_being_RS}}}.
	Suppose $c \colon \X \to X$ is RLS via the total preorder $\succsim$. 
	Let $A \in \X$ and $x \in X$ be such that $(A,A \cup x)$ is a switch, %violates $\alpha$
	 whence $c(A) =y$ and $c(A \cup x) = z \neq x,y$.
	Assume there is some $w\in A$ such that $w\succsim x$.
	This implies that $\max(A,\succsim)\subseteq \max(A \cup x,\succsim)$, hence, by normality, $c(A \cup x)=y$ or $c(A \cup x)=x$, which is false. 
	%Let $\{S_1,\ldots,S_n\}$ be the ordered partition of the menu $A \cup \{x\}$ in salience classes, that is, $S_i \neq \es$ for all $i$, $S_i \cap S_j = \es$ for $i \neq j$, $A \cup \{x\} = S_1 \cup \ldots \cup S_n$, and $S_1 \succ \ldots \succ S_n$. 
	%Further, let $\rhd_i$ be the linear order associated to the salience class $S_i$. 
	%By assumption, $z = c(A \cup \{x\}) = \max (A \cup \{x\},\rhd_1)$. 
	%Since $c(A) = y$ is different from both $x$ and $z$, we have $c(A) = \max(A,\rhd_2)$.  
	%It follows that $S_1 =\{x\}$, and $S_2 \cup \ldots \cup S_n$ is a partition of $A$.
	We conclude  that $x \succ a$ for all $a \in A$, as claimed. 
%	By assumption, there are linear orders $\rhd_1$ and $\rhd_2$ on $X$ such that 
%	$$
%	\max(A,\rhd_1) = c(A) = y \neq z = c(A \cup \{x\}) = \max(A \cup \{x\}\,\rhd_2)\,.      
%	$$
%	Note that since $z \neq x$, the two linear orders $\rhd_1$ and $\rhd_2$ must be distinct. %we have $z \in A$ and $z \rhd_2 a$ for all $a \in A \setminus \{z\} \cup \{x\}$.  
%	Let $\{S_1,\ldots,S_n\}$ be the ordered partition of $A \cup \{x\}$ in salience classes, that is, $S_i \neq \es$ for all $i$, $S_i \cap S_j = \es$ for $i \neq j$, $A \cup \{x\} = S_1 \cup \ldots \cup S_n$, and $S_1 \succ \ldots \succ S_n$.  	
%	Toward a contradiction, suppose there is $a \in A$ such that $x \succ a$ is false.
%	Thus, we have either $a \sim x$ or $a \succ x$. 
%	Thus, there must be $a \in A$ such that $a \in S_1$, which in turn implies that $a$ also belongs to the class of most salient items in $A$. 
%	However, the last fact yields that $\rhd_1$ coincides with $\rhd_2$, a contradiction. 
\qed 
\bigskip

\noindent \underline{\textbf{\large Proof of Lemma~\ref{LEMMA:asymmetry implies acyclicity}}}.
%This technical result requires some preparatory facts.   
In what follows, we fix a choice $c \colon \X \to X$, and denote by $\vDash $ the relation of revealed salience.
%Furthermore, to keep notation simple, we shall write $c(xyz)$ in place of $c(\lbrace x,y,z\rbrace)$, $c(xyz \cup Axyz)$ in place of $c(\lbrace x,y,z\rbrace \cup A)$, etc.
We first prove three preliminary results. %Lemmata~\ref{LEMMA:switching choice},~\ref{LEMMA:switching choice2}, and~\ref{LEMMA:Choice from the triplet}. 

\begin{lemma} \label{LEMMA:switching choice}
Let $A \in \X$ and $x,y\in X$ be such that $x \neq y \in A$ and $x \not\vDash y$.\vs\vs
\begin{itemize} 
	\item[\rm (i)] If $x \notin A$, then adding $x$ to $A$ does not switch the choice, except maybe to $x$.\vs\vs 
	\item[\rm (ii)] If $x \in A - c(A)$, then removing $x$ from $A$ does not  affect the choice.
\end{itemize}
\end{lemma}

\noindent \textsc{Proof.}
By Definition~\ref{DEF:revealed salience}, $x \vDash y$ means that there is $A \in \X$ such that $y \in A$ and $c(A) \neq c(A \cup x) \neq x$.
Thus $x \not\vDash y$ means that for any $A \in \X$ containing $y$, $c(A \cup x)$ is equal to either $x$ or $c(A)$.
Now both (i) and (ii) readily follow. 
\qed

\begin{lemma}\label{LEMMA:switching choice2}
For any $A,A^{\prime},B \in \X$ and $x\in X$, if $A^{\prime} \subseteq A$, $A \not\vDash  x$ and $x \in B$, then $c(B \cup A^{\prime}) \in A^{\prime} \cup c(B)$.
%So if $c(B \cup A^{\prime}) \not\in A^{\prime} $, then $c(B) = c(B \cup A^{'})$.
\end{lemma}

\noindent \textsc{Proof.}
%Suppose the claim is false. 
Take $A,A',B \in \X$ and $x \in B$ such that Lemma~\ref{LEMMA:switching choice2} fails, where $A^{\prime}$ is a subset of $A$ that is minimal for this failure. 
Thus, $A \not\vDash  x$ and $c(B \cup A') \notin A' \cup c(B)$. 
If $A^{\prime} = \{y\}$ for some $y\in X$, then $y \not\vDash  x$ and $c(B \cup y) \notin \{c(B),y\}$. 
However, this is impossible by Lemma~\ref{LEMMA:switching choice}. 
Next, consider the case $\vert A' \vert \geqslant 2$. 
Choose $y \in A^{\prime}$, and set $A^{\prime\prime} := A^{\prime} - y \subseteq A$.
By the minimality of $A^{\prime}$, we get $c(B \cup A^{\prime\prime}) \in  A^{\prime\prime} \cup c(B) \subseteq A^{\prime} \cup c(B)$.
It follows that $c(B \cup A'') \neq c(B \cup A^{\prime}) = c(B \cup A^{\prime\prime} \cup y)$, which contradicts $y \not\vDash x$.
\qed 

%In other words, if $A \not\vDash x \in B$, then adding new items coming from the menu $A$ to the menu $B$ does not switch the choice, except possibly for one of these new items.
%Moreover, if $A \not\vDash x \in B$, then removing unchosen items from $A$ other than $x$ does not switch the choice.
%
%The next result is the key tool for the proof of Lemma~\ref{LEMMA:asymmetry implies acyclicity}.
%It establishes how the choice acts on sets with three items under suitable conditions on revealed salience. 

\begin{lemma}[Choice on triples] \label{LEMMA:Choice from the triplet}
Suppose $\vDash$ is asymmetric.
For any distinct $x,y,z \in X$, if $x \vDash y$ and $x \not\vDash z \not\vDash y$, then $c(xyz) \neq y$.
\end{lemma}

\noindent \textsc{Proof.}
Let $x,y,z$ be distinct elements of $X$ satisfying the hypothesis. 
Since $x \vDash y$, there is $A \in \X$ such that $x, y \not\in A$, $c(A \cup y) \neq c(A \cup xy) \neq x$. 
Thus, we have $x \vDash A \cup y$, which in turn implies $y \not\vDash x$ and $A  \not\vDash x$ by the asymmetry of $\vDash $.
Note also that $z \notin A$, since otherwise $x \vDash  z$, contradicting the hypothesis. 
Now we can make the following deductions:\vs\vs\vs
\begin{itemize}
%\item If $c(acP) \not\in P$, then $c(acP) = c(ac)$.
%\item If $c(acP) \neq a$, then $c(acP) = c(cP)$.
	\item[\rm (i)] if $c(A \cup yz) \neq z$, then $c(A \cup yz) = c(A \cup y)$ (since $z \not\vDash y$);\vs\vs\vs
	\item[\rm (ii)] if $c(A \cup xyz) \neq x$, then $c(A \cup xyz) = c(A \cup yz)$ (since $x \not\vDash z$);\vs\vs\vs
	\item[\rm (iii)] if $c(A \cup xyz) \neq y$, then $c(A \cup xyz) = c(A \cup xz)$ (since $y \not\vDash x$);\vs\vs\vs
	\item[\rm (iv)] if $c(A \cup xyz) \neq z$, then $c(A \cup xyz) = c(A \cup xy)$ (since $z \not\vDash y$);\vs\vs\vs
	\item[\rm (v)] if $c(A \cup xyz) \notin A$, then $c(A \cup xyz) = c(xyz)$ (by Lemma~\ref{LEMMA:switching choice2}, since $A \not\vDash x$).\vs\vs
\end{itemize}
Three cases: (1) $c(A \cup xyz) \in A$; (2) $c(A \cup xyz) = y$; (3) $c(A \cup xyz) \in \{x,z\}$. 

In case (1), the implications (ii), (iii), and (iv) yield $c(A \cup yz) = c(A \cup xz) = c(A \cup xy)$, hence these chosen items are all equal to some $a \in A$.
Now (i) applies, and so $c(A \cup y) = a$, which contradicts $c(A \cup xy) \neq c(A \cup y)$.

In case (2), the implications (ii), (iv), and (v) yield $c(A \cup yz) = c(A \cup xy) = c(xyz)$, hence these chosen items  are all equal to $y$. 
Now (i) applies, and so $c(A \cup y) = y$, which again contradicts $c(A \cup xy) \neq c(A \cup y)$.

It follows that case (3) holds, and so the implication (v) yields $c(A \cup xyz) = c(xyz)$.
This implies $c(xyz) \neq y$, thus completing the proof of Lemma~\ref{LEMMA:Choice from the triplet}.
\qed 

\medskip
We now proceed to the combinatorial proof of Lemma \ref{LEMMA:asymmetry implies acyclicity}.
Toward a contradiction, suppose $\vDash $ is asymmetric, but there is a $\vDash $-cycle of minimum length, say $a_1 \vDash  a_2 \vDash  \ldots  \vDash  a_n \vDash  a_1$, where $n \geqslant 3$ and all $a_i$'s are distinct; let $C=\{a_1,a_2,\ldots,a_n\}$ be the set of items involved in the cycle.  
To start, assume $n = 3$, that is, $a_1 \vDash  a_2 \vDash  a_3 \vDash  a_1$ and $C = \{a_1,a_2,a_3\}$. 
Using the asymmetry of $\vDash $ and applying Lemma~\ref{LEMMA:Choice from the triplet}, we get:\vs\vs\vs
\begin{itemize}
	\item[(1)] $a_1 \vDash  a_2$ and $a_1 \not\vDash  a_3 \not\vDash  a_2$, hence $c(a_1a_2a_3) \neq a_2$;\vs\vs\vs 
	\item[(2)] $a_2 \vDash  a_3$ and $a_2 \not\vDash  a_1 \not\vDash  a_3$, hence $c(a_1a_2a_3) \neq a_3$;\vs\vs\vs 
	\item[(3)] $a_3 \vDash  a_1$ and $a_3 \not\vDash  a_2 \not\vDash  a_1$, hence $c(a_1a_2a_3) \neq a_1$.\vs\vs\vs
\end{itemize}
Thus $c(C)$ is empty, a contradiction. 
Next, assume $n = 4$, i.e., $a_1 \vDash  a_2 \vDash  a_3 \vDash  a_4 \vDash  a_1$ and $C = \{a_1,a_2,a_3,a_4\}$.
Minimality yields $a_1 \not\vDash  a_3 \not\vDash a_1$ and $a_2 \not\vDash  a_4 \not\vDash  a_2$, and asymmetry entails $a_1 \not\vDash   a_4 \not\vDash   a_3 \not\vDash   a_2 \not\vDash   a_1$. 
Using again Lemma~\ref{LEMMA:Choice from the triplet}, we now make the following deductions:\vs\vs\vs
\begin{itemize}
	\item[(1)] $a_1 \vDash  a_2$ and $a_1 \not\vDash  a_3 \not\vDash  a_2$, hence $c(a_1a_2a_3) \neq a_2$;\vs\vs\vs 
	\item[(2)] $a_2 \vDash  a_3$ and $a_2 \not\vDash  a_1 \not\vDash  a_3$, hence $c(a_1a_2a_3) \neq a_3$;\vs\vs\vs 
	\item[(3)] $a_2 \vDash  a_3$ and $a_2 \not\vDash  a_4 \not\vDash  a_3$, hence $c(a_2a_3a_4) \neq a_3$;\vs\vs\vs
	\item[(4)] $a_3 \vDash  a_4$ and $a_3 \not\vDash  a_2 \not\vDash  a_4$, hence $c(a_2a_3a_4) \neq a_4$;\vs\vs\vs
	\item[(5)] $a_3 \vDash  a_4$ and $a_3 \not\vDash  a_1 \not\vDash  a_4$, hence $c(a_3a_4a_1) \neq a_4$;\vs\vs\vs
	\item[(6)] $a_4 \vDash  a_1$ and $a_3 \not\vDash  a_2 \not\vDash  a_1$, hence $c(a_3a_4a_1) \neq a_1$.\vs\vs\vs
\end{itemize}
Thus, we have $c(a_1 a_2 a_3) = a_1$, $c(a_2 a_3 a_4) = a_2$, and $c(a_3 a_4 a_1) = a_3$. 
In what follows, we derive again that $c(C)$ is empty, a contradiction.  
Indeed, $a_4 \not\vDash a_3$ implies that $c(C)$ is equal to either $a_4$ or $c(a_1a_2a_3)=a_1$.
Similarly, $a_1 \not\vDash a_4$ implies that $c(C)$ is equal to either $a_1$ or $c(a_2a_3a_4)=a_2$, and $a_2 \not\vDash a_1$ implies that $c(C)$ is equal to either $a_2$ or $c(a_3a_4a_1)=a_3$. 
Summarizing, we have\vs\vs\vs
\[
c(C) \in \{a_1,a_2\} \cap \{a_1,a_3\} \cap \{a_2,a_3\} = \es
\vs\vs\vs
\]
as claimed. 
In the general case, let $a_1 \vDash  a_2 \vDash  \ldots  \vDash  a_n \vDash  a_1$, with $C = \{a_1,a_2,\ldots, a_n\}$. 
By a similar argument (or induction), we get\vs\vs\vs
\[
c(a_1 a_2 \ldots a_{n-1}) = a_1\,,\; c(a_2 a_3 \ldots a_n) = a_2\,,\; c(a_3 a_4 \ldots a_1) = a_{3}\,.\vs\vs\vs
\]
Now, $a_n \not\vDash  a_{n-1}$ implies $c(C) \in \{a_1,a_n\}$, $a_1 \not\vDash  a_n$ implies $c(C) \in \{a_1,a_2\}$, and $a_2 \not\vDash  a_1$ implies $c(C) \in \{a_2,a_3\}$, and so $c(C) = \{a_1,a_{n-1}\} \cap \{a_1,a_2\} \cap \{a_2,a_3\} = \es$, which is impossible.  
This completes the proof of Lemma~\ref{LEMMA:asymmetry implies acyclicity}.\qed 

\bigskip
\bigskip

%%%%%%%%%%%%%%%%%%%%%%%%%%%%%%%%%%%
%%%%%%%%%%%%%%%%%%%%%%%%%%%%%%%%%%%

\noindent \underline{\textbf{\large Proof of Theorem~\ref{THM:main}}}.
Fix a choice $c \colon \X \to X$, and let $\vDash $ be the relation of salience revealed by $c$.
The proof that (ii), (iii), and (iv) are all equivalent statements is straightforward, and is left to the reader. 
%
%We prove the contrapositive for all implications.
% 
%(ii)$\:\Longrightarrow\:$(iii):  
%Suppose there exist two conflicting menus $A,B\in\X$. 
%By definition, there are $a\in A$ and $b\in B$ such that $(A,Ab)$ and $(B,Ba)$ are switches. 
%This implies $b\vDash a$ and $a\vDash b$, that is, $\vDash$ is not asymmetric.
%
%(iii)$\:\Longrightarrow\:$(ii):
%Suppose $\vDash$ fails to be asymmetric, hence $b\vDash a$ and $a\vDash b$ for some $a,b\in X$.
%By definition, there are $A,B \in \X$ such that $a\in A$, $b\in B$, and $(A,Ab)$ and $(B,Ba)$ are switches.
%It follows that $A$ and $B$ are conflicting menus.
%
%(iii)$\:\Longrightarrow\:$(iv):  
%Suppose WARP(S) fails for $c$.
%Therefore, there are $A,B\in\X$, $a \in A$, and $b \in B$ such that $b \neq c(Ab) \neq c(A)$ and $c(Ba) \neq c(B)$, but $c(Ba) \neq a$.
%By definition, the two menus $A$ and $B$ are conflicting.
%
%(iv)$\:\Longrightarrow\:$(ii):
%Suppose $\vDash$ is not asymmetric, hence there are $a,b\in X$ such that $a \vDash b$ and $b \vDash a$.
%By definition, there are menus $A,B \in\X$ such that $b \neq c(Ab)\neq c(A)$ and $a \neq c(Ba) \neq c(B)$.  
%This violates WARP(S).
%
To prove that (i) implies (ii), assume $c$ is rationalizable by salience by a total  preorder $\succsim$.
By Lemma~\ref{LEMMA:NC_for_being_RS} and Definition~\ref{DEF:revealed salience}, $\succ$ is an asymmetric extension of $\vDash$.
It follows that $\vDash$ is asymmetric as well.  

To complete the proof of Theorem~\ref{THM:main}, it remains to show that (ii) implies (i). 
We need some preliminary results, namely Lemmata~\ref{LEMMA:existence of a weak order including revealed salience},~\ref{LEMMA:Alpha coherence}, and~\ref{LEMMA:Transitivity and Asymmetry of revealed preference}.

\begin{lemma}\label{LEMMA:existence of a weak order including revealed salience}
If $\vDash$ is asymmetric, then there is a total preorder that extends the transitive closure of~$\vDash$.
\end{lemma}

\noindent \textsc{Proof.}
Asymmetry of $\vDash$ implies its acyclicity by Lemma~\ref{LEMMA:asymmetry implies acyclicity}.
By \cite{Szpilrajn1930}'s theorem, there is a total preorder extending the transitive closure of $\vDash$.
\qed 

\bigskip

\noindent \textsc{Notation:} In what follows, $\succsim$ denotes a total preorder that extends the transitive closure of $\vDash $, whereas $\succ$ is the strict part of $\succsim$.
Furthermore, for any $x\in X$, set $x^\downarrow := \lbrace y \in X : x \succsim y \rbrace.$ 

\medskip

%The asymmetry of $\vDash$ ensures that subsets of $x^\downarrow$ containing $x$ do not violate $\alpha$:

\begin{lemma}\label{LEMMA:Alpha coherence}
If $\vDash $ is asymmetric, then any pair $(A,B)$ of menus included in $x^\downarrow$ is not a switch, as long as $x$ belongs to both $A$ and $B$. %intersect $S$. 
\end{lemma}

\noindent \textsc{Proof.}
Suppose $\vDash $ is asymmetric.
%Denote with $\succ$ a weak order extending the transitive closure of $\vDash $.
Toward a contradiction, assume there are $x\in X$ and menus $A,B \in \X$, with $A \subsetneq B\subseteq x^\downarrow$, and $x\in A$, such that $(A,B)$ is a switch. 
By Lemma~\ref{LEMMA:minimal_violations_of_alpha}, there are $y \in X$ and $C \in \X$ such that $A\subseteq C \subsetneq C \cup y \subseteq B$ and $(C,C \cup y)$ is a switch. 
It follows that $y \vDash C \supseteq A$, and so, in particular, $y \succ A$, because $\succ$ extends $\vDash$.
We conclude that $y\succ x$, a contradiction. 
%
%We distinguish two cases: (1) $x \in (\leftarrow, S]_\succsim \setminus S$; (2) $x \in S$. 
% 
%In case (1), the definition of $(\leftarrow, S]$ yields $S \succ x$.
%Furthermore, since $A \cap S \neq \es$, there is $y \in (A \cap S)$ such that $x \succ y$.
%However, $y \in S$ implies $y \succ x$, contradicting the asymmetry of $\succ$.
%If $x\not\succ y$, then, since $y\in (\leftarrow, S)_\succ \setminus S $,  exists $z\in S$, $z\neq x$, such that $z\succ y$.
%Note that $z\not\succ x$, since $x,z\in S$, and $x\not\succ y$, but $z\succ y$, contradicting negative transitivity of $\succ$.
%
%In case (2), $A \cap S \neq \es$ implies that there is $z\in S$, $z\neq x$, belonging to $A$.
%However, $x,z \in S \in \mathcal{S}_\succ$ implies $x \sim z$, whereas $z \in A$ implies $x \succ z$, a contradiction. 
\qed
\medskip

Next, we define a binary relation $>_x$ for each $x\in X$.
It will turn out that each $>_x$ is the strict part of a partial order whenever the relation of revealed salience is asymmetric (see Lemma~\ref{LEMMA:Transitivity and Asymmetry of revealed preference}). 
For any $x\in X$ and distinct $y,z \in x^\downarrow$, define\vs\vs\vs
\begin{equation} \label{EQ:partial order}
y>_{x} z \qquad \Longleftrightarrow \qquad \text{there is } A \subseteq x^\downarrow \text{ such that } x,y,z \in A \text{ and } y = c(A).\vs\vs\vs
\end{equation}
Note that if either $y$ or $z$ (or both) does not belong to $x^\downarrow$, then we leave $y$ and $z$ incomparable. 
Observe also that $>_x$ is irreflexive by construction.
We shall abuse notation, and write $y >_{x} A$, whenever exists $A \subseteq x^\downarrow$ such that $x,y\in A$ and $y=c(A)$.  
The reason for this abuse of notation is that $y >_x A$ implies $y >_{x} z$ for any $z\in A\setminus \lbrace y\rbrace$.

%As announced, we have:

\begin{lemma}\label{LEMMA:Transitivity and Asymmetry of revealed preference}
If $\vDash$ is asymmetric, then $>_{x}$ is asymmetric and transitive for any $x\in X$.
\end{lemma}

\noindent \textsc{Proof.}
Assume $\vDash$ is asymmetric, and let $x \in X$. % and let $\succ$ be a weak order extending the transitive closure of $ \vDash $.
%We show that $>_{x}$ is asymmetric and transitive. 
%irreflexive by construction, and the joint satisfaction of irreflexivity and transitivity implies asymmetry, it suffices to show that $>_x$ is transitive.  
To prove that $>_x$ is asymmetric, suppose by way of contradiction that $y >_x z$ and $z >_x y$ for some $y,z \in X$. 
(Note that $y \neq z$, because $>_x$ is irreflexive by construction.) 
By the definition of $>_x$, there are $A,B \subseteq x^\downarrow$ such that $x,y,z\in A\cap B$, $y = c(A)$, and $z = c(B)$. 
Consider the menu $A \cap B$, which is included in $x^\downarrow$ and contains $x,y,z$. 
If $c(A \cap B) \notin \{y,z\}$, then $(A \cap B, A)$ is a switch, which contradicts Lemma~\ref{LEMMA:Alpha coherence}.
On the other hand, if $c(A \cap B) =y$ (resp.\ $c(A \cap B)=z$), then $(A \cap B,B)$ (resp.\ $(A \cap B,A)$) is a switch, which is again forbidden by Lemma~\ref{LEMMA:Alpha coherence}.
Thus $c(A \cap B)$ is empty, which is impossible.  

To prove $>_x$ is transitive, let $w,y,z\in X$ be such that  $w>_{x}y>_{x}z$.
By the definition of $>_x$, there are $A,B \subseteq x^\downarrow$ such that $x,y\in A\cap B$, $z\in B$, $w=c(A)$, and $y=c(B)$.
Consider the menu $A \cup B$, which is included in $x^\downarrow$ and contains $x$. 
We claim that $c(A \cup B) = w$. 
Indeed, if $c(A\cup B) \in (A\cup B) - \lbrace w,y\rbrace $, then either $(A, A\cup B)$ or $(B, A\cup B)$ is a switch, which contradicts Lemma~\ref{LEMMA:Alpha coherence}. 
Moreover, if $c(A\cup B)=y$, then $(A, A\cup B)$ is a switch, which is impossible by  Lemma~\ref{LEMMA:Alpha coherence}. 
This proves the claim. 
Now we get $w>_{x}(A\cup B)$, hence $w>_{x}z$, as wanted.
%This proves that $>_x$ is transitive, and therefore it is asymmetric.
%\footnote{Note that $w\neq z$, since otherwise either $(A, A\cup B)$ or $((A\cup B),B)$ would be a switch.}
\qed

\bigskip

Now we complete the proof of Theorem~\ref{THM:main}. 
Suppose $\vDash $ is asymmetric.
Let $\succsim$ be a total preorder extending the transitive closure of $\vDash $, which exists by Lemma~\ref{LEMMA:existence of a weak order including revealed salience}.
 %and denote by $\mathcal{S}_\succ$ the linear partition of $X$ associated to $\succ$.
For any $x\in X$, define the binary relation $>_{x}$ as in~\eqref{EQ:partial order}.
By Lemma~\ref{LEMMA:Transitivity and Asymmetry of revealed preference}, each $>_x$ is asymmetric and transitive, thus it is the strict part of a partial order.
For any $x\in X$, let $\rhd_{x}$ be a linear extension of $>_{x}$, which exists by \cite{Szpilrajn1930}'s Theorem. 
%
%Then we have: 
%
%\begin{lemma}\label{LEMMA:Maximization} 
%For any $x\in X$ and $S\in\mathcal{S}_\succ$, if $x>_{S}A$ for some $A \subseteq (\leftarrow, S]_\succsim$, then $x=\max(A,\rhd_{S})$.
%\end{lemma} 
%
%\noindent \textsc{Proof.}
%Let $x\in X$ and $S\in\mathcal{S}_\succ$, and suppose $x>_{S}A$ for some $A \subseteq (\leftarrow, S]_\succsim$.
%This implies $x >_{S} y$ for all $y \in A\setminus\lbrace x\rbrace$.
%Moreover, $>_{S}$ is asymmetric, thus $\max(A,>_S)$ is unique and is exactly $x$. 
%Since $\rhd_{S}$ is asymmetric and extend $>_{S}$, we conclude that $\max(A,\rhd_{S})=y$. 
%\qed
%
%For any $S\in\mathcal{S}_\succ$, define $(\leftarrow, S)_\succ:=\lbrace y\in X\,\vert \,\forall\, x\in S\, (x\,\succ y)\rbrace \cup S.$
%if there is $A\subseteq\, (\leftarrow, S)_\succ$, such that $x\in A$, $y=c(A)$, and $A$ contains some element of $S$.
Now let $A$ be an arbitrary menu, and denote by $x$ be an item belonging to $\max(A,\succsim)$.
% that is, $\max(A,\succ) \subseteq S_{A}$.
Note that $A \subseteq x^\downarrow$.
By construction, $c(A)>_{x}A$, hence we can conclude $c(A)=\max\left(A,\rhd_{x}\right)$. 
This proves that $c$ is rationalizable by salience.  
\qed
\bigskip

\noindent \underline{\textbf{\large Proof of Lemma~\ref{LEMMA:RLS_is_CLA}}}.
	Let $c \colon \X \to X$ be an RLS choice.
	By Theorem~\ref{THM:main}, revealed salience $\vDash$ is asymmetric and acyclic, hence so is its reverse $\widetilde P$.   
	To prove that $c$ is CLA, we show that also the relation $P$ is asymmetric and acyclic. 
	To that end, it suffices to prove that $P$ is included in $\widetilde P$.
	Indeed, for all distinct $x,y \in X$, we have\vs\vs\vs
	\begin{align*}
			x \vDash y \quad & \Longleftrightarrow\quad \text{there is a menu $A$ such that } y \in A \text{ and } (A,A \cup x) \text{ is a switch} \\	
			& \Longleftrightarrow\quad \text{there is a menu $A$ such that } y \in A \text{ and } c(A) \neq c(A \cup x) \neq x \\
			& \Longleftrightarrow\quad \text{there is a menu $A$ such that } y \in A \text{ and } x \neq c(A) \neq c(A -x).\vs\vs\vs
		\end{align*}
		Thus $\widetilde P$ is defined by \eqref{EQ:Q_in_RLS}. 
		Since \eqref{EQ:P_in_CLA} implies \eqref{EQ:Q_in_RLS}, we obtain $P \subseteq \widetilde P$, as claimed. \qed
\bigskip

%\noindent \underline{\textsc{Proof of Lemma~\ref{LEMMA:properties_of_CLA}}}.
%	Parts (i) and (ii) are proved by \citet[p.\,2202]{MasatliogluNakajimaOzbay2012}, thus we only show (iii). 
%	Let $(\Gamma,\rhd)$ be a rationalization of $c$. 
%	
%	To start, we prove $\rhd \in \mathscr E_P$. 
%	Toward a contradiction, suppose $\rhd$ does not extend $P$. 
%	Thus, there are distinct $x,y \in X$ such that $x P y$ but $y \rhd x$.  
%	By definition, there is $A \in \X$ such that $x = c(A) \neq c(A -y)$. 
%	Since $x = \max(\Gamma(A),\rhd)$, $y \in A$, and $y \rhd x$, we derive $y \notin \Gamma(A)$, whence $\Gamma(A) = \Gamma(A -y)$ because $\Gamma$ is an attention filter. 
%	However, this implies $c(A -y) = \max(\Gamma(A -y),\rhd) = \max(\Gamma(A),\rhd) = c(A)$, which is impossible.
%	
%	Next, we show that $\Gamma(A) \subseteq \Gamma_\rhd(A)$ for all $A \in \X$. 
%	Toward a contradiction, suppose there is $A \in \X$ such that $\Gamma(A) \nsubseteq \Gamma_\rhd(A) = \{x \in A : c(A) \rhd x \} \cup \{c(A)\}$. 
%	Note that $c(A)$ must belong to $\Gamma(A)$, since otherwise $c(A) \neq \max(\Gamma(A),\rhd)$. 
%	It follows that there is $y \in \Gamma(A) \setminus \Gamma_\rhd(A)$ such that $y \rhd c(A)$. 
%	However, this clashes with $c(A) = \max(\Gamma(A),\rhd)$. \qed
%	\bigskip

\noindent \underline{\textbf{\large Proof of Proposition~\ref{PROP:CSLA=RLS}}}. 
$(\Longrightarrow)$ 
Suppose $c:\X\to X$ is RLS. 
By Theorem~\ref{THM:main} and Lemma~\ref{LEMMA:asymmetry implies acyclicity}, $\widetilde{P}$ is acyclic and asymmetric. 
Let $\rhd$ be any \textit{linear extension} of $\widetilde{P}$ (i.e., $\rhd$ is a linear order and contains $\widetilde{P}$).
Denoted $x^\downarrow  := \{y \in X : x \rhd y \text { or } y = x\}$ for any $x \in X$, define a choice correspondence $\Gamma_\rhd \colon \X \to \X$ as follows for all $A \in \X$:\vs\vs\vs
\begin{equation} \label{EQ:def_Gamma}
\Gamma_\rhd (A) := c(A)^\downarrow \cap A.\vs\vs\vs 	
\end{equation}
%\{a \in A : c(A) \rhd a\} \cup \{c(A)\} \quad\text{ for all $A \in \X$.}\vs\vs 
%
We claim that (i) $c(A) = \max (\Gamma_\rhd(A),\rhd)$ for all $A \in \X$, and (ii) $\Gamma_\rhd$ is a salient attention filter: this will show that $c$ is a CSLA. 
The first claim readily follows from the definition of $\Gamma_\rhd$.
To prove (ii), let $B \in \X$ and $x \in X$. 
We deal separately with the two possible cases: (1) $x \notin \Gamma_\rhd(B)$, and (2) $x \in \Gamma_\rhd(B)$, but $x \neq \min(B,\rhd), \max (\Gamma_\rhd(B),\rhd)$.\vs\vs
%Toward a contradiction, suppose there is $x \in B$, distinct from $\min(\Gamma(B),\rhd)$ and $\max(\Gamma(B),\rhd)$, such that $\Gamma(B) \neq \Gamma(B - x)$.
\begin{description}
	\item[\rm \textsc{Case 1:}] By \eqref{EQ:def_Gamma}, we get $x \rhd c(B)$.
	Since $\rhd$ extends $\widetilde{P}$, $\widetilde{P}$ is the converse of $\vDash$, and $\vDash$ is asymmetric, we derive that $x \vDash c(B)$ fails to hold, and so there is no menu $D \in \X$ such that $c(B) \in D$ and $x \neq c(D) \neq c(D - x)$. 
	It follows that we must have $c(B) = c(B -x)$, since otherwise $D := B$ would be a menu witnessing $x \vDash c(B)$, which is impossible. 
	Now the definition of $\Gamma_\rhd$ and the hypothesis $x \notin \Gamma_\rhd(B)$ yield $\Gamma_\rhd(B) - x = \Gamma_\rhd(B) = \Gamma_\rhd(B -x)$, as claimed.\vs\vs 
%	(Note that this shows that $\Gamma$ is an attention filter, and so $c$ is CLA.)	
%
	\item[\rm \textsc{Case 2:}] %$x\in\Gamma_\rhd(B)$ and $x \not\in \{\min(B,\rhd), \max (\Gamma_\rhd(B),\rhd)\}$. 
	Since $x \neq \max (\Gamma_\rhd(B),\rhd)$, formula~\eqref{EQ:def_Gamma} gives $c(B) \rhd x$. 
	We claim that $c(B) = c(B -x)$. 
	Indeed, we have:\vs\vs\vs
	\begin{align*}
		c(B) \neq c(B -x) \; 
		& \Longrightarrow \; x \neq c(B) \neq c(B -x) \\
		& \Longrightarrow \; (B-x,B) \text{ is a switch} & \text{\small (by the definition of a switch)}\\
		& \Longrightarrow \; x \vDash (B -x)  & \text{\small (by the definition of $\vDash$)} \\
		& \Longrightarrow \; (B-x) \widetilde{P} x & \text{\small (because $\widetilde{P}$ is the converse of $\vDash$)} \\ 
		& \Longrightarrow \; d (B-x) \rhd x & \text{\small (because $\rhd$ extends $\widetilde{P}$)} \\
		& \Longrightarrow \; x = \min(B,\rhd) 
		\vs\vs\vs
	\end{align*}
	which contradicts the hypothesis $x \neq \min(B,\rhd)$. 
	Now \eqref{EQ:def_Gamma} and the claim yield\vs\vs\vs
	$$
	\Gamma(B) - x = (c(B)^\downarrow \cap B) - x = c(B- x)^\downarrow \cap (B -x) = \Gamma(B -x),\vs\vs\vs
	$$
	as wanted. 
	This completes the proof of necessity.
\end{description} 

$(\Longleftarrow)$\footnote{We thank Davide Carpentiere for providing this simple proof.} 
Suppose $c\colon\X\to X$ is a CSLA.  
In what follows, we say that $(\Gamma,\rhd)$ \textsl{rationalizes} $c$ if $c(A) = \max (\Gamma(A),\rhd)$ for all $A \in \X$, where $\rhd$ is a linear order on $X$, and $\Gamma$ is a salient attention filter. 
Furthermore, we say that $(\Gamma,\rhd)$ \textsl{maximally rationalize $c$} if $(\Gamma,\rhd)$ rationalizes $c$, and there is no salient attention filter $\Gamma'\colon\X\to\X$ distinct from $\Gamma$ such that $(\Gamma',\rhd)$ rationalizes $c$ and $\Gamma(A)\subseteq\Gamma'(A)$ for all $A \in \X$.

\begin{lemma}\label{LEM:maximal_rationalizer_form}
	If $(\Gamma,\rhd)$ rationalizes $c$, then $(\Gamma_\rhd,\rhd)$ maximally rationalizes $c$.
\end{lemma}

\begin{proof}
	Suppose $(\Gamma,\rhd)$ rationalizes $c$. 
	To prove the claim, we show:\vs\vs\vs
	\begin{itemize}
		\item[(i)] $c(A) = \max (\Gamma_\rhd(A),\rhd)$ for all $A \in \X$;\vs\vs\vs
		\item[(ii)] $\Gamma_\rhd$ is a salient attention filter;\vs\vs\vs
		\item[(iii)] $\Gamma_\rhd$ is maximal.\vs\vs\vs
	\end{itemize} 
Part (i) readily follows from the definition \eqref{EQ:def_Gamma} of $\Gamma_\rhd$.
For (ii), let $B \in \X$ be any menu, and $x$ an item of $B$ different from both $\min(B,\rhd)$ and $\max(\Gamma_\rhd(B),\rhd)$.
Toward a contradiction, suppose $\Gamma_\rhd(B)-x \neq\Gamma_\rhd(B-x)$.
The definition of $\Gamma_\rhd$ yields $\left(c(B)^{\downarrow}\cap B\right)-x\neq c(B-x)^{\downarrow}\cap (B-x)$, hence $c(B)\neq c(B-x)$.
Moreover, we have $x\neq \max(\Gamma(B),\rhd)$. 
Since $\Gamma(B)-x=\Gamma(B-x)$ because $\Gamma$ is a salient attention filter, we obtain $c(B)\in\Gamma(B-x)$ and $c(B-x)\in\Gamma(B)$, which respectively yield $c(B-x)\rhd c(B)$ and $c(B)\rhd c(B-x)$, a contradiction. 
To prove (iii), suppose by way of contradiction that there is a salient attention filter $\Gamma^{\prime}$ such that $(\Gamma^{\prime},\rhd)$ rationalizes $c$ and $y\in \Gamma^{\prime}(D)-\Gamma_\rhd(D)$ for some $D \in \X$ and $y\in D$.
Since $y\notin \Gamma_\rhd(D)$, we get $y\rhd c(D)$.
On the other hand, since $y \in \Gamma'(D)$ and $(\Gamma^{\prime},\rhd)$ rationalizes $c$, we must have $c(D)\rhd y$ or $c(D) = y$, which is impossible. 
\end{proof}

%Note that Lemma~\ref{LEM:maximal_rationalizer_form}  implies that any CSLA has a maximal rationalization.

\begin{lemma}\label{LEM:maximal_rationale_includes_reverse_salience}
	If $(\Gamma_\rhd,\rhd)$ maximally rationalizes $c$, then $\rhd$ extends $\widetilde{P}$. 
\end{lemma}

\begin{proof}
Suppose $(\Gamma_\rhd,\rhd)$ maximally rationalizes $c$. 
To show that $\rhd$ extends $\widetilde{P}$, we prove that $\neg (x \rhd y)$ implies $\neg(x \widetilde{P} y)$, for distinct $x, y \in X$.
Suppose $\neg (x \rhd y)$, hence $y \rhd x$ by the completeness of $\rhd$. 
Since $\widetilde{P}$ is the converse of $\vDash$, we need show that $y\not\vDash x$. %that is, $(B-y,B)$ is not a switch for any $B\in \X$ such that $x,y \in B$.
Toward a contradiction, suppose $y \vDash x$, that is, $y\neq c(B)\neq c(B-y)$ for some menu $B\in\X$ containing both $x$ and $y$. 
Note that $y\neq \min(B,\rhd)$ (because $y\rhd x$) and $y \neq \max(\Gamma_\rhd(B),\rhd) = c(B)$. 
Since $c$ is a CSLA, we obtain $\Gamma_\rhd(B)-y\neq \Gamma_\rhd(B-y)$, which implies that $c(B)\in \Gamma_\rhd(B-y)$ and $c(B-y)\in\Gamma_\rhd(B)$.
Since $c(B) \neq c(B-y)$, condition~\eqref{EQ:def_Gamma} yields $c(B-y)\rhd c(B)$ and $c(B)\rhd c(B-y)$, which is impossible.
 \end{proof}

Lemma~\ref{LEM:maximal_rationalizer_form} and Lemma~\ref{LEM:maximal_rationale_includes_reverse_salience} readily yield

\begin{corollary}\label{COR:rationale_includes_reverse_salience}
	If $(\Gamma,\rhd)$ rationalizes $c$, then $\rhd$ extends $\widetilde{P}$.
	\end{corollary}
	
	Now we complete the proof of sufficiency.
	Suppose $(\Gamma,\rhd)$ rationalizes $c$. 
	By Corollary~\ref{COR:rationale_includes_reverse_salience}, $\rhd$ extends $\widetilde{P}$, hence $\widetilde{P}$ must be asymmetric.
	By Theorem~\ref{THM:main}, $c$ is RLS.  	
 \qed

\section{Appendix B: Proof of Theorem~\ref{THM:chaos_rules}}

\begin{figure}[htp!] \label{FIG:proof_THM_3}
\begin{center}
\psset{xunit=0.82} \psset{yunit=1}
\begin{pspicture}[showgrid=false](-0.2,0)(18.3,4) 
%
%
%%%%%%%%%%  DOWN  %%%%%%%%%%%%%%%%%
%
\rput(0,0){\small L\ref{LEMMA:Rodl_1985}}
\psline[arrowsize=6pt,linewidth=0.05]{->}(0.6,0)(1.5,0)
\rput(2,0){\small C\ref{COR:Betas}} 

\rput(0,1){\small L\ref{LEMMA:help_1}}
\psline[arrowsize=6pt,linewidth=0.05]{->}(0.6,1)(1.5,1)
\rput(2,1){\small L\ref{LEMMA:limit_applied}}

\psbrace(2.3,-0.5)(2.3,1.5){}
\psline[arrowsize=6pt,linewidth=0.05]{->}(3.3,0.5)(4.4,0.5)

\rput(5,0.5){\small C\ref{COR:Rodl_applied}}
\psline[arrowsize=6pt,linewidth=0.05]{->}(5.6,0.5)(6.5,0.5)
\rput(7,0.5){\small C\ref{COR:Rodl_applied_new}}
\psline[arrowsize=6pt,linewidth=0.05]{->}(7.6,0.5)(8.5,0.5)
\rput(9,0.5){\small C\ref{COR:Rodl_applied_2}}
\psline[arrowsize=6pt,linewidth=0.05]{->}(9.6,0.5)(10.5,0.5)
\rput(11,0.5){\small C\ref{COR:Rodl_partition}}
\psline[arrowsize=6pt,linewidth=0.05]{->}(11.6,0.5)(12.5,0.5)
\rput(13,0.5){\small C\ref{COR:Rodl_partition_final}}
\psline[arrowsize=6pt,linewidth=0.05]{->}(13.6,0.5)(14.5,0.5)
\rput(15,0.5){T\ref{THM:tail_fail_weakly_hereditary_asymptotically_fails}}

%%%%%%%%%% UP %%%%%%%%%%%%%%%%%%%%

\rput(8,3){\small L\ref{LEM:selecting_always_a_b_x1_xP} }
\psline[arrowsize=6pt,linewidth=0.05]{->}(8.6,3)(9.5,3)
\rput(10,3){\small L\ref{LEM:a,b-Ramsey_p*-hscrambled_contradiction}}

\rput(8,2){\small L\ref{LEM:constant_Ramsey_signatures_characterizes_Ramsey_set}}
\psline[arrowsize=6pt,linewidth=0.05]{->}(8.6,2)(9.5,2)
\rput(10,2){\small L\ref{LEM:existence_of_ab_Ramsey set}}

\psbrace(10.3,1.5)(10.3,3.5){} 
\psline[arrowsize=6pt,linewidth=0.05]{->}(11.4,2.5)(12.4,2.5)

\rput(13,2.5){\small L\ref{LEM:weakly_flipped_choice_moody}}
\psline[arrowsize=6pt,linewidth=0.05]{->}(13.6,2.5)(14.5,2.5)
\rput(15.05,2.5){L\ref{LEM:not_moody_tail_fail_weakly_hereditary}}

%%%%%%%%%%%%%%%%%%%%%%%%%%%%%%%%

\rput(18,1.5){\large T\ref{THM:chaos_rules}}
\psline[arrowsize=6pt,linewidth=0.05]{->}(16.4,1.5)(17.5,1.5) 
\psbrace(15.3,-0.5)(15.3,3.5){} 

\end{pspicture}
\end{center}
\caption{The proof of Theorem~\ref{THM:chaos_rules}: `L', `C', and `T' stand for, respectively,  `Lemma', `Corollary', and `Theorem'; an arrow from $A$ to $B$ signals that $A$ is  used to prove $B$.}
\end{figure}
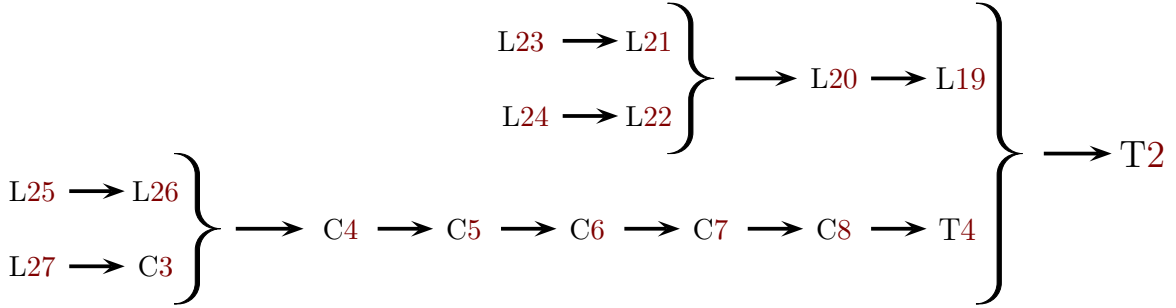

We shall obtain Theorem~\ref{THM:chaos_rules} as a corollary of a more general result, namely Theorem~\ref{THM:tail_fail_weakly_hereditary_asymptotically_fails}, which states that certain categories of properties of choice functions (called TFLH) occur almost never when the size of the ground set tends to infinity. 
Then, upon showing that being non-moody is a TFLH property (Lemma \ref{LEM:not_moody_tail_fail_weakly_hereditary}), we readily derive Theorem~\ref{THM:chaos_rules}. 
To ease the comprehension of the long and involved proof of Theorem~\ref{THM:chaos_rules}, we describe in Figure~\ref{FIG:proof_THM_3} all implications needed to achieve our claim. 
%\vs\vs\vs
%%%%%%%%%%%%%%%%%%%%%%%%%%%%%%%%
%%%%%%%%%%%%  Fig.2  %%%%%%%%%%%%%%%%
%%%%%%%%%%%%%%%%%%%%%%%%%%%%%%%%

We begin by defining TFLH properties.
%
%Below we state and prove Theorem~\ref{THM:tail_fail_weakly_hereditary_asymptotically_fails}. 

\begin{definition} \label{DEF:TFLH_properties}
	A property $\mathscr{P}$ of choice functions is:\footnote{A \textsl{property of choice functions} is a set $\mathscr{P}$ of choices that is closed under isomorphism. Equivalently, a property of choices is a formula of second-order logic, which involves quantification over elements and sets, has a symbol for choice, and is invariant under choice isomorphisms. Thus to say that a property $\mathscr{P}$ holds for $c$ means that $c' \in \mathscr{P}$ for all choices $c'$ isomorphic to $c$.}
	\begin{itemize} 
		\item \textsl{locally hereditary} if, when $\mathscr{P}$ holds for $c\colon \X \to X$, there are $x,y\in X$ such that, for any $Y\subseteq X$ with $x,y \in Y$, there is a choice $c'\colon \mathscr{Y} \to Y$ satisfying $\mathscr{P}$;\vs
	\item \textsl{tail-fail} if, for any $k\in\mathbb{N}$, there is a set $X$ of size $k$ and a choice $c$ on $X$ such that $\mathscr{P}$ fails for any choice $c'$ on $X$ satisfying $c'(A)=c(A)$ for any $A \in \X$ of size at least $k$.\vs
\end{itemize}
Then $\mathscr{P}$ is a \textsl{tail-fail locally hereditary} \textsl{(TFLH)} property if it is both tail-fail and locally hereditary.\footnote{In \cite{GiarlottaPetraliaWatson2022a}, we introduce a notion of `hereditary  property' to prove that bounded rationality according to most models present in the literature is rare. Specifically, a property $\mathscr{P}$ is \textsl{hereditary} if whenever it holds for a choice, it also holds for any of its subchoices. TFLH properties obviously comprise hereditary properties as very special cases.}
Moreover, we say that $\mathscr{P}$ is \textsl{asymptotically rare} if the fraction of choices on $X$ satisfying $\mathscr{P}$ tends to zero as the size of $X$ tends to infinity.     
\end{definition}

The two results needed to prove Theorem~\ref{THM:chaos_rules} are the following (see Figure~\ref{FIG:proof_THM_3}):\vs

\begin{lemma}\label{LEM:not_moody_tail_fail_weakly_hereditary} 
Being non-moody is a TFLH property.\vs 	
\end{lemma}

\begin{theorem} \label{THM:tail_fail_weakly_hereditary_asymptotically_fails}
	Any TFLH property of choices is asymptotically rare.
\end{theorem} 

%Below we show that Lemma~\ref{LEM:not_moody_tail_fail_weakly_hereditary} and Theorem~\ref{THM:tail_fail_weakly_hereditary_asymptotically_fails} hold true. 

%%%%%%%%%%%%%%%%%%%%%%%%%%%%%%%%%%%%%%%
%%%%%%%%%%%%%%%%%%%%%%%%%%%%%%%%%%%%%%%

\subsection*{\bf Proof of Lemma~\ref{LEM:not_moody_tail_fail_weakly_hereditary}}  

To start, we define a more articulated notion of flipped choice.\vs 

\begin{definition} \label{DEF:homogeneous_scrambled}
Let $(X,\lessdot)$ be a linearly ordered set of size $\vert X\vert = n \geq 12$.
List the items of any $A=\{x_1,\ldots,x_p\} \in \X$ in $\lessdot$-increasing order, i.e., $x_1 \lessdot x_{2} \lessdot \ldots \lessdot x_p$.  
Then a choice $c \colon \X \to X$ is \textsl{$p^*$-homogeneous scrambled} \textsl{(w.r.t.\;$\lessdot$)} if there are six distinct integers $p_1,p_2,p_3,p_4,p_5,p_6 \in \{7,8,\ldots,n\}$ such that $p^* = \max\{p_1,\ldots,p_6\}$, and the following properties hold for any $A \in \X$:\vs\vs
\begin{itemize}
	\item if $\vert A\vert = p_1$, then $c(A)=\max(A, \lessdot) = x_{p_1}$ (the best w.r.t.$\:\lessdot$);\vs\vs\vs
	\item if $\vert A\vert = p_2$, then $c(A)=\max(A \setminus \{x_{p_2}\}, \lessdot)= x_{p_2 -1}$ (the second best);\vs\vs\vs
	\item if $\vert A\vert = p_3$, then $c(A) = \max (A\setminus\{x_{p_3},x_{p_3 -1}\}, \lessdot) = x_{p_3 -2}$ (the third best);\vs\vs\vs
	\item if $\vert A\vert = p_4$, then $c(A)=\min(A, \lessdot) = x_1$ (the worst);\vs\vs\vs 
	\item if $\vert A\vert = p_5$, then $c(A)=\min(A \setminus \{x_{1}\}, \lessdot) = x_2$ (the second worst);\vs\vs\vs
	\item if $\vert A\vert = p_6$, then $c(A)= \min (A \setminus\{x_{1},x_{2}\}, \lessdot) = x_3$ (the third worst).
\end{itemize} 
%Let $p^* := \max\{p_1,p_2,p_3\}$.\footnote{This notation is improper, because it hides the dependence of the parameter $p^*$ from the values $p_1,p_2,p_3$ and from the choice $c$. However, this abuse of notation should cause no confusion in context.} %Note also that there may exist other triplets $(q_1,q_2,q_3)$ that behave as in Definition~\ref{DEF:homogeneous_scrambled} (and so other values of $q^*$). Again, this does not affect our analysis by any means.
%    In this case, we say that the $6$-tuple $\vec{p} = (p_1,p_2,p_3,p_4,p_5,p_6)$ \textsl{witnesses} that $c$ is homogeneous scrambled, and denote this fact by $\vec{p} \circlearrowleft c$. 
%    Let $P^*_c$ be the following set of integers: 
%    $$
%    P_c^* := \big\{ \max\{p_1,p_2,p_3,p_4,p_5,p_6\} : \vec{p} \circlearrowleft c \big\}\,.
%    $$   
\end{definition}	

%Observe that any flipped choice is weakly flipped: \textbf{\magenta (change)} use, for instance, the set of parameters $\{p_1,p_2,p_3\}$ given by $p_1:=6$, $p_2:=3$, and $p_3:=4$ (or $p_3 := 5$).
The next result says that any $p^*$-homogeneous scrambled choice defined on a sufficiently large set is moody: 
 
\begin{lemma}\label{LEM:weakly_flipped_choice_moody}
%A homogeneous scrambled choice $c$ on a set of size $\geq 18 p^*$, with $p^* \in P^*_c$, is context-sensitive moody.  
For any $p^* \geqslant 12$, there is an integer $N >p^*$ such that any $p^*$-homogeneous scrambled choice $c\colon\X\to X$ on a set $X$ of size $\vert X\vert\geq N$ is moody.
%For any $h\in \mathbb{N}$, there is $l\in\mathbb{N}$ and a $(h,l)$-function $f$ such that any choice generated by $f$ is moody.
\end{lemma}

\noindent {\textsc{Proof of Lemma~\ref{LEM:weakly_flipped_choice_moody}}}.
Lemma \ref{LEM:weakly_flipped_choice_moody} will be an immediate consequence of two results, namely Lemma \ref{LEM:a,b-Ramsey_p*-hscrambled_contradiction} and Lemma~\ref{LEM:existence_of_ab_Ramsey set}.
In order to state them, we need an additional notion, which obviously shares features with properties typically defined in \textsl{Ramsey Theory} (whence the terminology): 

\begin{definition}
	Let $c\colon\X\to X$ be a non-moody choice, where $X$ is endowed with a linear order $\lessdot$.
	Let $\mathscr{L} = \{\rhd_x : x \in X\}$ be a family of linear orders on $X$ rationalizing $c$ by salience. 
	Then, there are $a,b\in X$ such that $\rhd_{a}=\rhd_b$ (denote this linear order by $\rhd_{ab}$).
	A menu $Y\subseteq X$ containing $a$ and $b$ is $\{a,b\}$-\textsl{Ramsey} whenever the following conditions of `homogeneity' hold (to simplify notation, we set $Y' := Y \setminus \{a,b\}$):\vs\vs 
	\begin{enumerate}
		\item[(R1)] if there is $x\in Y'$ such that $b\rhd_x x$, then $b\rhd_x x$ for all $x\in Y'$;\vs\vs\vs
		\item[(R2)] If there is $x\in Y'$ such that $b\rhd_x a$, then $b\rhd_{x}a$ for all $x\in Y'$;\vs\vs\vs
		\item[(R3)] if there is $x\in Y'$ such that $a\rhd_x x$, then $a\rhd_x x$ for all $x\in Y'$;\vs\vs\vs
		\item[(R4)] if there is $x\in Y'$ such that $a\lessdot x$, then $a\lessdot x$ for all $x\in Y'$;\vs\vs\vs
		\item[(R5)] if there is $x\in Y'$ such that $b\lessdot x$, then $b\lessdot x$ for all $x\in Y'$;\vs\vs\vs
		\item[(R6)] if there are $x,x^{\prime} \in Y'$ such that $x\lessdot x^{\prime}$ and $b\rhd_{x}x^{\prime}$, then $b\rhd_{x}x^{\prime}$ for all $x,x^{\prime}\in Y'$ such that $x\lessdot x^{\prime}$;\vs\vs\vs 
		\item[(R7)] if there are $x,x^{\prime} \in Y'$ such that $x\lessdot x^{\prime}$ and $b\rhd_{x^{\prime}}x$, then $b\rhd_{x^{\prime}}x$ for all $x,x^{\prime}\in Y'$ such that $x\lessdot x^{\prime}$;\vs\vs\vs
		\item[(R8)] if there are $x,x^{\prime} \in Y'$ such that $x\lessdot x^{\prime}$ and $a\rhd_{x}x^{\prime}$, then $a\rhd_{x}x^{\prime}$ for all $x,x^{\prime}\in Y'$ such that $x\lessdot x^{\prime}$;\vs\vs\vs
		\item[(R9)] if there are $x,x^{\prime} \in Y'$ such that $x\lessdot x^{\prime}$ and $a\rhd_{x^{\prime}}x$, then $a\rhd_{x^{\prime}}x$ for all $x,x^{\prime}\in Y'$ such that $x\lessdot x^{\prime}$;\vs\vs\vs
		\item[(R10)] if there is $x\in Y'$ such that $x\rhd_{ab}a$, then $x\rhd_{ab} a$ for all $x\in Y'$;\vs\vs\vs
		\item[(R11)] if there is $x\in Y'$ such that $x\rhd_{ab}b$, then $x\rhd_{ab} b$ for all $x\in Y'$.\vs\vs
 \end{enumerate}
 \end{definition}
 
 We can now state the two technical results which imply Lemma~\ref{LEM:weakly_flipped_choice_moody}.  
 
 \begin{lemma}\label{LEM:a,b-Ramsey_p*-hscrambled_contradiction}
 	Let $c\colon\X\to X$ be a non-moody choice such that $\vert X\vert=N$ for some $N\in\mathbb{N}$.
 	Further, let $a,b\in X$ be two distinct items such that $\rhd_{a}=\rhd_{b}$.
 	If there is some $\{a,b\}$-Ramsey set $Y\subseteq X$ with $\vert Y\vert\geq p^*$ for some $p^*< N$, then $c$ is not $p^*$-homogeneous scrambled.
 \end{lemma}
 
 \begin{lemma} \label{LEM:existence_of_ab_Ramsey set}
For any $p^*\in\mathbb{N}$, there is an integer $N > p^*$ such that, for any non-moody choice $c\colon \X \to X$ on a set $X$ of size $\vert X\vert\geq N$, there are two items $a,b\in X$ and an $\{a,b\}$-Ramsey set $Y\subseteq X$ of cardinality $\vert Y\vert\geq p^*$. 
\end{lemma}
 
Next, we prove Lemmata~\ref{LEM:a,b-Ramsey_p*-hscrambled_contradiction} and~\ref{LEM:existence_of_ab_Ramsey set}. 
\bigskip
  
 \noindent {\textsc{Proof of Lemma}~\ref{LEM:a,b-Ramsey_p*-hscrambled_contradiction}}.
 We need the following preliminary result: 
 
 \begin{lemma}\label{LEM:selecting_always_a_b_x1_xP}
 	Let $c\colon\X\to X$ be a non-moody $p^*$-homogeneous scrambled choice (w.r.t.\ $\lessdot$). 
 	For any distinct $a,b\in X$ such that $\rhd_{a}=\rhd_{b}$, if there is an $\{a,b\}$-Ramsey set $Y\subseteq X$ with $\vert Y\vert\geq p^*$, then there are distinct $p^{\prime}, p^{\prime\prime}, p^{\prime\prime\prime}, p^{\prime\prime\prime\prime}\leq p^*$ satisfying the following properties for any $A\in\X$ such that $\{a,b\}\subseteq A\subseteq Y$:\vs\vs
 	\begin{enumerate} 		
 		\item if $\vert A\vert=p^{\prime}$, then $c(A)=a$;\vs\vs\vs
 		\item if $\vert A\vert=p^{\prime\prime}$, then $c(A)=b$;\vs\vs\vs
 		\item if $\vert A\vert=p^{\prime\prime\prime}$, then $c(A)=\min(A\setminus\{a,b\},\lessdot)$;\vs\vs\vs
 		\item if $\vert A\vert=p^{\prime\prime\prime\prime}$, then $c(A)=\max(A\setminus\{a,b\},\lessdot)$. 
 		\end{enumerate}
 	 \end{lemma}

\noindent \textsc{Proof of Lemma~\ref{LEM:selecting_always_a_b_x1_xP}.}
Let $a,b$ be distinct elements in $X$ such that $\rhd_{a}=\rhd_{b}$, and let $Y\subseteq X$ be an $\{a,b\}$-Ramsey set such that $\vert Y\vert\geq p^* \geqslant 12$.  
Set $Y^{\prime}:=Y\setminus\{a,b\} = \{x_1,\ldots,x_p\}$ (which is, as usual, listed in increasing order w.r.t.\ $\lessdot$).
Since $Y$ is $\{a,b\}$-Ramsey, by (R4) and (R5) exactly one of the following cases must hold:\vs\vs\vs
\begin{itemize}
	\item $a\lessdot b\lessdot Y^{\prime}$, or\vs\vs\vs
	\item $a\lessdot Y^{\prime} \lessdot b$, or\vs\vs\vs
	\item $Y^{\prime}\lessdot a \lessdot b$, or\vs\vs\vs
	\item $b\lessdot a\lessdot Y^{\prime}$, or\vs\vs\vs
	\item $b\lessdot Y^{\prime} \lessdot a$, or\vs\vs\vs
	\item $Y^{\prime}\lessdot b \lessdot a$.\vs\vs\vs
\end{itemize}
Note that, for each of the six cases above, $a,b,x_1,x_p$ are among the first best, second best, third best, worst, second worst, or third worst positions in $A$.
Thus the claim readily follows from the fact that $c$ is $p^*$-homogeneous scrambled. 
\qed
\medskip

We now complete the proof of Lemma~\ref{LEM:a,b-Ramsey_p*-hscrambled_contradiction}.
Toward a contradiction, suppose $c$ is a non-moody choice on a set $X$ of size $\vert X\vert\geq N$ for some $N\in\mathbb{N}$, $c$ is $p^*$-homogeneous scrambled for some $p^* < N$, and there is some $\{a,b\}$-Ramsey set $Y$ of size $\vert Y\vert\geq p^*$, where $a,b \in X$ are distinct and such that $\rhd_a = \rhd_b = \rhd_{ab}$. 
By (R10) and (R11), exactly one of the following cases holds for $Y^{\prime}$:\vs\vs\vs
\begin{itemize}
	\item[(A)]  $Y^{\prime}\rhd_{ab} a\rhd_{ab} b\,$;\vs\vs\vs
	\item[(B)] $a\rhd_{ab} Y^{\prime}\rhd_{ab}b\,$;\vs\vs\vs  
	\item[(C)] $a\rhd_{ab} b\rhd_{ab}Y^{\prime}$;\vs\vs\vs 
	\item[(D)] $Y^{\prime}\rhd_{ab} b\rhd_{ab} a\,$;\vs\vs\vs
	\item[(E)] $b\rhd_{ab} Y^{\prime}\rhd_{ab}a\,$;\vs\vs\vs
	\item[(F)] $b \rhd_{ab} a\rhd_{ab}Y^{\prime}$.\vs\vs
\end{itemize}
By Lemma~\ref{LEM:selecting_always_a_b_x1_xP}, there are distinct $p^{\prime}, p^{\prime\prime}, p^{\prime\prime\prime}, p^{\prime\prime\prime\prime}\leq p^*$ such that any $A\subseteq Y$ containing $a$ and $b$ satisfies properties 1--4 in Lemma~\ref{LEM:selecting_always_a_b_x1_xP}. 
Denote $A = \{x_1,\ldots,x_p\}\cup\{a,b\} \subseteq Y$, where $x_1 \lessdot \ldots \lessdot x_p$. 
Further, set $[p] := \{1,2,\ldots,p\}$.
%$p\in\{1,\ldots,p^*\}$ depends on the cardinality of $A$.  

\begin{description}
 \item[Case 1:] Suppose $\vert A \vert = p'$, hence $c(A) = a$. 
   In cases (A), (D), (E), and (F), we have\vs\vs\vs
\begin{equation}\label{COND:case1_first_step}
	a\rhd_{x_i} \!b\,,\;\; a\rhd_{x_i}\!x_i \,, \;\; a \rhd_{x_i}\!x_j \vs\vs\vs
\end{equation}
for some $i\in [p]$ and for all $j\in [p] \setminus \{i\}$. 
By (R2) and (R3), it follows\vs\vs\vs
\begin{equation}\label{COND:case1_second_step} 
	a\rhd_{x}b \;\;\text{ and }\;\; a\rhd_{x}x \vs\vs\vs
\end{equation}
for any $x\in Y^{\prime}$.
Moreover, if $i=1$, then \eqref{COND:case1_first_step} and (R8) imply\vs\vs\vs
\begin{equation}\label{COND:case1_third_step}
 	a\rhd_{x}x^{\prime}\vs\vs\vs
 \end{equation}  
 for any $x,x^{\prime}\in Y^{\prime}$ such that $x\lessdot x^{\prime}$.
If $i\neq 1$, then \eqref{COND:case1_first_step} and (R9) imply\vs\vs\vs
 \begin{equation}\label{COND:case1_fourth_step}
 	b\rhd_{x^{\prime}} x \vs\vs\vs
 \end{equation}
 for all $x,x^{\prime}\in Y^{\prime}$ such that $x\lessdot x^{\prime}$.
 \item[Case 2:]
Suppose $\vert A\vert=p^{\prime\prime}$, hence $c(A)=b$.
In cases (A), (B), (C), and (D), we have\vs\vs\vs
\begin{equation}\label{COND:case2_first_step}
	b\rhd_{x_i}\!a\,,\;\; b\rhd_{x_i}\!x_i\,,\;\; b \rhd_{x_i}\!x_j\vs\vs\vs
\end{equation}
for some $i\in [p]$ and for all $j\in[p]\setminus \{i\}$.
By (R1) and (R2), it follows\vs\vs\vs
\begin{equation}\label{COND:case2_second_step}
	b\rhd_{x}a \;\;\text{ and } \;\;b\rhd_{x}x\vs\vs\vs
\end{equation}
 for any $x\in Y^{\prime}$.
 Moreover, if $i=1$, then \eqref{COND:case2_first_step} and (R6) imply\vs\vs\vs
 \begin{equation}\label{COND:case2_third_step}
 	b\rhd_{x}x^{\prime}\vs\vs\vs
 \end{equation}  
 for any $x,x^{\prime}\in Y^{\prime}$ such that $x\lessdot x^{\prime}$.
 If $i\neq 1$, then \eqref{COND:case2_first_step} and (R7) imply\vs\vs\vs
 \begin{equation}\label{COND:case2_fourth_step}
 	b\rhd_{x^{\prime}}x\vs\vs\vs
 \end{equation}
 for all $x,x^{\prime}\in Y^{\prime}$ such that $x\lessdot x^{\prime}$.
 \item[Case 3:] Suppose $\vert A\vert=p^{\prime\prime\prime}$, so $c(A)=x_1$.
  In cases (B), (C), (E), and (F), we have\vs\vs\vs
 \begin{equation}\label{COND:case3_first_step}
	x_1\rhd_{x_i}\!a\,,\;\; x_1\rhd_{x_i}\!b\,,\;\; x_1 \rhd_{x_i}\!x_j \vs\vs\vs
	 \end{equation}
for some $i\in [p]$ and for all $j\in [p]\setminus \{1\}$.
If $i=1$, then (R1) and (R3) imply\vs\vs\vs
\begin{equation}\label{COND:case3_second_step}
	x\rhd_{x}a\;\;\text{and}\;\; x\rhd_{x}b\vs\vs\vs
\end{equation}
for all $x\in Y^{\prime}$.
On the other hand, if $i\neq 1$, then \eqref{COND:case3_first_step}, (R7) and (R9) imply \vs\vs\vs
\begin{equation}\label{COND:case3_third_step}
	x\rhd_{x^{\prime}}a\;\;\text{and}\;\; x\rhd_{x^{\prime}}b \vs\vs\vs
\end{equation}
for all $x,x^{\prime}\in Y^{\prime}$ such that $x\lessdot x^{\prime}$. 
\item[Case 4:]
Suppose $\vert A\vert=p^{\prime\prime\prime\prime}$, so $c(A)=x_p$.
In cases (B), (C), (E), and (F), we have \vs\vs\vs
 \begin{equation}\label{COND:case4_first_step}
	x_p\rhd_{x_i}\! a\,,\;\; x_p\rhd_{x_i}\!b\,,\;\;x_p \rhd_{x_i}\!x_j \vs\vs\vs
	 \end{equation}
for some $i\in [p]$ and for all $j\in [p]\setminus \{p\}$. 
 If $i=p$, by (R1) and (R3) we get\vs\vs\vs 
\begin{equation}\label{COND:case4_second_step}
	x\rhd_{x}a\;\;\text{and}\;\; x\rhd_{x}b \vs\vs\vs
\end{equation}
for all $x\in Y^{\prime}$.
On the other hand, if $i\neq p$, then \eqref{COND:case4_first_step}, (R6) and (R8) imply \vs\vs\vs
\begin{equation}\label{COND:case4_third_step}
	x^{\prime}\rhd_{x}a\;\;\text{and}\;\; x^{\prime}\rhd_{x}b \vs\vs\vs
\end{equation}
for all $x,x^{\prime}\in Y^{\prime}$ such that $x\lessdot x^{\prime}$. 
\end{description}
Next, we use Cases 1, 2, 3, and 4 to derive a contradiction.  
(Whenever two conditions $\mathscr{C}$ and $\mathscr{C}^{\prime}$ cannot simultaneously hold, we write $\mathscr{C}\perp \mathscr{C}^{\prime}$.)\vs\vs\vs

\begin{itemize}
	\item[(i)] We have $\eqref{COND:case1_fourth_step}\perp \eqref{COND:case3_third_step}$ and $\eqref{COND:case1_third_step}\perp \eqref{COND:case4_third_step}$.
	Since one among \eqref{COND:case1_third_step} and \eqref{COND:case1_fourth_step} must happen, we conclude that $\eqref{COND:case1_first_step}\perp \left(\eqref{COND:case3_third_step}\wedge \eqref{COND:case4_third_step}\right).$\vs\vs\vs
	 \item[(ii)] We have $\eqref{COND:case2_third_step}\perp\eqref{COND:case4_third_step}$ and $\eqref{COND:case2_fourth_step}\perp \eqref{COND:case3_third_step}$.
	 Since one among \eqref{COND:case2_third_step} and \eqref{COND:case2_fourth_step} must happen, we conclude that $\eqref{COND:case2_first_step}\perp (\eqref{COND:case3_third_step}\wedge \eqref{COND:case4_third_step})$.\vs\vs\vs
  	 \item[(iii)] By (i), we know that $\eqref{COND:case1_first_step}\perp \left(\eqref{COND:case3_third_step}\wedge \eqref{COND:case4_third_step}\right).$
           Since we have $\eqref{COND:case1_second_step}\perp \eqref{COND:case3_second_step}$ and $\eqref{COND:case1_second_step}\perp \eqref{COND:case4_second_step}$, a simple computation yields $\eqref{COND:case1_first_step}\perp(\eqref{COND:case3_first_step}\wedge\eqref{COND:case4_first_step})$.\vs\vs\vs
	 \item[(iv)] By (ii), we know that $\eqref{COND:case2_first_step}\perp \left(\eqref{COND:case3_third_step}\wedge \eqref{COND:case4_third_step}\right).$
	      Since we have $\eqref{COND:case1_second_step}\perp \eqref{COND:case3_second_step}$ and $\eqref{COND:case1_second_step}\perp \eqref{COND:case4_second_step}$, a simple computation yields $\eqref{COND:case2_first_step}\perp(\eqref{COND:case3_first_step}\wedge\eqref{COND:case4_first_step})$.\vs\vs\vs
        \item[(v)] Since $\eqref{COND:case1_second_step}\perp\eqref{COND:case2_second_step}$, we conclude $\eqref{COND:case1_first_step}\perp\eqref{COND:case2_first_step}$.\vs\vs\vs
\end{itemize}

Note that at most one among $\eqref{COND:case1_first_step}$, $\eqref{COND:case2_first_step}$, and $(\eqref{COND:case3_first_step}\wedge\eqref{COND:case4_first_step})$ can hold.
However, for each of the cases (A), (B), (C), (D), (E), and (F), two of the above conditions must simultaneously hold, and this is impossible.
This proves Lemma~\ref{LEM:a,b-Ramsey_p*-hscrambled_contradiction}.
\qed
\bigskip

\noindent {\textsc{Proof of Lemma}~\ref{LEM:existence_of_ab_Ramsey set}}.
We need the following notion:

\begin{definition}\label{DEF:Ramsey_signatures}
Let $c\colon \X \to X$ be a non-moody choice on a set $X$ endowed with a linear order $\lessdot$, and let $a,b\in X$ be two items such that $\rhd_{a}=\rhd_{b}=\rhd_{ab}$.
We call \textsl{$\{a,b\}$-Ramsey $1$-signature} the map  $r\colon X\setminus\{a,b\}\to\{0,1\}^7$ which assigns to any $x\in X\setminus\{a,b\}$ a vector $(r_1(x),\ldots,r_7(x)) \in \{0,1\}^7$ (which is one of $2^7=128$ possible `colors') according to the following rules:\vs\vs\vs
\begin{itemize}
	\item $r_1(x)=0 \;\Longleftrightarrow\; b\rhd_{x} x\,$,\vs\vs\vs 
	\item $r_2(x)=0 \;\Longleftrightarrow\; b\rhd_{x} a\,$,\vs\vs\vs
 	\item $r_3(x)=0 \;\Longleftrightarrow\; a\rhd_{x} x\,$,\vs\vs\vs
	\item $r_4(x)=0 \;\Longleftrightarrow\; a\lessdot x\,$,\vs\vs\vs
	\item $r_5(x)=0 \;\Longleftrightarrow\; b\lessdot x\,$,\vs\vs\vs
	\item $r_6(x)=0 \;\Longleftrightarrow\; x\rhd_{ab} a\,$,\vs\vs\vs
	\item $r_7(x)=0 \;\Longleftrightarrow\; x\rhd_{ab} b\,$.\vs\vs
\end{itemize}
Moreover, we call \textsl{$\{a,b\}$-Ramsey $2$-signature} the map $\widehat{r}\colon \left[X\setminus\{a,b\}\right]^2\to\{0,1\}^4$ which assigns to any unordered pair $\{x,x^{\prime}\} \in \left[X\setminus\{a,b\}\right]^2$ such that $x \lessdot x'$ a vector $(\widehat{r}_1(x,x^{\prime}),\ldots,\widehat{r}_4(x,x^{\prime})) \in \{0,1\}^4$ (which is one of $2^4=16$  possible `colors') according to the following rules:\footnote{Recall that the symbol $[A]^2$ stands for $\{B \subseteq A : \vert B \vert =2\}$.}\vs\vs\vs
\begin{itemize}
	\item $\widehat{r}_1(x,x^{\prime})=0 \;\Longleftrightarrow\; b\rhd_{x}x^{\prime}\,$,\vs\vs\vs
	\item $\widehat{r}_2(x,x^{\prime})=0 \;\Longleftrightarrow\; b\rhd_{x^{\prime}}x\,$,\vs\vs\vs
	\item $\widehat{r}_3(x,x^{\prime})=0 \;\Longleftrightarrow\; a\rhd_{x}x^{\prime}\,$,\vs\vs\vs
	\item $\widehat{r}_4(x,x^{\prime})=0 \;\Longleftrightarrow\; a\rhd_{x^{\prime}}x\,$.\vs\vs 
\end{itemize}
\end{definition} 

The next result characterizes $\{a,b\}$-Ramsey sets in terms of the signature maps; its proof is straightforward, and is left to the reader.

\begin{lemma}\label{LEM:constant_Ramsey_signatures_characterizes_Ramsey_set}
Let $c\colon \X \to X$ be a non-moody choice on a set $X$ endowed with a linear order $\lessdot$, and let $a,b\in X$ be two items such that $\rhd_{a}=\rhd_{b}$.
The following conditions are equivalent for any set $Y\subseteq X$ containing $a$ and $b$:\vs\vs\vs 
\begin{itemize}
	\item[\rm (i)] $Y$ is $\{a,b\}$-Ramsey;\vs\vs\vs
	\item[\rm (ii)] the maps $r_{\upharpoonright Y\setminus\{a,b\}}$ and $\widehat{r}_{\upharpoonright[Y\setminus\{a,b\}]^2}$ are constant.\footnote{We denote by $r_{\upharpoonright Y\setminus\{a,b\}}$ and $\widehat{r}_{\upharpoonright [Y\setminus\{a,b\}]^2}$ the restrictions of $r$ and $\widehat{r}$ to $Y\setminus\{a,b\}$ and $[Y\setminus\{a,b\}]^2$.}\vs  
\end{itemize}
\end{lemma}

We now prove Lemma \ref{LEM:existence_of_ab_Ramsey set}.
Let $p^*$ be an integer $\geqslant 12$, and set $N^{*} :=128(p^*-1)+1$. 
By Ramsey's theorem, there is $N^{**}\in\mathbb{N}$ such that, for any edge coloring with $16$ colors on a graph of $N^{**}$ vertices, there is a monochromatic subgraph on $N^{*}$ vertices.\footnote{A graph is a pair $G=(V,E)$, where $V$ is a finite set of elements (\textsl{vertices}) and $E$ is a set of unordered pairs of vertices (\textsl{edges}).
A graph $G=(V,E)$ is \textsl{complete} if $E$ contains all possible pairs of distinct vertices. 
Given a graph $G=(V,E)$, a \textsl{subgraph} of $G$ is a graph $G'= (V^{\prime},E^{\prime})$ such that $V^{\prime}\subseteq V$ and $E^{\prime}\subseteq E$. 
A complete subgraph of a graph is called a \textsl{clique}. 
Given a set $K=\{1,\ldots,k\}$ of $k \geqslant 1$ labels (the `colors'), an \textsl{edge coloring} is a map $\gamma \colon E\to K$ that assigns a color in $K$ to each edge.
Then the pair $(G,\gamma)$ is a \textsl{colored graph}, which is \textsl{monochromatic} whenever $\gamma$ is constant.
In its general form, \textsl{Ramsey's theorem} states that for any given number $k$ of colors and any given integers $n_1, \ldots, n_k$, there is an integer $R(n_1,\ldots,n_k)$ such that if the edges of a complete graph $G$ on $R(n_1,\ldots,n_k)$ vertices are colored with $k$ different colors, then there is a color $i \in \{1,\ldots, k\}$ such that $G$ has a monochromatic clique on $n_i$ vertices whose edges are all colored with $i$.} 
We claim that the integer $N:=N^{**}+2 > p^*$ is the one we are looking for. 

Let $c\colon\X\to X$ be a non-moody choice on a set $X$ of cardinality $\vert X\vert=N$, and let $a,b\in X$ be distinct items such that $\rhd_{a}=\rhd_{b}$.
Fix a linear order $\lessdot$ on $X$, and let $\widehat{r} \colon \left[X\setminus\{a,b\}\right]^2\to\{0,1\}^4$ be the associated $\{a,b\}$-Ramsey $2$-signature (which is an edge coloring with $16$ colors on $\left[X\setminus\{a,b\}\right]^2$). 
 It follows that there is some (monochromatic) set $Y^{*}\subseteq X\setminus\{a,b\}$  of size $\vert Y^* \vert=N^*$ such that $\widehat{r}_{\upharpoonright \left[Y^{*}\right]^2}$ is constant.
 Now let $r \colon X \setminus \{a,b\} \to \{0,1\}^7$ be the $\{a,b\}$-Ramsey $1$-signature associated to $\lessdot$. 
 Note that for any $x\in Y^*$, $r_{\upharpoonright Y^{*}}$ has $128$ possible values.
 Since $N^* = 128(p^* -1) +1$, by the pigeon principle there is $Y^{**}\subseteq Y^{*}$ such that $\vert Y^{**}\vert \geqslant p^*$ and $r_{\upharpoonright Y^{**}}$ is constant.
 By Lemma~\ref{LEM:constant_Ramsey_signatures_characterizes_Ramsey_set}, $Y := Y^{**} \cup \{a,b\}$ is an   $\{a,b\}$-Ramsey set, as required.   
\qed
\bigskip

Lemma~\ref{LEM:weakly_flipped_choice_moody} readily follows from Lemmata~\ref{LEM:a,b-Ramsey_p*-hscrambled_contradiction} and~\ref{LEM:existence_of_ab_Ramsey set}. \qed 
\bigskip

We can finally prove Lemma~\ref{LEM:not_moody_tail_fail_weakly_hereditary}.  
Let $\mathscr{P}^*$ be the property of being non-moody.
We first show that $\mathscr{P}^*$ is locally hereditary.
Suppose $c \colon \X \to X$ is non-moody.
By definition, there are distinct $a,b \in X$ such that $\rhd_a = \, \rhd_b$. 
The elements $a$ and $b$ are the ones we are seeking to prove that $\mathscr{P}^*$ is locally hereditary. 
Indeed, take any $Y \subseteq X$ such that $a,b \in Y$.
Then the (sub)choice $c' \colon \mathscr{Y} \to Y$ on $Y$, defined by $c'(B) := c(B)$ for any $B \in \mathscr{Y}$, is non-moody. 

To prove that $\mathscr{P}^*$ is a tail-fail property, for any $k\in\mathbb{N}$ take six integers $p_1,p_2,p_3,p_4,p_5,p_6$, having maximum $p^*$ and minimum $p_*$, such that  $k<p_*$.
By Lemma~\ref{LEM:weakly_flipped_choice_moody}, there is $N\in\mathbb{N}$ such that any $p^*$-homogeneous scrambled choice $c$ on a ground set $X$ of size at least $N$ is moody.
Moreover, any choice $c'$ on $X$ such that $c'(A)=c(A)$ for any $A\in\X$ of size at least $k$ is $p^*$-homogeneous scrambled, and so, by Lemma~\ref{LEM:weakly_flipped_choice_moody}, it is moody.  
This completes the proof. 

%%%%%%%%%%%%%%%%%%%%%%%%%%%%%%%%%%
%%%%%%%%%%%%%%%%%%%%%%%%%%%%%%%%%%

\subsection*{\bf Proof of Theorem~\ref{THM:tail_fail_weakly_hereditary_asymptotically_fails}}  

To start, we make some computations based on calculus, namely Lemmata~\ref{LEMMA:help_1} and~\ref{LEMMA:limit_applied}.
%Next, we prove some straightforward preliminary facts what follows we prove some preliminary facts, which will be used to show that all TFHL properties are asymptotically rare (). 
%Then, applying Lemma~\ref{LEM:not_moody_tail_fail_weakly_hereditary}, we derive our main claim, that is, being non-moody is an asymptotically rare property (Theorem~\ref{THM:chaos_rules}). 

\begin{lemma} \label{LEMMA:help_1}
For any $\delta,\alpha \in \mathbb{R}$ such that $0< \delta < 1$ and $\alpha > 0$, we have\vs\vs\vs
\[
\lim_{n\to \infty}\big(1-(1-\delta)^{n^{\alpha}}\big)^{n^{2}}=1\,.
\]	
\end{lemma}

\noindent \textsc{Proof.}
Replacing variables ($n^2$ by $m$),  it suffices to show that\vs
\[
\lim_{m\to \infty}\Big(1-\left(1-\delta\right)^{m^{\frac{\alpha}{2}}}\Big)^{m}=1,\vs
\]
that is, taking logs of both sides,\vs
\begin{equation}\label{EQ:log_limit}
\lim_{m\to \infty}\log\left(1-(1-\delta)^{m^{\frac{\alpha}{2}}}\right)^{m}=0.\vs
\end{equation}
It is straightforward to check that \eqref{EQ:log_limit} holds. 
\qed 

%By de L'H\^{o}pital's rule, we get 
%\begin{align*}
%\lim_{m\to \infty} \frac{\log\left(1-(1-\delta)^{m^{\frac{\alpha}{2}}}\right)}{m^{-1}} 
%& =
%\lim_{m\to \infty} \frac{\frac{-(1-\delta)^{m^{\frac{\alpha}{2}}} \frac{\alpha}{2}m^{\frac{\alpha}{2}-1} \log(1-\delta)}{1-(1-\delta)^{m^{\frac{\alpha}{2}}}}}{-m^{-2}} \\\\
%& = \frac{\alpha}{2} \log(1-\delta)\lim_{m\to \infty}\tfrac{(1-\delta)^{m^{\frac{\alpha}{2}}}m^{\frac{\alpha}{2}+1}}{1-(1-\delta)^{m^{\frac{\alpha}{2}}}}  & = 0\,.
%\end{align*} 
%
%%The following Corollary is an immediate application of Theorem \ref{THM:Hereditary_finitely_witnessed_property}.
%Since $\lim_{m\to \infty} \left( 1-(1-\delta)^{m^{\frac{\alpha}{2}}} \right) =1$, we get \eqref{EQ:log_limit}. 
%%
%$$\frac{\alpha}{2}(1-\delta)\lim_{m\to \infty}\frac{(1-\delta)^{m^{\frac{\alpha}{2}}}m^{\frac{\alpha}{2}-1}}{1-(1-\delta)^{m^{\frac{\alpha}{2}}}}=\frac{\alpha}{2}(1-\delta)\lim_{m\to \infty}(1-\delta)^{m^{\frac{\alpha}{2}}}m^{\frac{\alpha}{2}-1}=\frac{\alpha}{2}(1-\delta)\lim_{n\to \infty}(1-\delta)^{n^{\alpha}}n^{\alpha-1}=0.$$
%\begin{corollary}
%Almost all choices on a finite set are not rationalizable by a trivial quasi moody GRS.\qed
%\end{corollary}

\begin{lemma}\label{LEMMA:limit_applied}
Let $0<\delta<1$.
Suppose there exists $\beta>2$ such that for some function $h \colon \mathbb{N} \to \mathbb{R}$, it holds $h(n)>n^{\beta}$ for all but finitely many $n$.
Then\vs\vs\vs
\[
\lim_{n\to \infty}\left(1-(1-\delta)^{\frac{h(n)}{n^2}}\right)^{\!n^{2}}=1\,.
\]
\end{lemma}

\noindent \textsc{Proof.}
 Setting $\alpha:=\beta-2$ in Lemma~\ref{LEMMA:help_1}, we get\vs\vs
 \begin{align*}
 1 &\geq \lim_{n\to \infty}\left(1-(1-\delta)^{\frac{{h(n)}}{n^2}}\right) \geq
 \lim_{n\to \infty}\left(1-(1-\delta)^{\frac{{h(n)}}{n^2}}\right)^{\!\!n^{2}} \\ 
 & \geq \lim_{n\to \infty}\left(1-(1-\delta)^{\frac{n^{\beta}}{n^2}}\right)^{\!\!n^{2}} \!= \lim_{n\to \infty}\left(1-(1-\delta)^{n^{\alpha}}\right)^{n^{2}}=1\,,\vs\vs 
 \end{align*} 
which proves the claim. 
\qed

%The next notion plays a major role in the proof of Theorem~\ref{THM:tail_fail_weakly_hereditary_asymptotically_fails}.

\begin{definition} \label{DEF:q-sparse}
	Let $X$ be a set of cardinality $n \geqslant 2$.	
	Given an integer $p <n$, a family $\mathscr{F}$ of subsets of $X$ is \textsl{$m$-uniform} if $\vert F \vert = m$ for all $F \in \mathscr{F}$.
	If, in addition, $\vert F \cap G \vert \leqslant k$ for all $F,G \in \mathscr{F}$, where $1 \leqslant k < m$, then $\mathscr{F}$ is called  \textsl{$(m,k)$-sparse}. 
	We denote by $S(n,m,k)$ the maximum size of a $(m,k)$-sparse family on a set of size $n$. 
	Moreover, we denote by $T(n,m,k)$ the maximum size of a $(m,k)$-sparse family $\mathscr{F}$ on a set of size $n$ such that $\vert \mathscr{F} \vert$ is a multiple of $n^2$.
\end{definition}

In what follows we derive some results about $S(n,m,k)$, which will be then extended to $T(n,m,k)$. 
The following combinatorial result is well-known: 

\begin{lemma}[Rodl, 1985]\label{LEMMA:Rodl_1985} \it
	For any positive integers $m , k \in \mathbb{N}$ such that $m >k$,\vs\vs
	\[
	\lim_{n\to \infty}S(n,m,k){m\choose k}{n\choose k}^{\!-1}=1\,.
	\]
\end{lemma}
%
%We now derive some facts from Rodl's result. % some facts, namely Corollaries~\ref{COR:Betas}--\ref{COR:Rodl_partition_final}.
%Only the last of these consequences (Corollary~\ref{COR:Rodl_partition_final}) will be used in the proof of Theorem~\ref{THM:tail_fail_weakly_hereditary_asymptotically_fails}.
% To start, we rewrite Lemma~\ref{LEMMA:Rodl_1985} in our context, where $p,q$ are fixed, and so $S(n,p,q)$, henceforth denoted by $h_{pq}(n)$, \textit{only} depends on $n$.

\begin{corollary}\label{COR:Betas} 
For any $k,m \in \mathbb{N}$ with $5 \leq k < m$, there is $\beta>2$ such that $S(n,m,k)>n^\beta$ for all but many finitely integers $n$.
\end{corollary}

\noindent \textsc{Proof.}
Fix $k,m \in \mathbb{N}$ such that $5 \leq k < m$.  
Lemma~\ref{LEMMA:Rodl_1985} yields\vs\vs\vs 
\[
S(n,m,k) \geq \frac{1}{2} {m \choose k}^{\!-1}\! {n \choose k}\vs\vs\vs
\]
for almost all $n \in \mathbb{N}$.
Since $k \geq 5$ implies that ${n \choose k} \geq n^3$ for almost all $n$, we get\vs\vs\vs  
\[
S(n,m,k) \geq \frac{1}{2} {m \choose k}^{\!-1} \!n^3 = c n^3 \vs\vs\vs
\]
for almost all $n$ and for some $c >0$.
Take any $\beta$ such that $2 < \beta < 3$.   
Since $cn^3>n^\beta$ if and only if $n^{3-\beta}>\frac{1}{c}$, 
% \quad \iff \quad cn^{\beta}n^{3-\beta}>n^\beta \quad \iff \quad cn^{3-\beta}>1 \quad \iff \quad n^{3-\beta}>\frac{1}{c}\,, \vs
we obtain $S(n,m,k) \geq cn^3>n^\beta$ for almost all $n$. 
%Thus $S(n,m,k+1) \geq cn^3 >n^\beta$ holds for almost all $n$.   
%Thus, for any $2<\beta<3$, 
%$$h(n)\geq cn^3\geq cn^\beta$$
%holds for almost all $n$, {\magenta which implies, for large values of $n$, that 
%$$h(n)\geq n^3\geq n^\beta.$$}
\qed 

\begin{corollary}
\label{COR:Rodl_applied}
For any $k,m \in \mathbb{N}$ and $\delta \in \mathbb{R}$ such $5 \leq k < m$ and $0 < \delta < 1$, 
% where $k$ is sufficiently large (likely $k \geq 5$).
%If $h(n) = S(n,m,k)$, then 
\vs\vs\vs 
\[
\lim_{n \rightarrow \infty} \left(1 - (1- \delta)^{\frac{S(n,m,k)}{n^2}}\right)^{\!n^2} = 1\,.
\]
\end{corollary}

\noindent \textsc{Proof.}
Apply Lemma~\ref{LEMMA:limit_applied} and Corollary~\ref{COR:Betas}.	
\qed

\begin{corollary}
\label{COR:Rodl_applied_new}
For any $k,m \in \mathbb{N}$ and $\delta \in \mathbb{R}$ such $5 \leq k < m$ and $0 < \delta < 1$, 
% where $k$ is sufficiently large (likely $k \geq 5$).
%If $h(n) = S(n,m,k)$, then 
\vs\vs\vs 
\[
\lim_{n \rightarrow \infty} \left(1 - (1- \delta)^{\frac{T(n,m,k)}{n^2}}\right)^{\!n^2} = 1\,.
\]
\end{corollary}

\noindent \textsc{Proof.}
We use Corollary~\ref{COR:Rodl_applied} and a sandwich argument. 
Let $k,m \in \mathbb{N}$ and $\delta \in \mathbb{R}$ such $5 \leq k < m$ and $0 < \delta < 1$.
We first prove\vs\vs\vs 
\begin{equation}\label{EQ:Rodl_new1}
	\frac{S(n,m,k)}{2} \;\leq\; S(n,m,k) - n^2 \;\leq\; T(n,m,k) \;\leq\; S(n,m,k)\,.	\vs\vs\vs  
\end{equation}	
The last two inequalities are an immediate consequence of the definition of $S(n,m,k)$ and $T(n,m,k)$.
Therefore, it suffices to show that the first holds as well. 
Toward a contradiction, suppose $S(n,m,k)/2 > S(n,m,k) - n^2$.
Then, we have\vs\vs\vs 
\begin{align*}
	S(n,m,k) < 2n^2 \quad &\Longrightarrow \quad (1- \delta)^{\frac{S(n,m,k)}{n^2}} > (1- \delta)^{\frac{2n^2}{n^2}} \\ 
%	& \Longrightarrow \quad 1 - (1- \delta)^{\frac{S(n,m,k)}{n^2}} < 1 - (1- \delta)^{2} \\
	& \Longrightarrow \quad \left(1 - (1- \delta)^{\frac{S(n,m,k)}{n^2}}\right)^{n^2} < \left(1 - (1- \delta)^{2}\right)^{n^2}.\vs\vs\vs
\end{align*}
However, by Corollary~\ref{COR:Rodl_applied}, we get\vs\vs\vs 
\[
1 \; = \; \lim_{n \rightarrow \infty} \left(1 - (1- \delta)^{\frac{S(n,m,k)}{n^2}}\right)^{\!n^2} \; \leq \; 
\lim_{n \rightarrow \infty} \left(1 - (1- \delta)^{2} \right)^{n^2} \;=\; 0\,,\vs\vs\vs
\]
which is impossible. 
Next, we prove\vs\vs\vs
\begin{equation} \label{EQ:Rodl_new2}
	\lim_{n \rightarrow \infty} \left(1 - (1- \delta)^{\frac{S(n,m,k)}{2n^2}}\right)^{n^2} = 1\,.\vs\vs\vs
\end{equation}
Since there is $0 < \sigma < 1$ such that  $1 - \sigma = (1- \delta)^{\frac{1}{2}}$, Corollary~\ref{COR:Rodl_applied} readily yields\vs\vs\vs
\begin{align*}
\lim_{n \rightarrow \infty} \left(1 - (1- \delta)^{\frac{S(n,m,k)}{2n^2}}\right)^{n^2} &= \;\; \lim_{n \rightarrow \infty} \left(1 - (1- \sigma)^{\frac{S(n,m,k))}{n^2}} \right)^{n^2} =\;\; 1.\vs\vs\vs
\end{align*} 
%We now complete the proof of Corollary~\ref{COR:Rodl_applied_new}.  
Since $\frac{S(n,m,k)}{2n^2} \leq \frac{T(n,m,k)}{n^2} \leq \frac{S(n,m,k)}{n^2}$ by \eqref{EQ:Rodl_new1}, we get\vs\vs\vs 
\[
	1- (1-\delta)^{\frac{S(n,m,k)}{2 n^2}} \leq 1- (1-\delta)^{\frac{T(n,m,k)}{n^2}} \leq 1- (1-\delta)^{\frac{S(n,m,k)}{n^2}} \vs\vs\vs
\]
hence\vs\vs\vs
\[
	\left(1- (1-\delta)^{\frac{S(n,m,k)}{2 n^2}}\right)^{\!n^2} \leq \left(1- (1-\delta)^{\frac{T(n,m,k)}{n^2}}\right)^{\!n^2} \leq \left(1- (1-\delta)^{\frac{S(n,m,k)}{n^2}}\right)^{\!n^2}\vs\vs\vs  
\]
and so, by \eqref{EQ:Rodl_new2} and Corollary~\ref{COR:Rodl_applied},\vs\vs\vs 
\[
1 \!=\! \lim_{n \to \infty} \left(1\!-\! (1\!-\!\delta)^{\frac{S(n,m,k)}{2 n^2}}\right)^{\!n^2}\!\!\!\leq\! 
\lim_{n \to \infty} \left(1\!-\! (1\!-\!\delta)^{\frac{T(n,m,k)}{n^2}}\right)^{\!n^2} \!\!\!\leq \!\lim_{n \to \infty} \left(1 \!-\! (1 \!-\!\delta)^{\frac{S(n,m,k)}{n^2}}\right)^{\!n^2}\!\! = \!1.\vs\vs\vs
\]
This completes the proof.
\qed

\begin{corollary} \label{COR:Rodl_applied_2}
Let $k, m \in \mathbb{N}$ and $\delta, \epsilon \in \mathbb{R}$ be such that $5 \leq k < m$, $0 < \delta < 1$, and $\epsilon > 0$.
Then there exist positive integers $n$ and $h$, a set $X$ of size $n$, and an $(m,k)$-sparse family $\mathscr{F}$ of $\vert \mathscr{F} \vert =h$ subsets of $X$ such that $h$ is divisible by $n^2$ and\vs\vs\vs
\begin{equation} \label{EQ:Rodl_new} 
\left(1 - (1- \delta)^{\frac{h}{n^2}}\right)^{\!n^2} > 1 - \epsilon. 
\end{equation}
\end{corollary}

\noindent \textsc{Proof.}
Apply Corollary~\ref{COR:Rodl_applied_new}.   	
\qed

\begin{corollary}\label{COR:Rodl_partition}
Let $k, m \in \mathbb{N}$ and $\delta, \epsilon \in \mathbb{R}$ be such that $5 \leq k < m\!-\! 2$, $0 < \delta < 1$, and $\epsilon > 0$.
Then there exist positive integers $n$ and $h$, a set $X$ of size $n$, an $(m,k+2)$-sparse family $\mathscr{G} = \{G_i : i \in I\}$ of $\vert \mathscr{G} \vert= h$ subsets of $X$, and a partition $\mathscr{I}=\{I_{x,y} : x, y \in X\}$ of $I$ in sets having all the same size $\vert I_{x,y}\vert= h/n^2$ such that\vs\vs\vs  
\begin{itemize}
 	\item[\rm(i)] $\big(1 - (1- \delta)^{h/n^2}\big)^{n^2} > 1 - \epsilon\,$, and\vs\vs\vs  
 	\item[\rm(ii)] $i \in I_{x,y}$ implies $x, y \in G_i$ for any $i\in I$.
\end{itemize}
%Fix $5 \leq k < m-2$, $0 < \delta< 1$, and $\epsilon > 0$.
%There is $n$, $h$, $X$, 
% where $k$ is sufficiently large (likely $k \geq 5$).
%and a family $\{B_i| i \in I\}$ of subsets of $X$ where $|I| = h, |X| = n$
%and a partition $I = \cup\{I_{x,y}| x, y \in X\}$
%so that each $|I_{x,y}| = h/n^2$, $|B_i \cap B_j| \leq k+2$, each $|B_i| = m$,
%$ (1 - (1- \delta)^{h/n^2})^{n^2} > 1 - \epsilon$ and $i \in I_{x,y}$ implies
%$x, y \in B_i$.
\end{corollary}

\noindent \textsc{Proof.}
Apply Corollary~\ref{COR:Rodl_applied_2} to $k$, $m-2$, and $\delta$ to get an integer $n$ and an $(m-2,k)$-sparse family $\mathscr{F} = \{F_i : i \in I\}$ of size $h$ such that \eqref{EQ:Rodl_new} holds.
Define a partition $\mathscr{I}=\{I_{x,y} : x,y \in X\} $ of $I$ such that (ii) holds. 
Finally, for any $i\in I$, define $G_i := F_i \cup \{x,y\}$ when $i \in I_{x,y}$. 
\qed

%\bigskip
%The last consequence of Rodl's result is the following:  

\begin{corollary}\label{COR:Rodl_partition_final}
Let $0 < \delta < 1$ and $\epsilon > 0$.
Then there are positive integers $n$, $m$, and $h$, a set $X$ of size $n$, an $(m,7)$-sparse family $\mathscr{G} = \{G_i : i \in I\}$ of $\vert I \vert = h$ subsets of $X$, and a partition $\mathscr{I}=\{I_{x,y} : x, y \in X\}$ of $I$ in sets having all the same size $\vert I_{x,y}\vert= h/n^2$ such that\vs\vs\vs 
\begin{itemize}
 	\item[\rm(i)] $\big(1 - (1- \delta)^{h/n^2}\big)^{n^2} > 1 - \epsilon\,$, and\vs\vs\vs 
 	\item[\rm(ii)] $i \in I_{x,y}$ implies $x, y \in G_i$ for any $i\in I$.
\end{itemize}
\end{corollary}

\noindent \textsc{Proof.}
Apply Corollary~\ref{COR:Rodl_partition} for $k:=5$.
\qed 

%\bigskip
%We introduce a last notion. 

\begin{definition}
	Two choice correspondences $c \colon \X \to \X$ and $c' \colon \X' \to \X'$, respectively having $X$ and $X'$ as ground sets, are \textsl{isomorphic}, denoted by $c \simeq c'$, if there is a bijection $\sigma \colon X \to X'$ (called an \textsl{isomorphism}) such that $\sigma(c(A)) = c'(\sigma(A))$ for any $A\in\X\,$, where $\sigma(A)$ is the set $\{\sigma(a): a \in A\}$. 
\end{definition}

%\begin{definition}
%Let $c \colon \X \to X$ be a choice function.
%For any $A \in X$, let $\mathscr{A}$ be the family of all nonempty subsets of $A$.
%The choice \textsl{induced by $c$ on $A$} is the function $c_{\upharpoonright A} \colon \mathscr{A} \to A$, defined by $c_{\upharpoonright A}(B) = c(B)$ for any $B \in \mathscr{A}$.
%\end{definition}
%
%We are ready to show that, when the size of the ground set goes to infinity, the fraction of choices satisfying any TFLH property tends to zero.
We are ready to prove Theorem~\ref{THM:tail_fail_weakly_hereditary_asymptotically_fails}. 
Let $\mathscr{P}$ be a TFLH property.
We shall show that $\mathscr{P}$ is asymptotically rare, that is, as the number of items in the ground set tends to infinity, the fraction of choices satisfying $\mathscr{P}$ tends to zero. 
Notation: if $c$ and $c'$ are choices on ground sets of the same size, then we write $c \approx c'$ to mean that $c$ and $c'$ are isomorphic if restricted to menus of cardinality at least $8$. %(that is, there exists a bijection $\sigma \colon X \to X'$ such that $\sigma(c(A)) = c'(\sigma(A))$ for any $A\in\X$ having size $\vert A\vert\geq 8$). 

Since $\mathscr{P}$ is tail-fail, there is a choice $c_0$ on a set of size $m \geq 8$ 
%which fails to have property P and 
such that, for any choice $c$ defined on a set of the same size $m$, if $c \approx c_0$, then $c$ does not satisfy $\mathscr{P}$.
%fails to have property P. 
%\\Since property P is invariant under $\approx$. So we will speak of a choice restricted to
%menus of size $\geq 8$ as satisfying or not satisfying P.
Let $\delta$ be the probability that a random choice $c$ on a set of size $m$ be such that $c\approx c_0$; thus, $0 < \delta < 1$. 

Fix $\epsilon > 0$. 
Apply Corollary \ref{COR:Rodl_partition_final} to get integers $n,m,h$, a set $X$ of size $n$, an $(m,7)$-sparse family $\mathscr{G} = \{G_i : i \in I\}$ of subsets of $X$ having maximum size $\vert I\vert = h$, and a partition $\mathscr{I} = \{I_{x,y} : x, y \in X\}$ of $I$ such that $\vert I_{x,y}\vert = h/n^2$ for any $ I_{x,y}\in\mathscr{I}$ with the properties that $i \in I_{x,y}$ implies $x, y \in G_i$, and $(1 - (1- \delta)^{h/n^2})^{n^2} > 1 - \epsilon$.

Let $c$ be a random choice on $X$.
%Let $c_0^* =0 | [Z]^{\geq 8}$.
For any $i\in I$, let $\mathscr{T}_i := [G_i]^{\geq 8}$ be the family of all subsets of $G_i$ of size at least $8$.
Note that $\mathscr{T}_i \cap \mathscr{T}_j = \es$ for any distinct $i,j\in I$.
We conclude that $c_{\upharpoonright G_i}$ are independent random variables as $i$ varies, as long as we only look at menus of size at least eight.\footnote{Here by $c_{\upharpoonright G_i}$ we denote the choice restricted to the family of all nonempty subsets of $G_i$.}
%
%It follows that $Pr(c\approx  c_{\upharpoonright  G_i},c\approx  c_{\upharpoonright  G_j})=Pr(c\approx  c_{\upharpoonright  G_i})Pr(c\approx c_{\upharpoonright  G_j})$ for any distinct $i,j\in I$. %\footnote{As a consequence $Pr(c\approx c_{\upharpoonright  B_i},c\approx c_{\upharpoonright  B_i})=Pr(c\approx c_{\upharpoonright  B_i})Pr(c\approx c_{\upharpoonright  B_j})$.}
Since $Pr\big(c_{\upharpoonright G_i} \not\approx c_0\big) = 1 - \delta$ for any $i\in I$, and all $c_{ \upharpoonright G_i}$'s are independent, we get\vs\vs\vs 
\[
Pr\!\left((\forall i \in I_{x,y})\;  c_{ \upharpoonright G_i} \not\approx c_0\right) \leq (1 - \delta)^{\frac{h}{n^2}}\vs\vs\vs
\] 
for all $x,y \in X$, hence\vs\vs\vs
\[
Pr\!\left((\exists i \in I_{x,y})\; c _{\upharpoonright G_i} \approx c_0\right) \geq 1 - (1 - \delta)^{\frac{h}{n^2}}\vs\vs\vs
\]
for all $x,y\in X$. 
We conclude\vs\vs\vs  
\[
Pr\!\left((\forall x,y \in X)(\exists i \in I_{x,y})\; c_{\upharpoonright G_i} \approx c_0\right) \geq \left(1 - (1 - \delta)^{\frac{h}{n^2}}\right)^{\!n^2} > 1 - \epsilon \vs\vs\vs
\] 
and so\vs\vs\vs 
\begin{equation} \label{EQ:final}
	Pr\!\left((\exists x,y \in X)(\forall i \in I_{x,y})\; c_{\upharpoonright G_i} \not\approx c_0\right)  < \epsilon.\vs	
\end{equation}

Now suppose $c$ satisfies $\mathscr{P}$. 
Since $\mathscr{P}$ is locally hereditary, there are $x,y\in X$ such that $c_{\upharpoonright Y}$ satisfies $\mathscr{P}$ for any $Y \subseteq X$ containing $x$ and $y$.
Thus, since $i \in I_{x,y}$ implies $x, y \in G_i$, we conclude that there are $x,y \in X$ such that $c_{\upharpoonright G_i}$ satisfies $\mathscr{P}$ for any $i\in I_{x,y}$, hence $c_{\upharpoonright G_i}\not\approx c_0$ for any $i\in I_{x,y}$. 
Thus, there are $x,y\in X$ such that $c_{\upharpoonright G_i} \not\approx c_0$ for any $i\in I_{x,y}$.
Now \eqref{EQ:final} yields that the probability that $c$ satisfies $\mathscr{P}$ is lower than $\epsilon$.  
By the arbitrariness of $\epsilon$, the proof of Theorem~\ref{THM:tail_fail_weakly_hereditary_asymptotically_fails} is complete.  

%%%%%%%%%%%%%%%%%%%%%%%%%%%%%%%%%%%%%%%%
%%%%%%%%%%%%%%%%%%%%%%%%%%%%%%%%%%%%%%%%
%%%%%%%%%%%%%%%%%%%%%%%%%%%%%%%%%%%%%%%%
%%%%%%%%%%%%%%%%%%%%%%%%%%%%%%%%%%%%%%%%

\section*{Appendix C: Independence}

Here we show that choice by linear salience is independent of some models of bounded rationality. 
Recall that $c\colon\X\to X$ satisfies \textsl{Always Chosen} when for any $A\in\X$, if $x\in A$ is such that $x=c(\{x,y\})$ for all $y\in A$, then $c(A)=x$.
 \cite{ManziniMariotti2007} show that Always Chosen is a necessary condition for the sequential rationalizability of a choice.
In particular, a choice is a \textsl{rational shortlist method} if it is sequentially rationalizable by two rationales.
Theorem~1 in \cite{ManziniMariotti2007} characterizes rational shortlist methods by the satisfaction of two properties, namely \textsl{Expansion Consistency} (called Axiom$\:\gamma$ by \citealp{Sen1971}) and \textsl{Weak WARP}.
Expansion Consistency requires that any item selected from two menus is also selected from their union.
Weak WARP says that for any $A,B \in X$ and $x,y \in X$ such that $\{x,y\}\subseteq A \subseteq B$, if $c(xy)=x=c(B)$, then $c(A)\neq y$.

\begin{example}[\it Independence of sequential rationalizability]
	The choice in Example~\ref{EX:Luce&Raiffa} is RLS, but not sequentially rationalizable, because Always Chosen fails. %(chicken is selected in the two menus of size two in which is present, but fails to be chosen in the whole ground set).   
Conversely, define on $X=~\lbrace w,x,y,z\rbrace$ a choice $c \colon \X \to X$ by\vs\vs\vs 
\[
\underline{w}xyz \,,\qquad \underline{w}xy \,,\; wx\underline{z} \,,\; w\underline{y}z \,,\; \underline{x}yz\,,\qquad \underline{w}x \,,\; w\underline{y} \,,\; w\underline{z}\,,\; \underline{x}y \,,\; x\underline{z} \,,\; \underline{y}z\,.\vs\vs\vs 
\]
%
%
%see Section~I.b in \cite{ManziniMariotti2007} for an example of a pathological behavior that cannot be explained by this model. 
%On the other hand, there are also some `reasonable' choices that fail to be sequentially rationalizable:  
%
%However, there configurations of choices violating $\alpha$ are not justified by this model.
%In particular the authors show that choices violating a property called `Always Chosen' are not sequentially rationalizable.
%Always Chosen requires that, for any $A\in\X$, if exists an $x\in A$ such that, for each $y\in A\setminus \lbrace x \rbrace$, $x=c(\{x,y\})$, then $c(A)=x$. 
%
%
%As one can see, the choice discussed in the Example \ref{LuceRaffaexample} violates Always Chosen and is rationalizable by salience. 
%
%\begin{equation}\label{choice function--Manzini e Mariotti}
%s\underline{y}z \,,\; \underline{t}xy \,,\; \underline{s}y\,,\; t\underline{x}\,,\;  \underline{y}z\,.
%y\underline{z}t\,,\; xy\underline{t}\,,\; s\underline{y}z \,,\; \underline{y}z\,,\; \underline{x}t\,,\; \underline{s}y\,.
%\end{equation}
%We consider a choice domain that lacks some subsets of $X$.
%This modification does not affect our analysis.
This choice is not RLS, because revealed salience is not asymmetric: indeed, we have $x \vDash y \vDash x$, since $(wy,wyx)$ and $(xz,xzy)$ are switches. 
It is easy to check that $c$ satisfies Axiom$\:\gamma$ and Weak WARP, and so it is a rational shortlist method.
\end{example}

A choice $c \colon \X \to X$ is \textsl{consistent with basic rationalization theory} if there are a choice correspondence $\psi \colon \X \to \X$ satisfying Axiom$\:\alpha$ and an asymmetric relation $\succ$ on $X$ such that $c(A) = \max(\psi(A),\succ)$ for all $A \in \X$.\footnote{If $\succ$ is a linear order, then $c$ is consistent with \textsl{order rationalization theory}.}
%Given a choice $c$ on $X$, \citet[p.\,779]{CherepanovFeddersenSandroni2013} call \textit{anomalous} a pair $\langle A,B \rangle$ of menus violating Axiom$\:\alpha$, that is, $A \subseteq B$ and $c(A) \neq c(B) \in A$. 
%The existence of anomalies reveals preferences: for any $x,y \in X$, let $x\,\textsf{Rev}\,y$ if there is an anomalous pair $\langle A,B\rangle$ such that $x=c(A)$ and $y=c(B)$. 
%Then $c$ satisfies \textsl{No Binary Chain Cycles} if $\textsf{Rev}$ is acyclic. 
Proposition~1 in \cite{CherepanovFeddersenSandroni2013} characterizes consistency with basic rationalization theory by the satisfaction of Weak WARP. 
The same axiom characterizes the model \textsl{categorize-then-choose} of \citep{ManziniMariotti2012}.
In this model, there are (1) an asymmetric relation $\succ^*$ on $\X$ (\textsl{shading} relation), and (2) an asymmetric complete relation $\succ$ on $X$ such that $c(A)=\max(\max(A,\succ^*),\succ)$ for all $A\in\X$, where $\max(A,\succ^*)=\bigcup\{\max(\mathscr{A},\succ^*)\}$.\footnote{Given a binary relation $\succ^*$ on $\mathscr{X}$ and a menu $A\in \X$, the set $\max(\mathscr{A},\succ^*)$ is the collection of (non-dominated) menus $\{B\subseteq A :  B'\succ B \;\text{for no}\; B'\subseteq A\}$.}

\begin{example}[\it Independence of basic rationalization theory and categorize-then-choose]
Any choice defined on $3$ items satisfies Weak WARP; however, the choice in Example~\ref{EX:acyclic_non-asymmetric_revealed_salience} is not RSL. 
Conversely, define on $X=\lbrace w,x,y,z\rbrace$ a choice $c \colon\X\to X$ by\vs\vs\vs
\[
wx\underline{y}z \,,\qquad wx\underline{y} \,,\; wx\underline{z} \,,\; w\underline{y}z \,,\; \underline{x}yz\,,\qquad \underline{w}x \,,\; w\underline{y} \,,\; w\underline{z}\,,\; x\underline{y} \,,\; \underline{x}z \,,\; y\underline{z}\,.\vs\vs\vs 
\] 
Weak WARP does not hold for $c$, because $wx\underline{y}z$, $\underline{x}yz$, and $x\underline{y}\,$.  
%Note that $z\,\textsf{Rev}\,y$ (because $y\underline{z}$ and $wx\underline{y}z$), $y\,\textsf{Rev}\,x$ (because $x\underline{y}$ and $\underline{x}yz$), and $x\,\textsf{Rev}\,z$ (because $\underline{x}z$ and $wx\underline{z}$).
%We conclude that $z\,\textsf{Rev}\,y\,\textsf{Rev}\,x\,\textsf{Rev}\,z$, and so $c$ is not consistent with order rationalization theory, because No Binary Chain Cycle fails.
However, $c$ is RLS. 
%as one may easily check. 
%: take $\langle \succsim,\L \rangle$, where $\succsim$ is  the total preorder on $X$ defined by $w\succ z\succ x$, $w\succ z \succ y$, and $x\sim y$, whereas $\L$ is the family $\{\rhd_w,\rhd_{x},\rhd_{y},\rhd_z\}$ of linear orders defined by $y\rhd_{w\!}z\rhd_{w\!}w\rhd_{w\!}x$, $\;y\rhd_{x\!}x\rhd_{x\!}w\rhd_{x\!}z$, and $w\rhd_{z\!}x\rhd_{z\!}z\rhd_{z\!}y$, with $\rhd_{y\!}=\rhd_{x}$.  
\end{example}

\vs

%%%%%%%%%%%%%%%%%%%%%%%%%%%%%%%%%%%%%%%%%%%%%%%%%%%%%%
%%%%%%%%%%%%%%%%%%%%%%%%%%%%%%%%%%%%%%%%%%%%%%%%%%%%%%
%%%%%%%%%%%%%%%%%%%%%%%%%%%%%%%%%%%%%%%%%%%%%%%%%%%%%%
%%%%%%%%%%%%%%%%%%%%%%%%%%%%%%%%%%%%%%%%%%%%%%%%%%%%%%

%%%%%%%%%%%%%%%%%%%%%%%%%%%%%%%%%%%%%%%%%%%%
%%%%%%%%%%%%%%%%%%%%%%%%%%%%%%%%%%%%%%%%%%%%
%%%%%%%%%%%%%%%%%%%%%%%%%%%%%%%%%%%%%%%%%%%%
%%%%%%%%%%%%%%%%%%%%%%%%%%%%%%%%%%%%%%%%%%%%

\end{document}